\documentclass[%
 aip,
 amsmath,amssymb,
 reprint,%
]{revtex4-1}

\usepackage[utf8]{inputenc}
\usepackage[T1]{fontenc}
\usepackage{mathptmx}
\usepackage{etoolbox}
\usepackage{dcolumn}
\usepackage{bm}
\usepackage{graphicx,xcolor}
\usepackage{amsfonts}
\usepackage{mathtools}
\usepackage{subfig}
\usepackage{float}
\usepackage{calrsfs}
\usepackage{multirow}

\DeclareMathAlphabet{\pazocal}{OMS}{zplm}{m}{n}

\newcommand{\be}[1]{\begin{equation}\label{#1}}
\newcommand{\ee}{\end{equation}}
\newcommand{\ba}[1]{\begin{eqnarray}\label{#1}}
\newcommand{\ea}{\end{eqnarray}}
\newcommand{\rf}[1]{(\ref{#1})}
\newcommand{\nn}{\nonumber}

\newcommand{\dd}{{\rm{d}}}

\newcolumntype{C}[1]{>{\centering\arraybackslash}p{#1}}

\def\mathrlap{\mathpalette\mathrlapinternal}

\def\mathrlapinternal#1#2{\rlap{$\mathsurround=0pt#1{#2}$}}

\makeatletter
\def\@email#1#2{%
 \endgroup
 \patchcmd{\titleblock@produce}
  {\frontmatter@RRAPformat}
  {\frontmatter@RRAPformat{\produce@RRAP{*#1\href{mailto:#2}{#2}}}\frontmatter@RRAPformat}
  {}{}
}%
\makeatother

\begin{document}

\preprint{AIP/123-QED}

\title[Radiation-induced instability of a finite-chord Nemtsov membrane]{Radiation-induced instability of a finite-chord Nemtsov membrane}

\author{Joris Labarbe}
\email{joris.LABARBE@univ-amu.fr}
\affiliation{Institut de Recherche sur les Phénomènes Hors \'Equilibre, UMR 7342, CNRS--Aix-Marseille Université, 49 rue F. Joliot Curie, 13384 Marseille, CEDEX 13, France}%

\author{Oleg N. Kirillov}
\email{oleg.kirillov@northumbria.ac.uk}
\affiliation{Northumbria University, Newcastle upon Tyne NE1 8ST, United Kingdom}%

\date{\today}
             
             
\begin{abstract}
    We consider a problem of stability of a membrane of an infinite span and a finite chord length, submerged in a uniform flow of finite depth with free surface. In the shallow water approximation, Nemtsov (1985) has shown that an infinite-chord membrane is susceptible to flutter instability due to excitation of long gravity waves on the free surface if the velocity of the flow exceeds the phase velocity of the waves and related this phenomenon with the anomalous Doppler effect. In the present work we derive a full nonlinear eigenvalue problem for an integro-differential equation for the finite-chord Nemtsov membrane in the finite-depth flow. In the shallow- and deep water limits we develop a perturbation theory in the small added mass ratio parameter acting as an effective dissipation, to find explicit analytical expressions for the frequencies and the growth rates of the membrane modes coupled to the surface waves. We find an intricate pattern of instability pockets in the parameter space and describe it analytically. The case of an arbitrary depth flow with free surface requires numerical solution of a new non-polynomial nonlinear eigenvalue problem. We propose an original approach combining methods of complex analysis and residue calculus, Galerkin discretization, Newton method and parallelization techniques implemented in MATLAB to produce high-accuracy stability diagrams within an unprecedentedly wide range of system's parameters. We believe that the Nemtsov membrane plays the same paradigmatic role for understanding radiation-induced instabilities as the  Lamb oscillator coupled to a string has played for understanding radiation damping.
\end{abstract}

\maketitle


\section{Introduction}

Exactly 120 years ago Lamb (1900) had proposed a model of a one-dimensional harmonic oscillator without damping constrained to move in the vertical direction and coupled to a horizontally taut semi-infinite elastic string \cite{L1900}. Quite surprisingly, he had found that the emission of traveling waves in the continuum by the oscillating mass contributes an effective Rayleigh damping correction term to the oscillator equation yielding decay of its vertical motion \cite{L1900}. In the course of time the radiating \emph{Lamb oscillator} became paradigmatic for understanding the \emph{radiation damping} in open and damped subsystems of closed conservative systems and gave rise to a number of
abstract models of dispersion of energy from a `small', usually finite-dimensional, subsystem to a `large', infinite-dimensional wave field \cite{L1904,BN1960a,U1964,Nu1972,C1989,BC1994,SW1999,S2001,FG2014,WW2015,OS2019,B2006}.

Remarkably, deep understanding of the radiation damping (including Lamb's model) involves Lax-Phillips scattering theory \cite{BN1960a,U1964,Nu1972,BC1994,B2006,A2017} and the concept of resonance, quasimode or metastable (Gamow) state in the context of open systems \cite{Nu1972,SW1999,S2001,WW2015,RG2017}. Resonant interaction of bound states (eigenfunctions) and radiation (continuous spectral modes), leading to energy transfer from the discrete to continuum modes, is a universal mechanism describing even asymptotic stability of solitary type solutions, when radiation going away from the solitons to infinity leaves  them to move freely \cite{SW1999,S2001,WW2015}.

To illustrate radiating vibratory motions in dimensions higher than one Love (1904) presented several extensions of Lamb's model that included decay of electromagnetic oscillations of a perfectly conducting spherical antenna due to emission of electromagnetic waves and decay of mechanical
vibrations of an elastic sphere emitting acoustic waves  \cite{L1904,BN1960a}. In \cite{U1964,BC1994} in the frame of the resonant
scattering theory an important question of interaction between the vibrational modes of a submerged solid and the scattering functions in the fluid for both light and heavy fluid loading has been discussed. It was established that modes of the fluid-solid system at zero fluid-loading can be identified with the solid whereas at infinite fluid-loading they correspond to scattering frequencies of the fluid alone with a soft
boundary condition. The intermediate values of the fluid-loading parameter appear to be the most complicated as it is impossible to label the mode of the fluid-solid system as corresponding solely to a `solid mode' or a `fluid mode' \cite{BC1994}.

In \cite{HBW2003, BHRW04} gyroscopic versions of the Lamb model were proposed, rather artificial however, such as the spherical pendulum and a rigid body with internal rotors, coupled either to the classical non-dispersive wave equation or to a dispersive equation of Klein-Gordon form. In such systems, the gyroscopic Lamb oscillator is susceptible to instabilities induced by wave emission (the \emph{radiation-induced instabilities}), to which such physically important effects belong as the famous Chandrasekhar-Friedman-Schutz (CFS) instability of rotating stars caused by emission of gravitational waves \cite{C1970,LD1977,S1980,C1984,A2003}, acoustic version of CFS instability \cite{B2017}, and the instability of vortices in a stratified rotating fluid due to emission of internal gravity waves \cite{LDB2009} as it happens, e.g., in the events of coalescence of lenticular vortices observed in recent experiments \cite{Legal2020}.

In a recent work \cite{LK2020} attention was paid to an overlooked classical model that appears to be a perfect candidate for the role of the Lamb oscillator in the field of radiation-induced instabilities. This is the model, proposed by Nemtsov in 1985, of an elastic membrane resting on the bottom of a uniformly flowing fluid layer of finite depth and loosing its stability due to emission of surface gravity waves \cite{N1985}. Nemtsov's membrane having infinitely many modes of free vibrations plays the role of a `small' subsystem, the fluid with the free surface is the `large' continuum supporting propagation of dispersive surface gravity waves, and the motion of the flow can contribute to a gyroscopic coupling \cite{P2013}.

We remark that scattering of surface gravity waves even by rigid horizontal submerged plates already has numerous applications in marine and coastal engineering such as submerged breakwaters or underwater wave lenses that allow exchange of water and hardly disturb horizontal currents \cite{P2015,IKG2019,WM2012}. The need for light, inexpensive and rapidly deployable wave barriers requires taking into consideration submerged horizontal flexible plates and membranes \cite{WM2012,CK1998}. Recent applications in energy harvesting exploit fluid-structure interaction, leading to the excitation (flutter) of an elastic plate or membrane, usually referred to as a flag \cite{A2008,SZ2011}, due to radiation of surface gravity waves when immersed in a moving flow with a free surface \cite{MM2020}. Another relevant setting comes from the problem of turbulent friction reduction in a boundary layer by using compliant coatings. In particular, it involves studying propagation of waves in a layer of a viscoelastic material of finite thickness when a layer of an ideal incompressible fluid is moving over it \cite{V2016}.

The scattering theory formalism is efficient for analytical derivation of such important quantities as the reflection and transmission coefficients and displacement of the free surface of the flow \cite{WM2012}. However, investigation of stability of a radiating object requires different methods. In a related set of problems on the stability of oscillations of moving wave emitters (e.g. radiation of elastic waves in rails by high-speed trains \cite{M1994} and emission of internal or surface gravity waves by a spherical body on an elastic spring moving parallel to the interface of two liquids \cite{GG1983,AMN1986}), the theory of Cherenkov radiation for structureless particles \cite{FG2014,G1996} and its extension by Ginzburg and Frank \cite{GF1947} to the particles having internal degrees of freedom, provides important clues both for derivation of necessary and sufficient criteria for instability and for better understanding radiation-induced instabilities in the general physical context \cite{N1976}.

Originally, Cherenkov radiation has been the name for the phenomenon that a charged particle, moving relativistically through a dielectric non-dispersive medium at constant speed $v$ higher than the phase velocity of light $v_p<c$ in the medium, becomes a source of electromagnetic radiation  \cite{FG2014,Tamm1960,BS1998,CR2013,IDE2018}. If the source has a natural frequency $\omega_0$ in its own static frame, then in the observer static frame one receives the far-field angular distribution of radiation with frequency $\omega=\omega_0\gamma^{-1}/(1-vv_p^{-1}\cos \theta)$, where $\gamma$ is the Lorentz factor, that turns out to be concentrated in the forward direction on a conical surface making an
angle $\theta$ with the velocity vector \cite{CR2013,IDE2018}. The angle $\theta_{ch}=\arccos(v_p/v)$, which is possible only if $\omega_0=0$, defines the \emph{Cherenkov cone} with the angular aperture $2\theta_{ch}$, i.e. a locus in the space of wavenumbers of resonant modes into which the Cherenkov radiation from the structureless particle occurs \cite{Tamm1960,CR2013,IDE2018}. In this case, the radiated electromagnetic field is spatially concentrated on a wave front forming a Mach cone with the angular aperture $\pi- 2\theta_{ch}$ behind the source, which is the apex of the cone \cite{G1996,CR2013}.

A source oscillating with the natural frequency $\omega_0>0$ and moving at a velocity $v<v_p$ or at a superluminal velocity $v>v_p$ but emitting outside the Cherenkov cone, i.e. under the condition $\cos\theta<v_p/v$, experiences conventional Doppler effect with the increase in $\omega>0$ while approaching the observer \cite{CR2013,IDE2018}. If the superluminal oscillator emits inside the Cherenkov cone then the condition $\cos\theta>v_p/v$ implies $\omega<0$, which means that the source becomes excited by passing from a lower energy level to an upper one during
the emission process \cite{G1996,GF1947}. That is, the kinetic energy of the source supplies both the energy of the emitted photon and the positive
increase in the internal energy of the source \cite{Tamm1960,IDE2018}. Radiation inside the Cherenkov cone is known as the
\emph{anomalous Doppler effect} \cite{G1996,GF1947,N1976,BS1998}. Recent work \cite{IDE2018} further distinguishes between the range $v_p/v<\cos\theta<2v_p/v$ and the range $\cos\theta>2v_p/v$ inside the Cherenkov cone as resulting in the superlight (superluminal) normal and inverse Doppler shift, respectively. It turns out that the major part of the change in the kinetic energy of the source contributes to a positive increase in the internal energy of the source (photon) for the superlight inverse (normal) Doppler effect \cite{IDE2018}.

The same basic process of generalized Cherenkov emission is characteristic of all the emission processes that take place when a uniformly moving source is coupled to some excitation field, even in the presence of dispersion that provides an individual Cherenkov cone for each frequency \cite{G1996}: as soon as the source velocity exceeds the phase velocity of some mode of the field, the latter becomes continuously excited \cite{CR2013}. For the sources having internal degrees of freedom, this serves also as a necessary condition for the presence in the space of parameters of a domain of instability of the source due to the anomalous Doppler effect \cite{M1994}. A sufficient condition for the radiation-induced instability is the prevalence of reaction of waves emitted inside the Cherenkov cone over those emitted outside it \cite{M1994}. Therefore, it is not surprising that already Tamm in his Nobel lecture foresaw application of the anomalous Doppler effect to describe ``self-excitation of some particular modes of vibrations of a supersonic airplane'' \cite{G1996,Tamm1960}.

In \cite{N1985} Nemtsov considered stability of the membrane under the surface of the uniform flow in the limits of (i) shallow water and (ii) vanishing added mass ratio \cite{M1998} that measures coupling between the membrane and the flow and serves as an effective damping parameter.
Since the surface gravity waves are \emph{non-dispersive} in the shallow water approximation, the formalism of Cherenkov radiation and anomalous Doppler effect applied to the infinite-chord-length Nemtsov membrane predicts its instability in the range where the velocity of the flow is exceeding the phase velocity of the surface gravity waves (as a consequence, they appear to be traveling backward in the frame moving with the flow) and exactly when the phase of the induced surface gravity wave is equal to the phase of the elastic wave propagating in the membrane \cite{N1985}. However, Nemtsov's shallow water result, being effective in uncovering fundamental physical reasons for the membrane destabilization, could not answer to a question of practical importance, namely, what is the domain of instability when the parameters of the system are allowed to take arbitrary values?

In \cite{LK2020} we extended analysis of Nemtsov's membrane with infinite chord to the case of \emph{dispersive} surface gravity waves by allowing the fluid layer to have arbitrary depth and the added mass ratio parameter to take arbitrary non-negative values. New complete dispersion relation has been derived and analyzed with the perturbation theory for multiple roots of polynomials \cite{K2013dg} to obtain an explicit analytical approximation to the critical flutter velocity that is in excellent agreement with the numerical computation of the full stability map. Moreover, we have identified in \cite{LK2020} a new instability domain arising from a conical singularity in the parameter space that could not be detected in the restrictive assumptions of  \cite{N1985}. This new domain is associated with a low-frequency flutter for short wavelengths and corresponds to the case when the velocity of propagation of elastic waves in the membrane is much smaller than the velocity of the flow. Finally, an elegant and applicable explicit expression for the total averaged energy has been derived by means of the direct integration and its reduction to the Cairns form involving the derivatives of the dispersion relation with respect to the frequency of oscillations has been proven. It was demonstrated that the radiation-induced instability of the membrane is the result of collision of modes of positive and negative energy and can be interpreted in terms of wave emission in the domain of the anomalous Doppler effect \cite{LK2020}.

In the present work, we consider a model of Nemtsov's membrane in its entirety, with a flow of finite depth and a membrane of finite chord. We formulate the dimensionless boundary value problem for this system and by means of Fourier analysis recover an explicit expression for the velocity potential in the form of an improper integral. Determining the domain of dependence from the Cherenkov condition written for dispersive surface gravity waves, we extract an integro-differential equation for the membrane displacement in the presence of the flow, which is our main object of investigation.

Following \cite{N1985} we first present a rigorous general treatment of the integro-differential equation in the shallow-water limit by means of the Laplace transform and complex analysis. We obtain an integral eigenvalue relation and develop a systematic procedure for its analysis based on the perturbation theory with respect to the small added mass ratio parameter and residue calculus. As a result, we find explicit expressions for the frequencies and growth rates of membrane's modes coupled to the free surface as a series in the small parameter. Analysing the first-order approximation we can treat membrane's destabilization as a classical dissipation-induced instability \cite{K2013dg,KV2010,MK91,H1992,BKMR94}. Plotting neutral stability curves in the plane of velocity of the
flow versus the speed of propagation of elastic waves along the
membrane we uncover a new and intricate pattern of self-intersecting instability pockets reminding a similar phenomenon characteristic of some forms of the Hill equation \cite{BL1995} and describe it explicitly in an analytical form. We derive an integral expression for the free surface of the flow which allows us to find and explore the fluid dynamical analogue to the superlight normal and inverse Doppler effects \cite{IDE2018}.

The case of the finite depth of the fluid layer requires numerical solution of the boundary eigenvalue problem for the original integro-differential
equation. This investigation involves a thorough treatment of the improper integral using complex analysis with the subsequent Galerkin decomposition of the solution to generate an algebraic nonlinear and non-polynomial in the eigenvalue parameter eigenvalue problem. The non-polynomial dependence on the eigenvalue parameter arises in the coupling term with the added mass ratio parameter as a factor. Setting the latter parameter to zero, we obtain a standard quadratic eigenvalue problem determining the free membrane modes. Once taken into consideration, the coupling term `turns on' the radiative instability mechanism tending to excite the flutter of the membrane.

We notice that non-polynomial eigenvalue problems frequently occur in the studies of fluid-structure interactions, see e.g. \cite{A2008,V2012}, and are notoriously hard to solve even numerically. The methods for their solution are a hot topic in modern numerical mathematics and linear algebra communities, see e.g. \cite{MV2004,BH2015,GT2017,MOM2020}, where a broad range of approaches is discussed. Most of the methods presented are either based on Newton-Raphson iterative process or on contour integration and we are restricting ourselves to the former.

In order to reach an acceptable
convergence rate of the Newton method, we derive the Jacobian in an analytical form using residue calculus and complex analysis instead of approximating it numerically. Since the domain of integration in our problem is one-dimensional and we intend to keep high accuracy of our numerical scheme, we use the Legendre-Gauss-Lobatto quadrature rule to approximate the integrals in the Galerkin discretization. Nodes and weights of this spectral collocation method are recovered using the Golub-Welsch algorithm \cite{GW1969}, which is based on the inversion of a linear system obtained from the three-term recurrence relation for Legendre polynomials. The computed eigenvalues correspond to the quadrature points, while the eigenvectors are used to recover the weights. This spectral quadrature is able to reach computer accuracy with less than 20 nodes of discretization and is used all over our code that has been fully implemented and parallelized in MATLAB using the Parallel Computing Toolbox available from the software and run on the High Performance Cluster at Northumbria University.

To the best of our knowledge, the approach developed in our work is original and making use of it, we are able to recover the eigenfrequencies of the complete system and hence, to perform an exhaustive stability analysis of the finite-chord Nemtsov membrane.
From the numerically found growth rates we recover stability maps for the finite-chord membrane in the finite depth layer and compare with the shallow water approximation. We show that in the limit of infinite chord length the neutral stability boundaries perfectly correspond to the shallow water boundaries found in \cite{LK2020}. We establish that there is a critical chord length such that the shorter membranes cannot be destabilized. The most intriguing finding is, however, the chains of intertwining instability pockets, which
our method is able to resolve, thus confirming its excellent convergence and accuracy. We believe that our procedure is applicable to a broad class of fluid-structure interaction problems that require solving nonlinear eigenvalue problems.

\section{Mathematical formulation}

Following \cite{LK2020}, in a Cartesian coordinate system $OXYZ$, we consider an inextensible elastic rectangular membrane strip of constant thickness $h$, density $\rho_m$, and tension $T$ along the membrane chord in the $X$-direction. The membrane has infinite span in the $Y$-direction and is held at $Z=0$ at the leading edge $(X=0)$ and at the trailing edge $(X=L)$ by simple supports.

The membrane is initially still and flat, immersed in a layer of inviscid, incompressible fluid of constant density $\rho$, with free surface at the height $Z=H$. The two-dimensional flow in the layer is supposed to
be irrotational and moving steadily with velocity $v$ in the positive $X$-direction. Therefore, the system is solved under the potential theory as it is the case in the original paper by Nemtsov \cite{N1985}. The bottom of the fluid layer at $Z=0$ is supposed to be rigid
and flat for $X\in (-\infty,0] \cup [L,+\infty)$.

In contrast to Nemtsov \cite{N1985} who assumed that vacuum exists below the membrane, we suppose that a motionless incompressible medium of the same density $\rho$ is present below the membrane with a pressure that is the same as the unperturbed pressure of the fluid \cite{LK2020}. The system is in a uniform gravity field acting in the negative $Z$-direction with $g$ standing for the gravity acceleration.

Let $w(X,t)$, where $t$ is time, be a small vertical displacement of the membrane, $u(X,t)$ the free surface elevation, and $\varphi(X,Z,t)$ the potential of the fluid.

Following \cite{LK2020} we choose the height of the fluid layer, $H$, as a length scale, and $\omega_0^{-1}$, where $\omega_0=\sqrt{g/H}$, as a time scale to introduce the dimensionless time and coordinates
\be{dlp1}
\tau=t\omega_0,\quad x=\frac{X}{H}, \quad y=\frac{Y}{H}, \quad z=\frac{Z}{H},
\ee
the dimensionless variables
\be{dlp2}
\xi=\frac{w}{H},\quad \eta=\frac{u}{H},\quad \phi=\frac{\omega_0}{gH}\varphi,
\ee
the dimensionless parameters of the added mass ratio \citep{M1998} and membrane chord length
\be{dlp3}
\alpha=\frac{\rho H}{\rho_m h},\quad \Gamma=\frac{L}{H},
\ee
and two dimensionless numbers
\be{dlp4}
M_w=\frac{c}{\sqrt{g H}}, \quad M=\frac{v}{\sqrt{g H}},
\ee
where $c^2=T/(\rho_m h)$ is the squared speed of propagation of elastic waves in the membrane
and $\sqrt{g H}$ is the speed of propagation of long surface gravity waves in the shallow water approximation.
The chosen scale for the velocity explains our choice of notation in \rf{dlp4} because the Froude number $v/\sqrt{g H}$ can be treated as a Mach number in the non-dispersive shallow water limit which simplifies comparison of our results with that of the supersonic aerodynamics \cite{V2012}.

Denoting the fluid domain by $\Omega$, the free surface, membrane, and rigid wall borders by $\partial\Omega_0$, $\partial\Omega_1$, and $\partial\Omega_2$, respectively, and assuming the time dependence for the velocity potential $\phi$ and the membrane displacement $\xi$ in the form of $e^{-i\omega\tau}$ we arrive at the dimensionless boundary value problem \cite{LK2020}
\begin{subequations}
\label{eom}
\begin{align}
\bm{\nabla}^2\phi &= 0, & &\text{in} \, \Omega \label{eoma} \\
\bm{\nabla}\phi\cdot\bm{n} + \left( -i\omega + M\partial_x \right)^2\phi &= 0, & &\text{on} \, \partial\Omega_0 \label{eomb}\\
\bm{\nabla}\phi\cdot\bm{n} + ( -i\omega + M\partial_x )\xi &= 0, & &\text{on} \, \partial\Omega_1 \label{eomc}\\
\bm{\nabla}\phi\cdot\bm{n} &= 0, & &\text{on} \, \partial\Omega_2 \label{eomd}\\
\omega^2 \xi + M_w^2\partial_x^2 \xi + \alpha\left(-i\omega + M\partial_x \right)\phi &= 0, & &\text{on} \, \partial\Omega_1 \label{eome}\\
\xi(0) = \xi(\Gamma) &= 0, & &\text{on} \, \partial\Omega_1 \label{eomf}
\end{align}
\end{subequations}
where $\bm{n}$ is the vector of the outward normal to a surface and, for simplicity, we retain the same notation for the membrane displacement and the fluid potential after the separation of time.

As one can notice, the Laplace equation \rf{eoma} is supplemented with a combination of dynamic and kinematic free surface conditions \rf{eomb} and the impermeability conditions for the membrane \rf{eomc} and the walls \rf{eomd}. The nonhomogeneous wave equation \rf{eome} describes the physics along the membrane and is solved with the rigid boundary conditions \rf{eomf}. This set of equations represents a boundary eigenvalue problem for the complex eigenfrequency $\omega$ and will be used for the stability analysis of the finite-chord Nemtsov membrane.

\subsection{Velocity potential via inverse Fourier transform}

As in the previous study \cite{LK2020}, since the fluid layer is assumed to have an infinite extension in the $x$-direction, we can
write, respectively, the Fourier transform of the velocity potential $\phi$ and its inverse
\begin{align}
\label{fourier}
&\hat\phi(\kappa,z,\omega) = \int_{-\infty}^{+\infty} \phi(x,z,\omega) e^{-i\kappa x} \dd x, \nn \\
&\phi(x,z,\omega) = \frac{1}{2\pi}\int_{-\infty}^{+\infty} \hat{\phi}(\kappa,z,\omega) e^{i\kappa x} \dd\kappa,
\end{align}
where $\kappa$ is the dimensionless wavenumber.

With \rf{fourier} taken into account, the Laplace problem (\ref{eoma}--\ref{eomc}) yields the boundary value problem for $\hat{\phi}$ in the Fourier space
\begin{subequations}
\label{bvpf}
\begin{align}
\partial_z^2\hat{\phi} - \kappa^2\hat{\phi} &= 0, &&\text{in} \, \Omega \label{bvpfa} \\
\partial_z\hat \phi - \left( \omega -\kappa M \right)^2\hat\phi &= 0, &&\text{on} \, \partial\Omega_0 \label{bvpfb} \\
\partial_z\hat{\phi} + \int_{0}^{\Gamma} \left( i\omega - M\partial_{x'}\right) \xi(x') e^{-i\kappa x'} \dd x' &= 0, &&\text{on} \, \partial\Omega_1 \label{bvpfc}
\end{align}
\end{subequations}
where $x'$ is a curvilinear abscissa along the membrane. Expression \rf{bvpfc} is the Fourier transform of the impermeability condition \rf{eomc}.

The general solution to equation \rf{bvpfa} is known to be
\be{gph}
\hat{\phi}(\kappa,z,\omega) = A(\kappa,\omega)e^{\kappa z} + B(\kappa,\omega)e^{-\kappa z},
\ee
where the functions $A$ and $B$ are to be determined from the boundary conditions \rf{bvpfb} and \rf{bvpfc}. This yields an expression for $\hat{\phi}$ evaluated at $z=0$
\begin{align}
\label{phihat}
\hat{\phi}(\kappa,0,\omega) = &\frac{\kappa - (\omega - \kappa M)^2 \tanh{\kappa}}{\kappa[(\omega - \kappa M)^2 - \kappa\tanh{\kappa}]} \times \nn \\ 
&\int_{0}^{\Gamma} \left( -i\omega + M\partial_{x'}\right)\xi(x') e^{-i\kappa x'}  \dd x' .
\end{align}

Returning to the physical space by means of the inverse Fourier transform \rf{fourier} we recover an explicit form for the potential disturbance
\begin{align}
\label{phi}
\phi(x,0,\omega) =  &\frac{1}{2\pi} \int_{0}^{\Gamma} \left( -i\omega + M\partial_{x'}\right)\xi(x') \times \nn \\ 
&\int_{-\infty}^{+\infty} \frac{\kappa - (\omega - \kappa M)^2 \tanh{\kappa}}{\kappa[(\omega - \kappa M)^2 - \kappa\tanh{\kappa}]} e^{i\kappa(x-x')} \dd\kappa \, \dd x' .
\end{align}

\subsection{Integro-differential equation for membrane's deflection}

\begin{figure*}
\centering

\subfloat[]{
\includegraphics*[width=.5\textwidth]{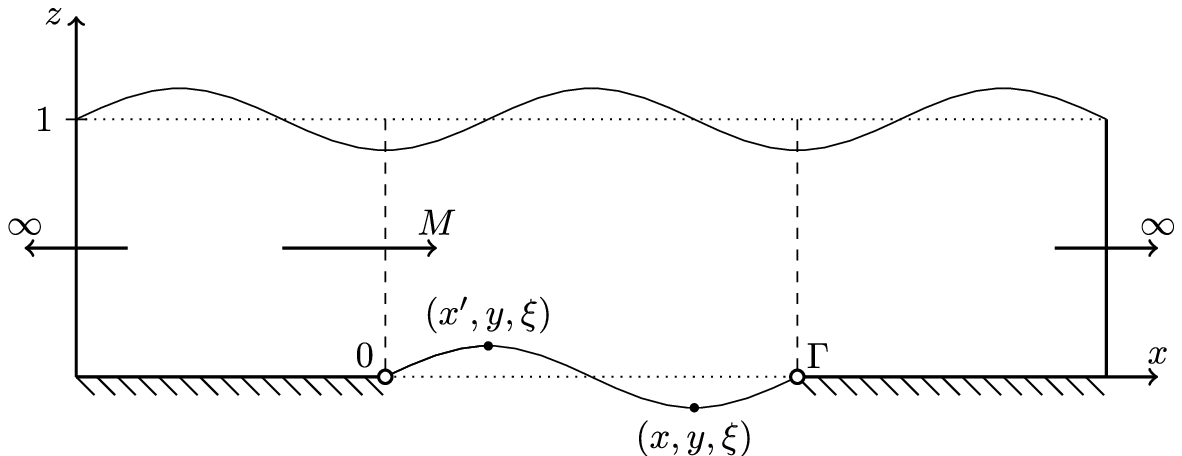}} 
\subfloat[]{
\includegraphics*[width=.5\textwidth]{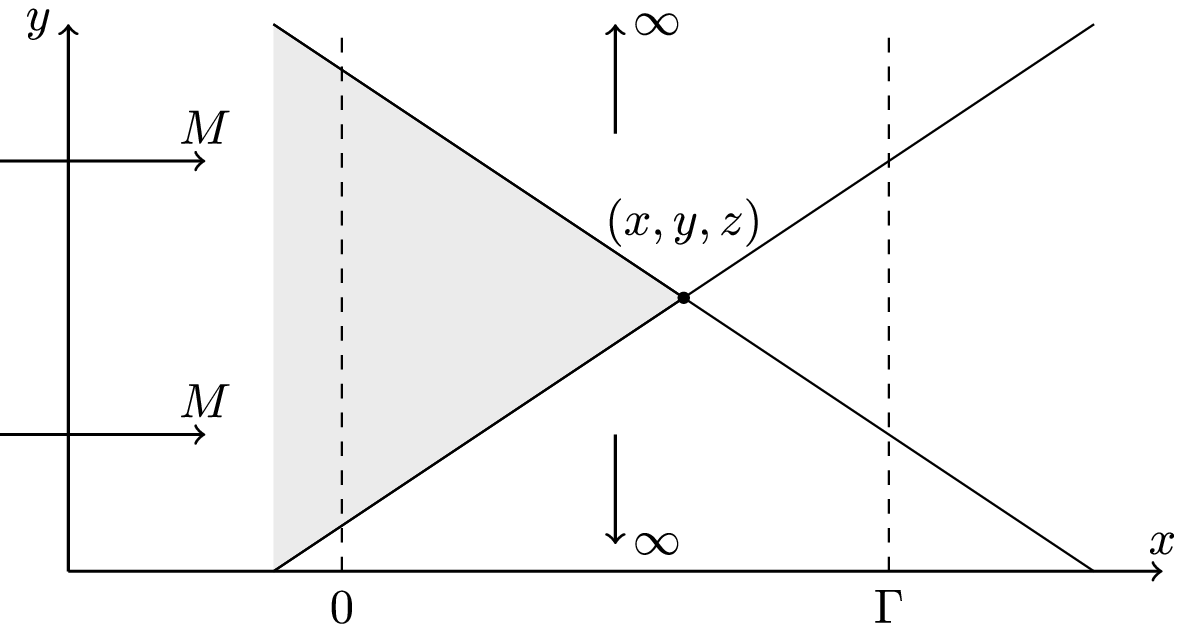}}

\caption{(a) Sketch of the Nemtsov system in the $(x,z)$-plane. (b) View of the system from above in the $(x,y)$-plane. A point obstacle at a position $(x,y,\xi)$ in a uniform flow moving in the positive $x$-direction with a velocity $M > \sigma(\kappa)$ affects only the pattern of the surface gravity waves with the wavenumber $\kappa$ in the (white) downstream Mach cone representing the domain of influence \cite{D2014,S1960,O2014}. The (gray) upstream half of the Mach cone is the domain of dependence \cite{D2014,S1960,O2014} at the point $(x,y,z)$.\label{fig:mach_cone}}
\end{figure*}

Surface gravity waves on a non-moving finite-depth layer are in general dispersive with the phase speed $\sigma(\kappa)=\sqrt{\kappa\tanh{\kappa}}/\kappa$, which is varying between $\sigma_{DW}=0$ when $\kappa\to +\infty$ (deep water) and $\sigma_{SW}=1$ when $\kappa\to 0$ (shallow water) \cite{LK2020}. Suppose we have a uniform flow in the positive $x$-direction with the supercritical speed $M>\sigma(\kappa)$ as shown in Fig.\ref{fig:mach_cone}.
Perturbation with the wavenumber $\kappa$ of the flow surface from a point $(x,y,\xi)$ of the membrane will spread from that point along concentric circles in the $(x,y)$ plane with the phase speed $\sigma(\kappa)$, which are washed downstream as $M>\sigma(\kappa)$ \cite{D2014,S1960}. The point obstacle therefore affects only the flow pattern in the conical domain of influence \cite{O2014}. The boundary of this downstream half of the Mach cone is the envelope of the moving and expanding circles centred at the points with the horizontal coordinates $x'\ge x$ \cite{D2014,S1960}.
The presence of the point obstacle at $x$ does not make itself felt at the points upstream $(x'<x)$ and outside of the half-cone \cite{D2014,S1960}.

On the other hand, the flow at the point $(x,y,z)$ depends only on the flow in the upstream half of the same Mach cone (domain of dependence \cite{O2014}), see Fig.~\ref{fig:mach_cone}(b) \cite{D2014,S1960}. Therefore, for $M>\sigma(\kappa)$ the perturbed potential \rf{phi} along the membrane with the finite chord length $\Gamma$ at a given $y$ is composed of all the single sources with coordinates $(x',\xi)$ where $x'\in [0,x]$ as shown in Fig.~\ref{fig:mach_cone}(a). Hence, the interval of integration in the first integral in \rf{phi} can be truncated from $x'\in[0,\Gamma]$ to $x'\in[0,x]$ to match the domain of dependence. Inserting the modified in this manner expression \rf{phi} into \rf{eome} yields the following integro-differential equation for the membrane displacement $\xi$
\begin{align}
\label{ide}
\omega^2 \xi + &M_w^2\partial_x^2 \xi = \frac{\alpha}{2\pi} \left(i\omega - M\partial_x \right) \int_{0}^{x} \left( -i\omega + M\partial_{x'}\right)\xi(x') \times \nn \\ 
&\int_{-\infty}^{+\infty} \frac{\kappa - (\omega - \kappa M)^2 \tanh{\kappa}}{\kappa[(\omega - \kappa M)^2 - \kappa\tanh{\kappa}]} e^{i\kappa(x-x')} \dd\kappa \, \dd x' .
\end{align}

The equation \rf{ide} has to be solved with respect to the eigenfrequency $\omega$ as a nonlinear eigenvalue problem, which is generally a highly challenging task due to the presence of an improper integral. By this reason, in the next section we begin the analysis of \rf{ide} in the shallow water limit that will allow us to apply perturbation theory and derive explicit approximation of the flutter domain. With the guidance provided  by the analytical solutions in the shallow- and deep-water limits we finally present a numerical method that eventually results in the solution of the full problem.

\section{Shallow water analysis of the finite-chord Nemtsov membrane}

The goal of this section is to extend the result of \cite{N1985} by presenting a rigorous and general treatment of expression \rf{ide} in the shallow water approximation and provide new physical interpretation of the instability mechanism for a membrane of the finite chord in the finite-depth flow.

\subsection{Velocity potential in the shallow water limit}

Introducing the phase speed $\sigma=\omega/\kappa$ and  re-writing the factor at $e^{i\kappa(x-x')}$ in the integrand of the improper integral in \rf{ide} as
\begin{align}
\label{pots}
 \frac{1}{\kappa^2}\frac{1 - \kappa(\sigma -  M)^2 \tanh{\kappa}}{[(\sigma -  M)^2 - \frac{\tanh{\kappa}}{\kappa}]} &= \frac{1}{\kappa^2}\left[\frac{1}{(\sigma -  M)^2 - 1}+O(\kappa^2)\right] \nn \\
 &= \frac{1}{(\omega -  \kappa M)^2 - \kappa^2}+O(1)
\end{align}
we find that in the long-wavelength (shallow-water) limit, $\kappa \rightarrow 0$, the velocity potential simplifies
\be{pot_lwl}
\phi_{SW}(x) = \frac{1}{2\pi} \int_{0}^{x} \left( -i\omega + M\partial_{x'}\right)\xi(x') \int_{-\infty}^{+\infty} \frac{e^{i\kappa(x-x')} \dd\kappa}{(\omega - \kappa M)^2 - \kappa^2} \, \dd x'
\ee
and, after factorizing the denominator in the integrand, it can be further expressed in the equivalent form
\be{imp_int_lwl}
\int_{-\infty}^{+\infty} \frac{e^{i\kappa(x-x')} \dd\kappa}{(\omega - \kappa M)^2 - \kappa^2} = \frac{1}{M^2 - 1} \int_{-\infty}^{+\infty} \frac{e^{i\kappa(x-x')}}{(\kappa - p_1)(\kappa - p_2)} \dd\kappa,
\ee
where the pole $p_1={\omega}/(M+ 1)$ corresponds to the wave travelling forward along the membrane and $p_2={\omega}/(M- 1)$ to the wave travelling backward.

\subsection{Explicit form of the velocity potential by means of residue calculus}

For $M>1$ the denominator in the expressions for the poles $p_{1,2}(\omega)$ always remains real and positive, only the (complex, in general) frequencies $\omega$ define the location of the poles in the complex $\kappa$-plane. In this configuration, we shall focus on the frequencies with positive imaginary parts ($\text{Im}(\omega)>0$) to define a contour in the upper-half plane and integrate the expression \rf{imp_int_lwl}.

We define the contour of integration $\mathcal{C}$ as a semi-circular and positively oriented curve, of radius $R$, closed with a segment along the real axis as follows
\be{contour}
\mathcal{C} = [-R , R] \cup \Delta_R, \quad \Delta_R= \{ Re^{it},0 \leq t \leq \pi \}.
\ee
Applying the Cauchy residue theorem around the contour \rf{contour} yields \cite{AF2003}
\be{cont_int_lwl}
\oint\limits_{\mathcal{C}} F(z) \dd z = \int_{-R}^{R} F(z) \dd z + \int_{\Delta_R} F(z) \dd z = 2\pi i \sum_{j=1}^2 \text{res} \left( F(z) , p_j \right),
\ee
where $F(z) = f(z)e^{iz(x-x')}$ and $f(z)=[(\omega - zM)^2 - z^2]^{-1}$.

According to Jordan's lemma, since the function $f(z)$ is continuous for any $z\in\mathcal{C}$, except at the poles $p_{1,2}(\omega)$, and that we have $\lim_{R\to +\infty} \vert f(Re^{i\theta}) \vert = 0$ for $\theta\in [0,\pi]$, the integral over $\Delta_R$ vanishes as we enlarge the radius. Therefore, the improper integral \rf{imp_int_lwl} reduces to \cite{AF2003}
\be{crt}
\lim\limits_{R\to +\infty} \int_{-R}^{R} F(\kappa) \dd\kappa = \int_{-\infty}^{+\infty} F(\kappa) \dd\kappa = 2\pi i \sum_{j=1}^2 \text{res}(F(\kappa),p_j).
\ee

In general, the residue $\text{res}(g(z),p)$ of a meromorphic function $g(z)$ having a simple pole $p$ and a factorized denominator for this pole, can be found as $\text{res}(g(z),p)=\lim_{z \rightarrow p}(z-p)g(z)$. Hence, for $p_1=\omega/(M+1)$, $p_2=\omega/(M-1)$ and $F(\kappa)$ in the form of \rf{imp_int_lwl}, we have \cite{AF2003}
\ba{res_lwl}
\text{res}(F(\kappa),p_1)&=&\lim_{\kappa \rightarrow p_1}\frac{1}{(M^2 - 1)} \frac{e^{i\kappa(x-x')}}{(\kappa - p_2)} \nn \\
&=&\frac{1}{(M^2 - 1)} \frac{e^{ip_1(x-x')}}{(p_1 - p_2)}=\frac{-e^{ip_1(x-x')}}{2\omega},\nn\\
\text{res}(F(\kappa),p_2)&=&\lim_{\kappa \rightarrow p_2}\frac{1}{(M^2 - 1)} \frac{e^{i\kappa(x-x')}}{(\kappa - p_1)} \nn \\
&=&
\frac{1}{(M^2 - 1)} \frac{e^{ip_2(x-x')}}{(p_2 - p_1)}=\frac{e^{ip_2(x-x')}}{2\omega}.
\ea
With the residues \rf{res_lwl} in expression \rf{crt}, the velocity potential \rf{pot_lwl} takes an explicit form
\ba{pot_lwl2}
\phi_{SW}(x) &=& \int_{0}^{x} \left( \omega + iM\partial_{x'}\right)\xi(x') \sum_{j=1}^2 \text{res}(F(\kappa),p_j) \dd x', \nn \\
&=& \frac{1}{2\omega} \int_{0}^{x} \left( \omega + iM\partial_{x'}\right)\xi(x') \left[ e^{ip_2(x-x')} - e^{ip_1(x-x')} \right] \dd x',\nn\\
&=& \frac{-i}{2\omega} \int_{0}^{x} V(x') \left[ e^{ip_1(x-x')} - e^{ip_2(x-x')} \right] \dd x',
\ea
which reproduces the result by Nemtsov, if we denote $V(x)=(-i\omega+M\partial_x)\xi(x)$ \cite{N1985}. Note that the term $e^{ip_1(x-x')} $ corresponds to the normal Doppler effect due to emission of surface gravity waves of positive energy whereas the term $e^{ip_2(x-x')}$ to the anomalous Doppler effect due to emission of surface gravity waves of negative energy.

\subsection{Explicit expression for the membrane displacement by means of Laplace transform}

Substituting solution \rf{pot_lwl2} with $u(x)=-iV(x)$ and $v(x)= e^{ip_2x} - e^{ip_1x}$ into the problem \rf{ide} yields
\be{lwl_eq0}
\omega^2\xi(x) + M_w^2\partial^2_x \xi(x) = - \frac{\alpha}{2\omega}\left( -i\omega + M\partial_x \right)\int_{0}^{x} u(x') v(x-x') \, \dd x',
\ee
which can be written as
\be{lwl_eq}
\omega^2\xi(x) + M_w^2\partial^2_x \xi (x) + \frac{\alpha}{2\omega}\left( -i\omega + M\partial_x \right)\left( u * v \right)(x) = 0,
\ee
where the symbol $*$ denotes the operator of convolution of two functions supported on the interval $[0,\infty)$.

Equation \rf{lwl_eq} supplemented with the boundary conditions \rf{eome} for the displacement $\xi(x)$ is suitable to solve by the Laplace method. We recall that in the general case, the Laplace transform $\mathcal{L}$ of an arbitrary function $g(x)$ is given as follows
\be{lt}
\bar{g}(s) \equiv \mathcal{L}\left[ g(x) \right] = \int_{0}^{+\infty} g(x) e^{-sx} \dd x, \quad s\in\mathbb{C}.
\ee

Also recall that if $h(x)$ is a convolution
$$
h(x)=(u*v)(x)=\int_0^x u(x')v(x-x') dx' ,
$$
then
\be{lconv}
\mathcal{L}[h(x)]=\bar{u}(s)\bar{v}(s).
\ee

Using Leibniz integral rule
\begin{align}
\label{leib}
\frac{d}{dx} h(x) &=\frac{d}{dx}\int_0^x u(x')v(x-x') dx' \nn \\
&=v(0)u(x)+\int_0^x\frac{d v(x-x')}{dx}u(x')dx' ,
\end{align}
we find the Laplace transform of the derivative of the convolution to be
\be{dconv}
\mathcal{L}[dh/dx]=\bar{u}(s)v(0)+\bar{u}(s)(s\bar{v}(s)-v(0))=s\bar{u}(s)\bar{v}(s).
\ee

Following the definition in \rf{lt}, applying the standard differentiation and integration properties of the Laplace transform to \rf{lwl_eq} and taking into account \rf{lconv} and \rf{dconv}, we find
\be{lt_lwl_eq}
\omega^2\bar{\xi}(s) + M_w^2\left[ s^2\bar{\xi}(s) - s\xi(0) - \xi'(0) \right] = \frac{i\alpha }{2\omega}\bar{u}(s)\bar{v}(s)(\omega +  isM),
\ee
where
\ba{lt_uv}
\bar{u}(s) &=& \omega\bar{\xi}(s) + iM\left[ s\bar{\xi}(s) - \xi(0) \right], \nn \\
\bar{v}(s) &=&  \frac{1}{s - ip_2} - \frac{1}{s - ip_1} = \frac{i(p_2 - p_1)}{(s-ip_1)(s-ip_2)} \nn \\ 
&=& - \frac{2i\omega}{(\omega + isM)^2 + s^2}.
\ea

Applying the boundary condition $\xi(0)=0$ to the expressions \rf{lt_lwl_eq} and \rf{lt_uv} yields
\be{lt_lwl_eq2}
\left( \omega^2 + s^2M_w^2 \right) \bar{\xi}(s) - \alpha \frac{\left( \omega + isM \right)^2}{(\omega + isM)^2 + s^2} \bar{\xi}(s) = M_w^2 \xi'(0).
\ee

Since the equation \rf{lt_lwl_eq2} is linear in $\bar{\xi}(s)$, we can isolate this term and invert the whole expression by means of Mellin's inverse formula to finally obtain
\begin{widetext}
\be{lwl_xi}
\xi(x) \equiv \mathcal{L}^{-1}\left[ \bar{\xi}(p) \right] = \frac{1}{2\pi i} \lim\limits_{T\to\infty} \int_{-T-i\nu}^{T-i\nu} \frac{ M_w^2 \xi'(0) [(\omega - pM)^2 - p^2] e^{ipx}}{(\omega^2 - p^2M_w^2)[(\omega - pM)^2 - p^2] - \alpha(\omega - pM)^2} \dd p,
\ee
\end{widetext}
where $p=-is$ and $\nu$ is a real number greater than the imaginary part of all the poles. The bounds in integral \rf{lwl_xi} define a line in the complex plane that is usually closed with a portion of a circle, thus delimiting a closed contour $\mathcal{C}_B$ (commonly known as the Bromwich contour \cite{AF2003}). Using the same argument of Jordan's lemma as for \rf{cont_int_lwl}, we prove that contribution of the circular integral is negligible in the limit of infinite radius. Then, another application of the Cauchy residue theorem to the contour integral allows an explicit computation of the inverse.

\subsection{Integral eigenfrequency relation}

Requiring $\xi(x)$ in the expression \rf{lwl_xi} to vanish at $x=\Gamma$ in accordance with \rf{eome}, we obtain the following eigenfrequency relation
\begin{widetext}
\be{eig_eq_lwl}
\xi(\Gamma) = \frac{M_w^2 \xi'(0)}{2\pi i} \oint_{\mathcal{C}_B} \frac{[(\omega - pM)^2 - p^2] e^{ip\Gamma}}{(\omega^2 - p^2M_w^2)[(\omega - pM)^2 - p^2] - \alpha(\omega - pM)^2} \dd p = 0,
\ee
\end{widetext}
which can be written as follows
\be{eig_eq_lwldr}
\xi(\Gamma) = \frac{M_w^2 \xi'(0)}{2\pi i} D(\omega,\alpha) = 0,
\ee
where
\be{Deigrel}
D(\omega,\alpha)=\oint_{\mathcal{C}_B} \frac{ e^{ip\Gamma}}{\mathcal{D}(\omega,\alpha,p)} \dd p ,
\ee
and
\be{drjfm}
\mathcal{D}(\omega,\alpha,p)=\omega^2 - p^2M_w^2 - \alpha\frac{(\omega - pM)^2}{(\omega - pM)^2 - p^2} ,
\ee
is nothing else but the shallow water dispersion relation of the membrane of the \textit{infinite} chord length in the case of a medium with constant pressure and the same density as that of the fluid being present below the membrane \cite{LK2020}.

An expression similar to \rf{eig_eq_lwldr}, derived earlier by Nemtsov for the membrane of the finite chord length with vacuum below the membrane \cite{N1985}, naturally has the corresponding shallow water dispersion relation in the denominator of the integrand, which slightly differs from the ours.

\subsection{Perturbation of eigenfrequencies}

In the case of $\alpha=0$, the eigenvalue relation \rf{eig_eq_lwldr} reduces to
\be{D0}
D(\omega,0)=\oint_{\mathcal{C}_B} \frac{ e^{ip\Gamma}}{\mathcal{D}(\omega,0,p)} \dd p=\frac{1}{M_w^2}\oint_{\mathcal{C}_B} \frac{ -e^{ip\Gamma}}{ p^2-p_0^2} \dd p = 0,
\ee
where $p_0=\omega/M_w$. Applying the residue theorem to the last integral in \rf{D0}, we find
$$
D(\omega,0)=\frac{1}{\omega M_w}\sin\left(\frac{\Gamma \omega}{M_w}\right)=0,
$$
which yields frequencies of the free (decoupled from the flow) membrane
\be{omn}
\omega_n=\pi n\frac{ M_w}{\Gamma} \ne 0,\quad n\in \mathbb{N}.
\ee

In contrast to Nemtsov, we develop a systematic approach based on the perturbation theory of simple eigenvalues $\omega_n$ to find how they are affected by a weak coupling to the flow with free surface \cite{MK91,H1992,BKMR94}.  Then, simple roots $\omega(\alpha)$ of the equation $D(\omega,\alpha)=0$ can be represented as a series in $\alpha$, $0<\alpha \ll 1$, as follows \cite{LK2020,K2013dg,MOM2020,GLO2020}
\be{oma}
\omega=\omega_n-\alpha\frac{\partial_{\alpha} D}{\partial_{\omega}D} + o(\alpha), \quad \omega(0)=\omega_n.
\ee

Computing the partial derivatives and evaluating them at $\alpha=0$ yields
\ba{der}
\partial_{\alpha} D
&=&\frac{1}{M_w^2(M^2 - 1)} \oint_{\mathcal{C}_B} \frac{e^{ip\Gamma}(\omega_n -pM)^2}{\left( p^2 - p_{0,n}^2 \right)^2 (p - p_{1,n})(p - p_{2,n})} \dd p,\nn\\
\partial_{\omega} D
&=&-\frac{2\omega_n}{M_w^4} \oint_{\mathcal{C}_B} \frac{ e^{i p \Gamma}}{(p^2-p_{0,n}^2)^2} \dd p,
\ea
where
$$
p_{0,n}=\frac{\pi n}{\Gamma}, \quad p_{1,n}=\frac{\pi n}{\Gamma}\frac{M_w}{M+1},\quad p_{2,n}=\frac{\pi n}{\Gamma}\frac{M_w}{M-1}.
$$

Applying the residue theorem to the integrals in \rf{der}, we find
\begin{widetext}
\begin{align}
\label{derres}
\partial_{\omega} D&= \frac{-i\Gamma^2}{n \pi M_w^3}(-1)^n,\nn\\
\partial_{\alpha} D&=
(-1)^n\frac{ i\Gamma^3}{2 \pi^2 n^2 M_w^4}\,\frac{(M_w^2-M^2)^2-M_w^2-M^2}{(M_w^2-M^2-1)^2-4M^2} \nn \\
&+ \frac{i\Gamma^3}{2 \pi^3 n^3 M_w^3}\left\{\frac{(M-1)^2\sin\left( \frac{\pi n M_w}{M-1}\right)}{(M_w^2-(M-1)^2)^2}-\frac{(M+1)^2\sin\left( \frac{\pi n M_w}{M+1}\right)}{(M_w^2-(M+1)^2)^2}\right\} \nn\\
&+ \frac{-\Gamma^3}{2 \pi^3 n^3 M_w^3}\left\{\frac{(M-1)^2 \left[(-1)^n-\cos\left( \frac{\pi n M_w}{M-1}\right)\right]}{ \left(M_w^2-(M-1)^2\right)^2}-\frac{(M+1)^2 \left[(-1)^n-\cos\left( \frac{\pi n M_w}{M+1}\right)\right]}{ \left(M_w^2-(M+1)^2\right)^2}\right\}.
\end{align}
\end{widetext}
With the derivatives \rf{derres} the series expansion \rf{oma} takes the form
\begin{widetext}
\ba{omafull}
\omega &=&\omega_n
+\frac{\alpha}{2 \omega_n }\,\frac{(M_w^2-M^2)^2-M_w^2-M^2}{(M_w^2-M^2-1)^2-4M^2}\nn\\
&+&(-1)^n\frac{\alpha \Gamma }{2 \pi^2 n^2}\left\{\frac{(M-1)^2\sin\left( \frac{\pi n M_w}{M-1}\right)}{\left(M_w^2-(M-1)^2\right)^2}-\frac{(M+1)^2\sin\left( \frac{\pi n M_w}{M+1}\right)}{\left(M_w^2-(M+1)^2\right)^2}\right\} \nn\\
&+&\frac{i \alpha\Gamma}{ 2 \pi^2 n^2}\left\{\frac{(M{-}1)^2 \left[1-(-1)^n\cos\left(\frac{\pi n M_w}{M-1}\right)\right]}{ \left(M_w^2-(M-1)^2\right)^2} -
\frac{(M{+}1)^2 \left[1-(-1)^n\cos\left(\frac{\pi n M_w}{M+1}\right)\right]}{ \left(M_w^2-(M+1)^2\right)^2}\right\}{+}o(\alpha).
\ea
\end{widetext}
In particular, from \rf{omafull} we easily obtain the growth rate of the perturbed simple real eigenvalue $\omega_n$
\begin{widetext}
\be{gr_lwlp}
\text{Im}\left(\omega\right) = \frac{\alpha  \Gamma}{ 2 \pi^2 n^2} \left\lbrace \frac{(M-1)^2\left[ 1 -  (-1)^n\cos{\left( \pi n\frac{M_w}{M-1} \right)} \right]}{\left( M_w^2 - (M-1)^2 \right)^2} - \frac{(M+1)^2\left[ 1 -  (-1)^n\cos{\left( \pi n\frac{M_w}{M+1} \right)} \right]}{\left( M_w^2 - (M+1)^2 \right)^2} \right\rbrace,
\ee
\end{widetext}
which, if re-written as follows
\begin{widetext}
\be{gr_lwl}
\text{Im}\left(\omega\right) = \frac{\alpha}{2\Gamma M_w^2} \left\lbrace \frac{p_{2,n}^2\left[ 1 - (-1)^n\cos{\left( p_{2,n}\Gamma \right)} \right]}{\left( p_{2,n}^2 - p_{0,n}^2 \right)^2} - \frac{p_{1,n}^2\left[ 1 -  (-1)^n\cos{\left( p_{1,n}\Gamma \right)} \right]}{\left( p_{1,n}^2 - p_{0,n}^2 \right)^2} \right\rbrace,
\ee
\end{widetext}
exactly reproduces the growth rate derived earlier by Nemtsov \cite{N1985}.

\subsection{Stability diagrams in the shallow water limit}

\begin{figure*}
\centering

\subfloat[]{
\includegraphics*[width=.5\textwidth]{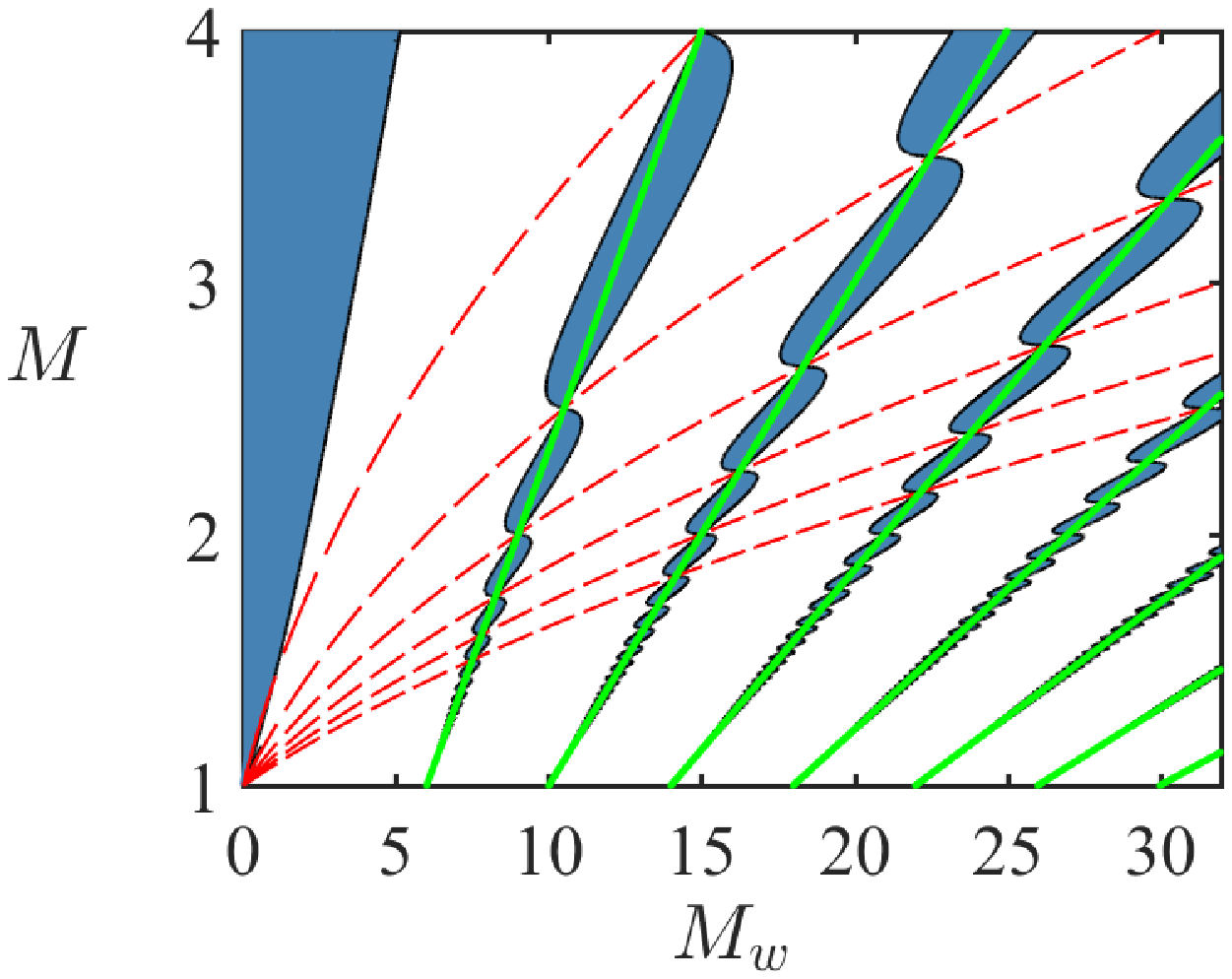} \label{fig2a}} \hspace*{-3em}
\subfloat[]{
\includegraphics*[width=.5\textwidth]{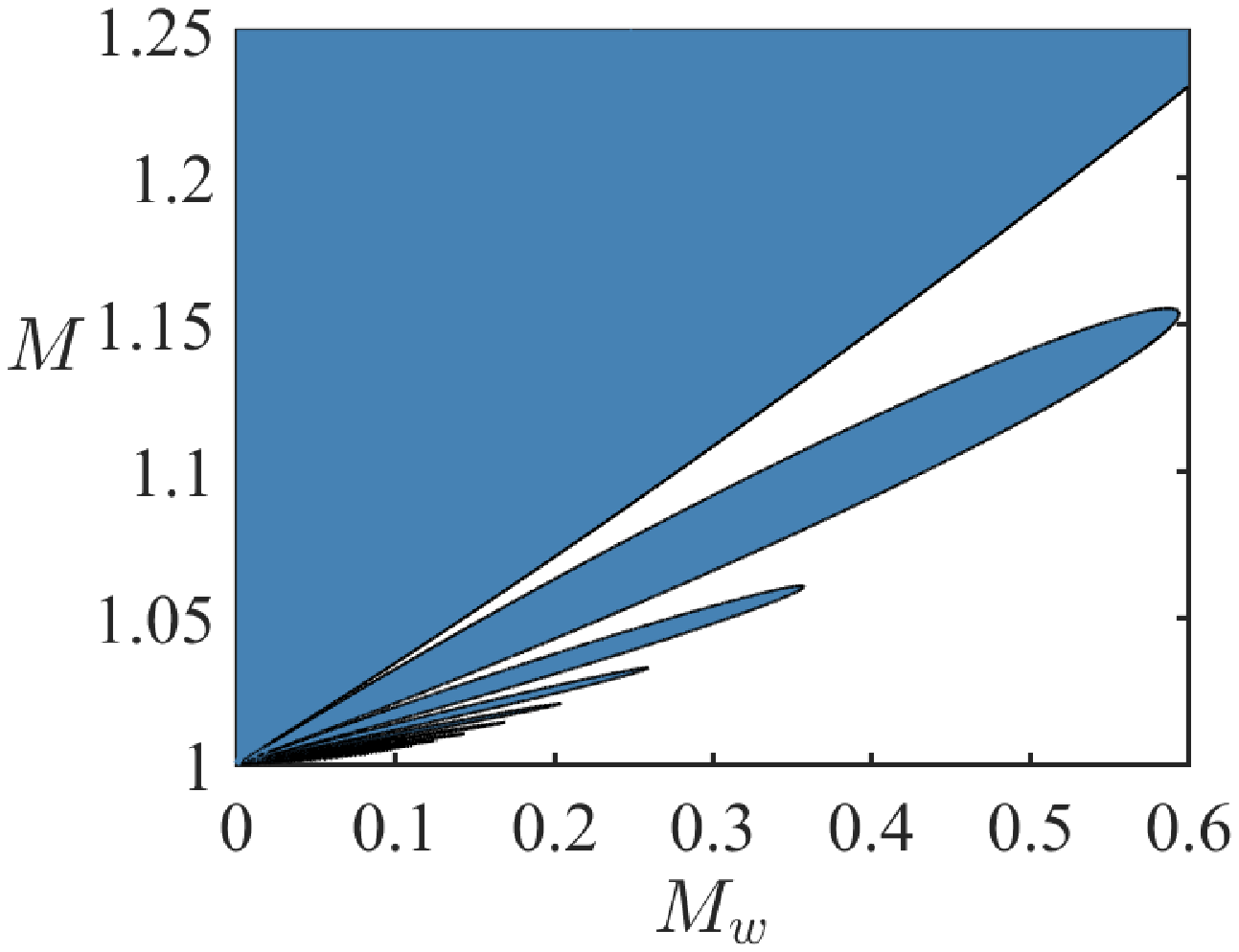} \label{fig2b}} \\ \vspace*{-1.5em}
\subfloat[]{
\includegraphics*[width=.5\textwidth]{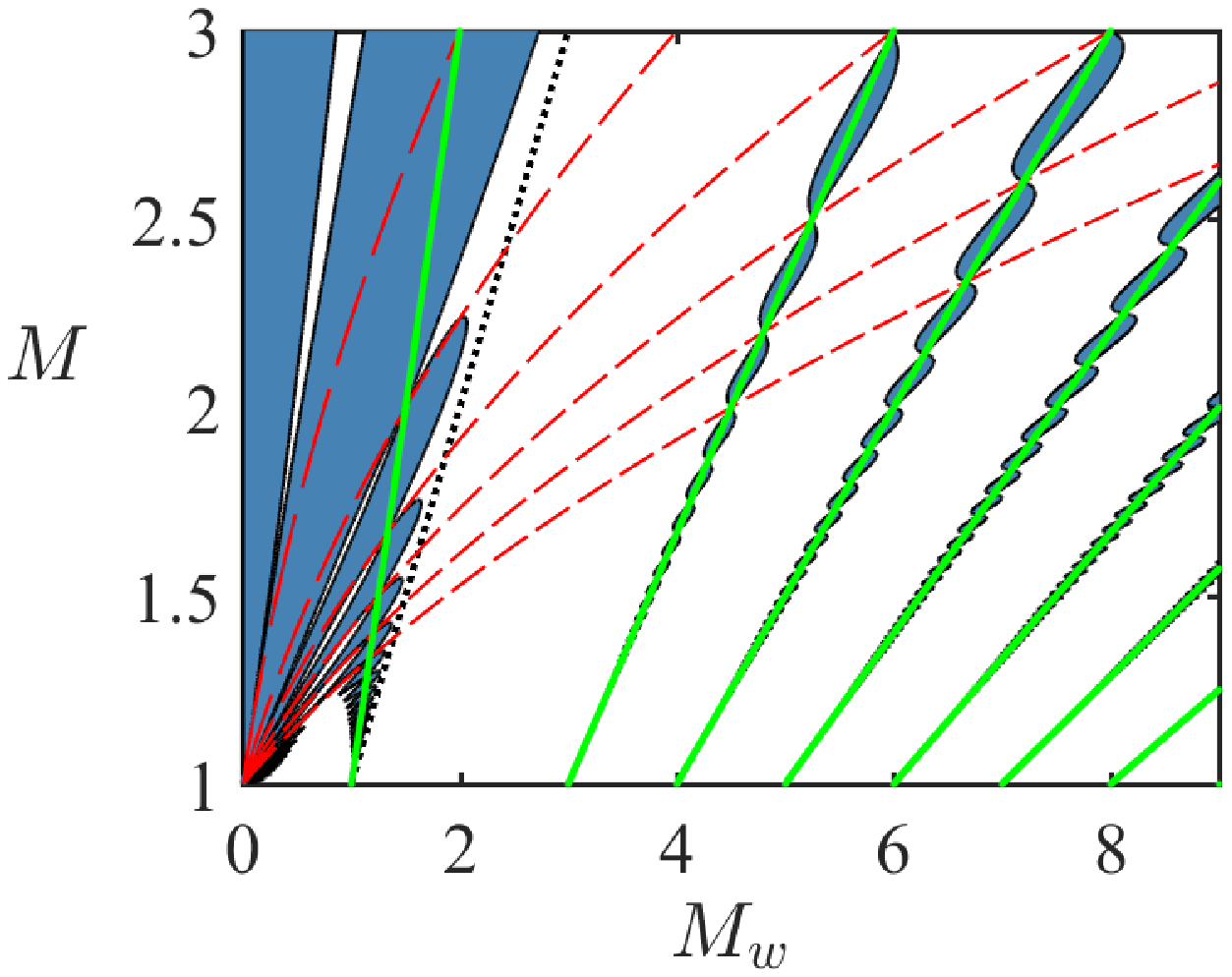} \label{fig2c}} \hspace*{-3em}
\subfloat[]{
\includegraphics*[width=.5\textwidth]{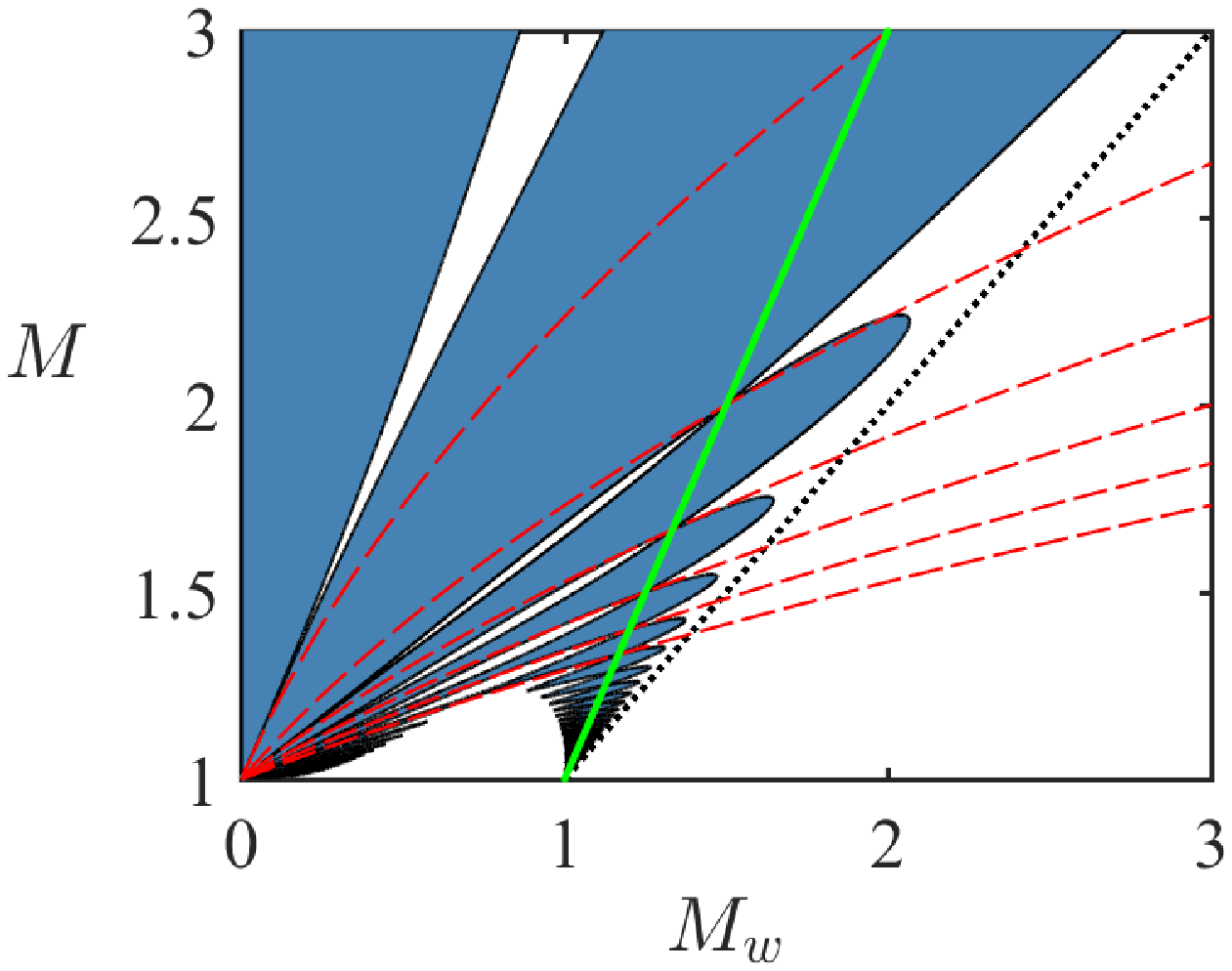} \label{fig2d}}

\caption{(Filled blue) Instability regions for (a,b) the first $(n=1)$ and (c,d) the fourth $(n=4)$ modes of the finite-chord Nemtsov membrane weakly coupled to the flow in the shallow water limit. The black dotted line is equation $M=M_w$. The green straight lines are given by $M_w=(2j/n+1)(M+1)$ for $j\in \mathbb{Z}$ and the red dashed curves by $M_w=k/n(M^2-1)$ for $k\in \mathbb{N}$. Notice absence of instability domains in a wide gap centered at $M_w=M+1$ and corresponding to $j=0$. \label{fig2}}
\end{figure*}

Setting to zero the linear in $\alpha$ approximation to the growth rate \rf{gr_lwl} of the $n$-th mode of the membrane,
we can find an approximation to the neutral stability curve for this mode that subdivides the plane of parameters $M$ and $M_w$ into the domains of stability and flutter instability, Fig.~\ref{fig2}.

\begin{figure*}
\centering

\subfloat[$n=1$]{
\includegraphics*[width=.8\textwidth]{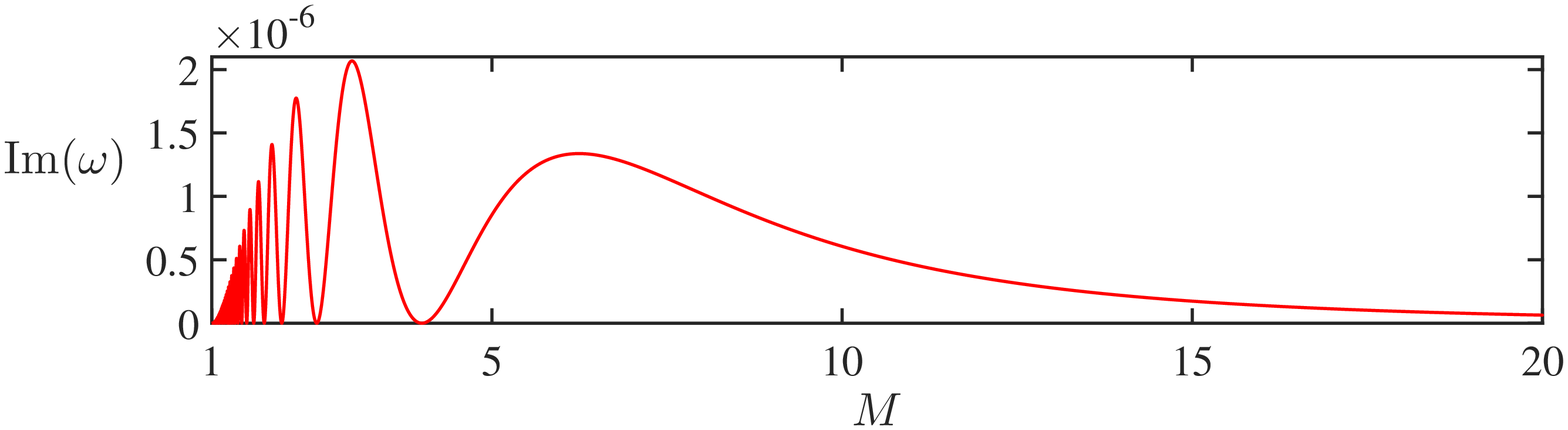} \label{fig3a}} \\ \vspace*{-1em}
\subfloat[$n=4$]{
\includegraphics*[width=.8\textwidth]{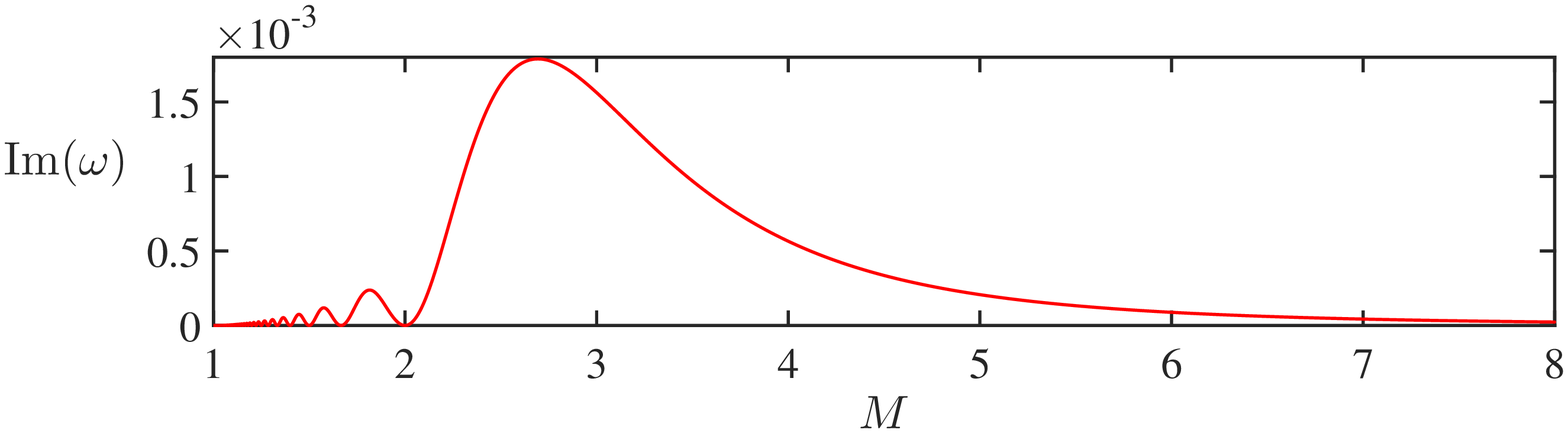} \label{fig3b}}

\caption{Growth rates of perturbed frequencies of the membrane along the line \rf{line} with $\alpha=10^{-3}$, $\Gamma=10$ and (a) $n=1$ and $j=1$ with zeros at $M=1+3/k$, $k\ge1$ and (b) $n=4$ and $j=-1$ with zeros at $M=1+2/k$, $k>1$. \label{fig3}}
\end{figure*}

First of all we observe a cluster of instability domains grouped in the region $M>M_w$ in Fig.~\ref{fig2}. The threshold $M=M_w$ is visible in
Fig.~\ref{fig2}(c,d) as a black dotted line. For the Nemtsov membrane of infinite chord length the physical meaning of this threshold is the equality of the velocity of the flow to the phase speed of elastic waves propagating along the membrane, which is a consequence of the Cherenkov condition \cite{LK2020,N1985}. However, in contrast to the infinite-chord membrane's stability map reported in \cite{LK2020}, there are infinitely many 'petals' of flutter instability for $M>M_w$ touching each other when $n>1$, see Fig.~\ref{fig2}(c,d). Quite surprisingly, the common points of the petals all belong to straight lines of the following form
\be{line}
M_w = \left( \frac{2j}{n} + 1 \right) \left( M+1 \right), \quad j\in\mathbb{Z},
\ee
where $j$ is a negative integer in the region $M>M_w$. For instance, $n=4$ and $j=-1$ yield $M=2M_w-1$, which is a line passing through the point with $M_w=M=1$ in Fig.~\ref{fig2}(c,d). The growth rate along this line presented in Fig.~\ref{fig3}(b) demonstrates vanishing to zero exactly at the common points of the instability regions. 

These points are located exactly at the intersections of the straight lines \rf{line} with the curves
\be{qsc}
M_w = \frac{k}{n}\left(M^2 - 1\right), \quad k\in\mathbb{N}
\ee
that are shown as red and dashed in Fig.~\ref{fig2}. Solving equations \rf{line} and \rf{qsc} we obtain the coordinates of the crossing points
\be{cp}
M = \frac{2j + k + n}{k}, \quad M_w = \frac{(2j + n)(2j + 2k + n)}{kn}.
\ee
For instance, for $j=-1$, $n=4$ this yields $M=1+2/k$ and $M_w=1+1/k$ and for $k=2$ results in $M=2$ and $M_w=3/2$, see Fig.~\ref{fig2}(c,d) and
Fig.~\ref{fig3}(b).

Notice absence of instability domains in a wide gap centered at $M_w=M+1$ and corresponding to $j=0$ in \rf{line}, which is clearly visible in Fig.~\ref{fig2}(a,c). In the case of an infinite-chord membrane with vanishing coupling parameter ($\alpha=0$) the relation $M_w=M+1$ corresponds to a crossing of dispersion curves of surface gravity waves and elastic waves in the membrane that unfolds into an avoided crossing (stability) for $\alpha>0$ \cite{LK2020}. We can conclude therefore that this very property of an infinite-chord membrane manifests itself as a stability gap at $M_w=M+1$ for the finite-chord membrane.

\begin{figure*}
\centering

\subfloat[$n=1$]{
\includegraphics*[width=.5\textwidth]{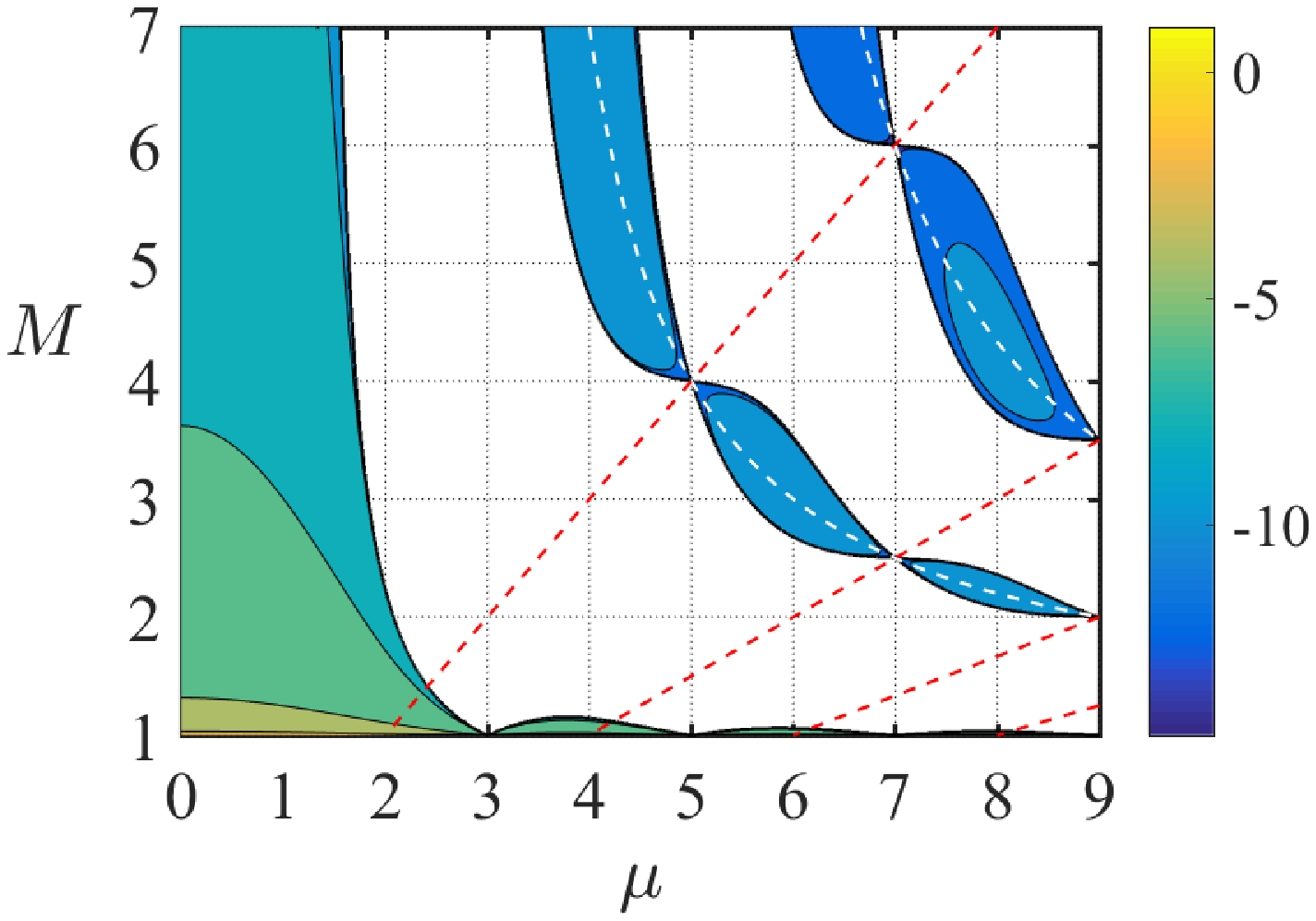} \label{fig4a}} \hspace*{-1em}
\subfloat[$n=2$]{
\includegraphics*[width=.5\textwidth]{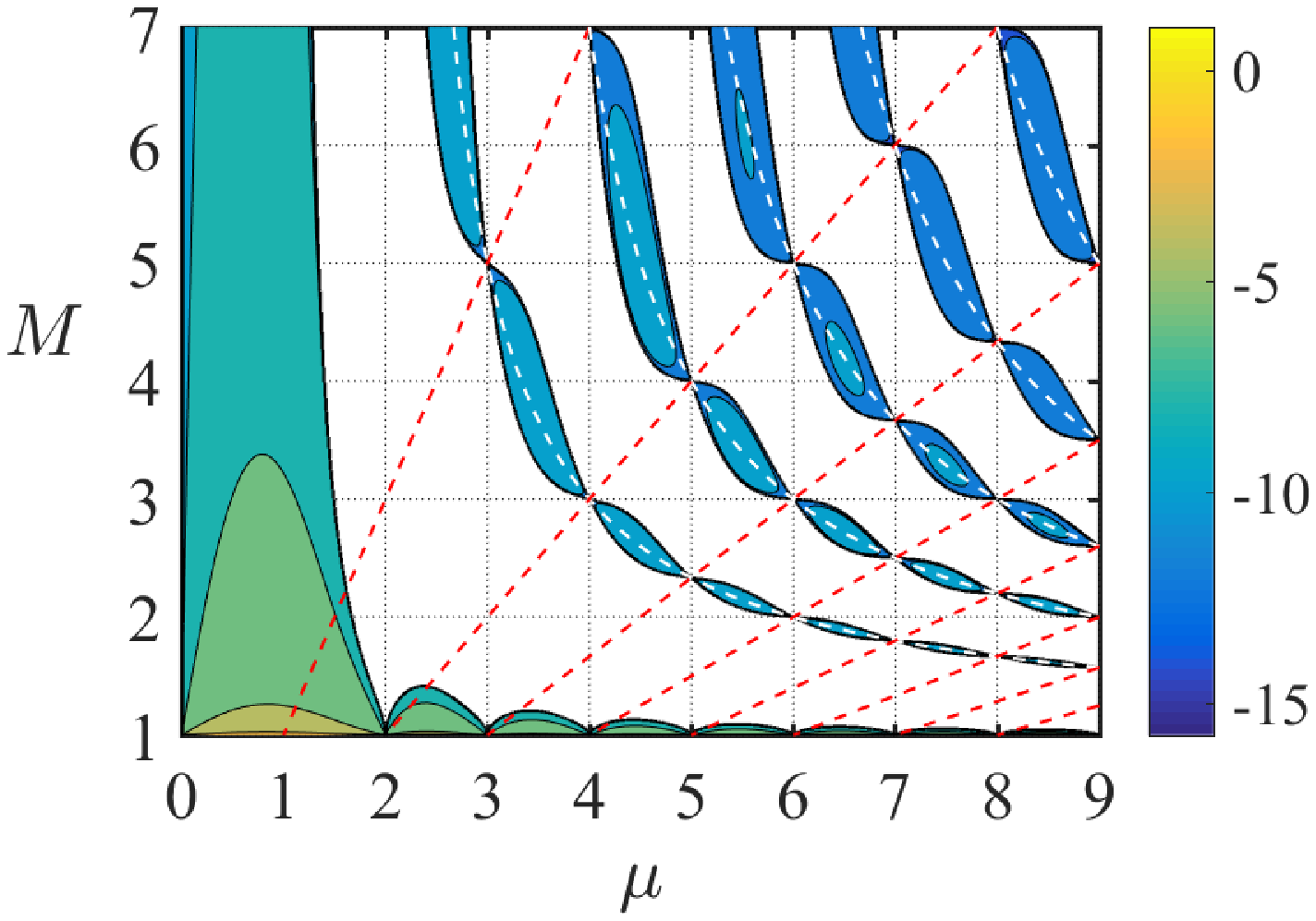} \label{fig4b}} \\ \vspace*{-1.5em}
\subfloat[$n=3$]{
\includegraphics*[width=.5\textwidth]{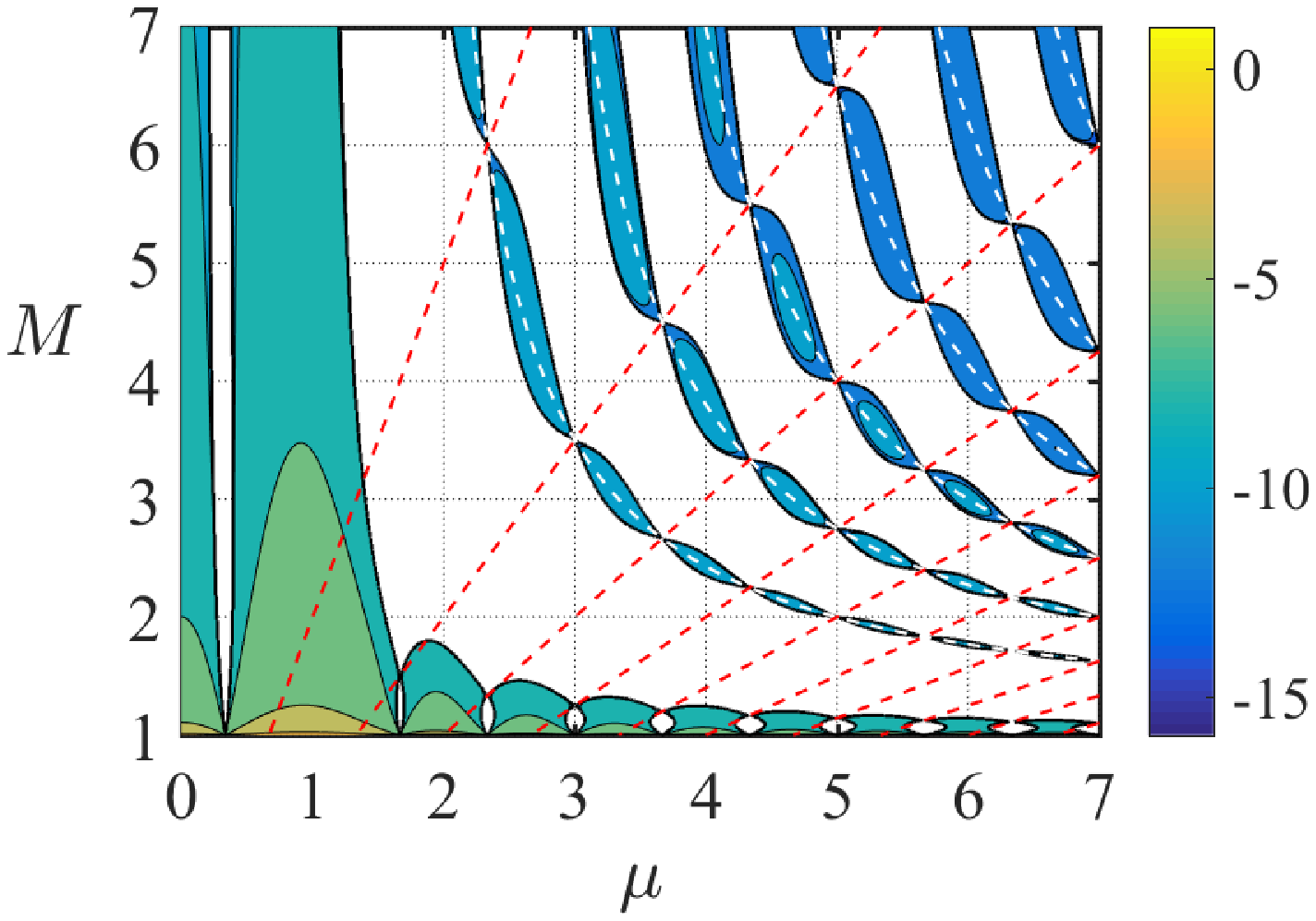} \label{fig4c}} \hspace*{-1em}
\subfloat[$n=4$]{
\includegraphics*[width=.5\textwidth]{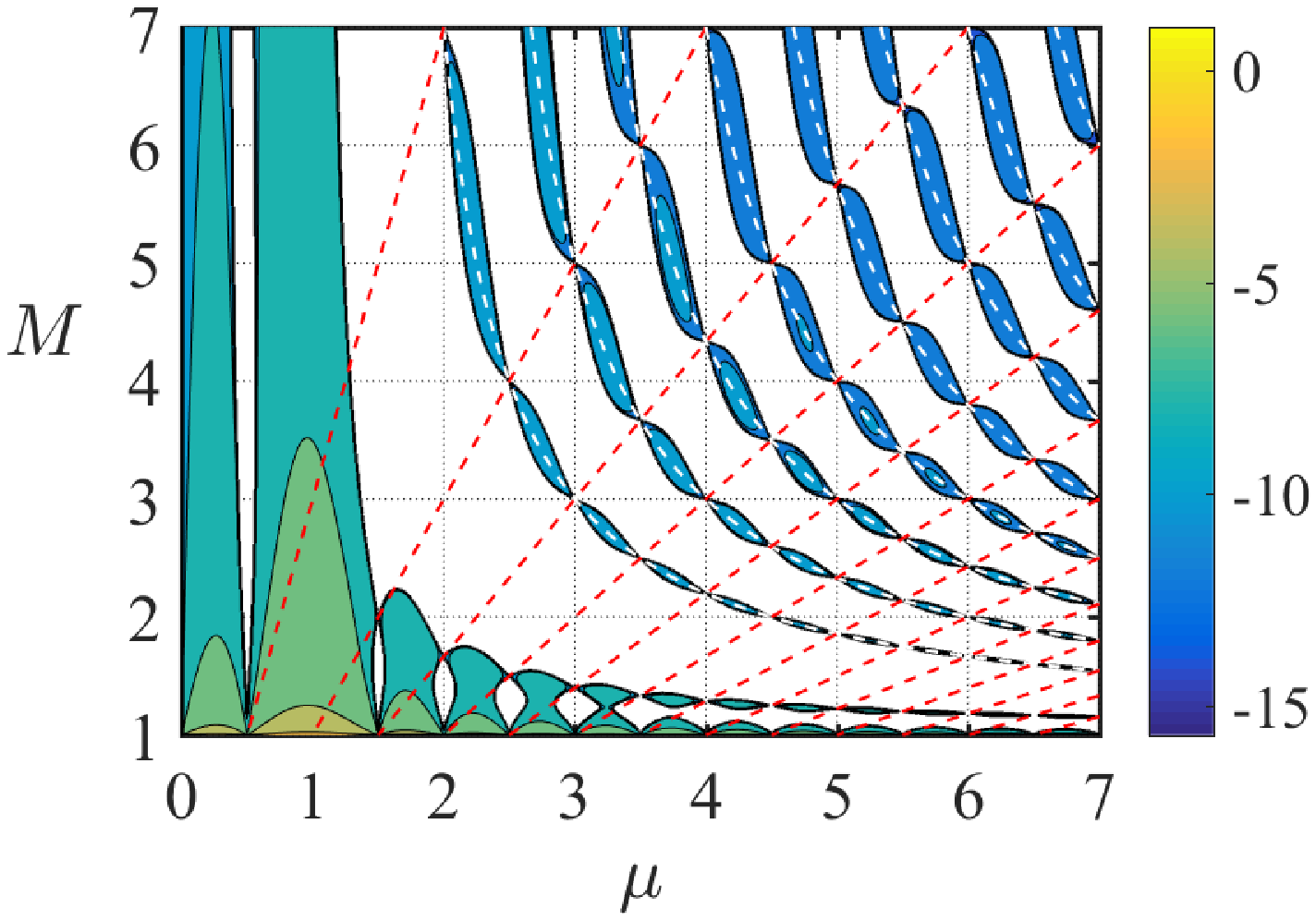} \label{fig4d}}

\caption{Stability maps with logarithmic scale for the growth rate \rf{gr_lwl} over $\mu=M_w/(M-1)$ and $M$. The dashed white curves and red lines represent the lines \rf{line} and curves \rf{qsc}, respectively, for $\Gamma=1$, $\alpha =10^{-4}$ and $n$ according to the caption. \label{fig4}}
\end{figure*}

In a strike contrast to the infinite-chord membrane, stability diagrams of Fig.~\ref{fig2} display a regular pattern of intertwined instability tongues, each centered along a line \rf{line} with $j\ge 1$. All the tongues commence at $M=1$ at the values of $M_w$ that are specified by \rf{line}.
For instance, if $n=1$, then the tongues grow from $M_w=6,10,14,\ldots 2(2j+1),\ldots$, see Fig.~\ref{fig2}(a). The growth rate along the line \rf{line} corresponding to the tongue with $n=1$ and $j=1$ is shown in Fig.~\ref{fig3}(a). The growth rate vanishes at $M=1+3/k$, $k\ge1$, i.e. exactly at the crossing points \rf{cp} that subdivide the instability tongue into a collection of infinitely many instability pockets, see Fig.~\ref{fig2}. Note that similar intertwined resonance tongues with instability pockets are known for the Hill equation with some specific forms of periodic excitation \cite{BL1995}.

In order to highlight the periodic pattern of the instability pockets, finally we plot the stability map in the new coordinates in Fig.~\ref{fig4} by projecting the growth rate onto the $(\mu,M)$-plane, where $\mu=M_w/(M-1)$. Then, the lines \rf{line} transform into the curves $\mu=(2j/n+1)(M+1)/(M-1)$ and the curves \rf{qsc} to the lines $\mu=k/n(M+1)$, see Fig.~\ref{fig4}.

\subsection{Exploring fluid dynamics analogy to superlight normal and inverse Doppler effects}

According  to  a  recent  study  \cite{IDE2018},  a  source  with  an  internal  structure  moving  in  a  medium  at a velocity that exceeds speed of light in the medium can be excited due to emission of electromagnetic waves (Ginzburg-Frank  anomalous  Doppler  effect  \cite{GF1947})  with  the  Doppler  shift  of  the  emitted  waves  remaining normal at small superluminal velocities and becoming inverted beyond some critical superluminal velocity. As it was emphasized in \cite{IDE2018}, virtually any wave system in nature, including classical wave systems such as acoustic waves and surface waves can exhibit the analogous phenomena. 

Non-dispersive character of surface gravity waves in the shallow water limit implies $M=1$
as a critical value for the flow velocity to exceed the speed of surface gravity waves and thus as a necessary condition for existence of the anomalous Doppler effect \cite{LK2020,N1985}. Indeed, flutter instability tongues in Fig.~\ref{fig2} and Fig.~\ref{fig4} exist at $M>1$. We need to verify that successive surface gravity waves radiated by the Nemtsov membrane carry wavelengths larger (smaller) than a characteristic value when $M>M_c>1$ ($1<M<M_c$).

Using the Bernoulli integral at the free surface $(z=1)$, retaining only
linear in $\phi$ terms \cite{LK2020}, and assuming the time dependence $\exp(-i\omega \tau)$, we obtain
\be{bern}
\eta(x) = -(-i\omega + M\partial_x) \phi(x,z=1) ,
\ee
where the velocity potential $\phi(x,z)$ is obtained as the inverse Fourier transform \rf{fourier} of the solution of the boundary value problem \rf{bvpf}, then evaluated at $z=1$. Indeed, from the general expression \rf{gph}, we find
\begin{align}
\label{phifsfd}
\hat{\phi}(\kappa,1,\omega) = &\frac{1}{(\omega - \kappa M)^2 \cosh{\kappa} - \kappa \sinh{\kappa}} \times \nn \\
&\int_{0}^{\Gamma} \left( i\omega - M \frac{\dd}{\dd x'} \right) \xi (x') e^{-i\kappa x'} \dd x' .
\end{align}
Inserting solution \rf{phifsfd} into expression \rf{fourier} and considering the shallow water limit ($\kappa\to 0$), we arrive at the fluid potential at the free surface $(z=1)$
\begin{align}
\label{phifs}
\phi(x,1,\omega) = &- \frac{1}{2\pi} \int_{0}^{\Gamma} \left( -i\omega + M \frac{\dd}{\dd x'} \right) \xi (x') \times \nn \\
&\int_{-\infty}^{+\infty} \frac{e^{i\kappa (x-x')}}{(\omega - \kappa M)^2 - \kappa^2} \dd\kappa \dd x' ,
\end{align}
where membrane's frequency $\omega$ and displacement $\xi$ are provided by expressions \rf{omafull} and \rf{lwl_xi}, respectively. 

With the help of \rf{phifs} equation \rf{bern} yields
\begin{align}
\label{eta}
\eta(x) = &- \frac{i}{2\pi} \int_{0}^{\Gamma} \left( -i\omega + M \frac{\dd}{\dd x'} \right) \xi (x') \times \nn \\
&\int_{-\infty}^{+\infty} \frac{(\omega - \kappa M) e^{i\kappa (x-x')}}{ (\omega - \kappa M)^2 - \kappa^2 } \dd\kappa \dd x'.
\end{align}
Following a procedure that we used previously to derive expression \rf{pot_lwl}, we apply the Cauchy residue theorem to \rf{eta} with exactly the same poles as in \rf{imp_int_lwl} and recover the surface wave `mode' (for a given $\omega$) as
\begin{widetext}
\be{eta2}
\eta(x) = - \frac{1}{2(M^2-1)} \int_{0}^{\Gamma} \left( -i\omega + M \frac{\dd}{\dd x'} \right) \xi (x') \left[ (M+1) e^{i\kappa^{-} (x-x')} + (M-1) e^{i\kappa^{+} (x-x')} \right] \dd x' .
\ee
\end{widetext}

\begin{figure*}
\centering

\subfloat[$M=1.2<M_c$]{
\includegraphics*[width=.33\textwidth]{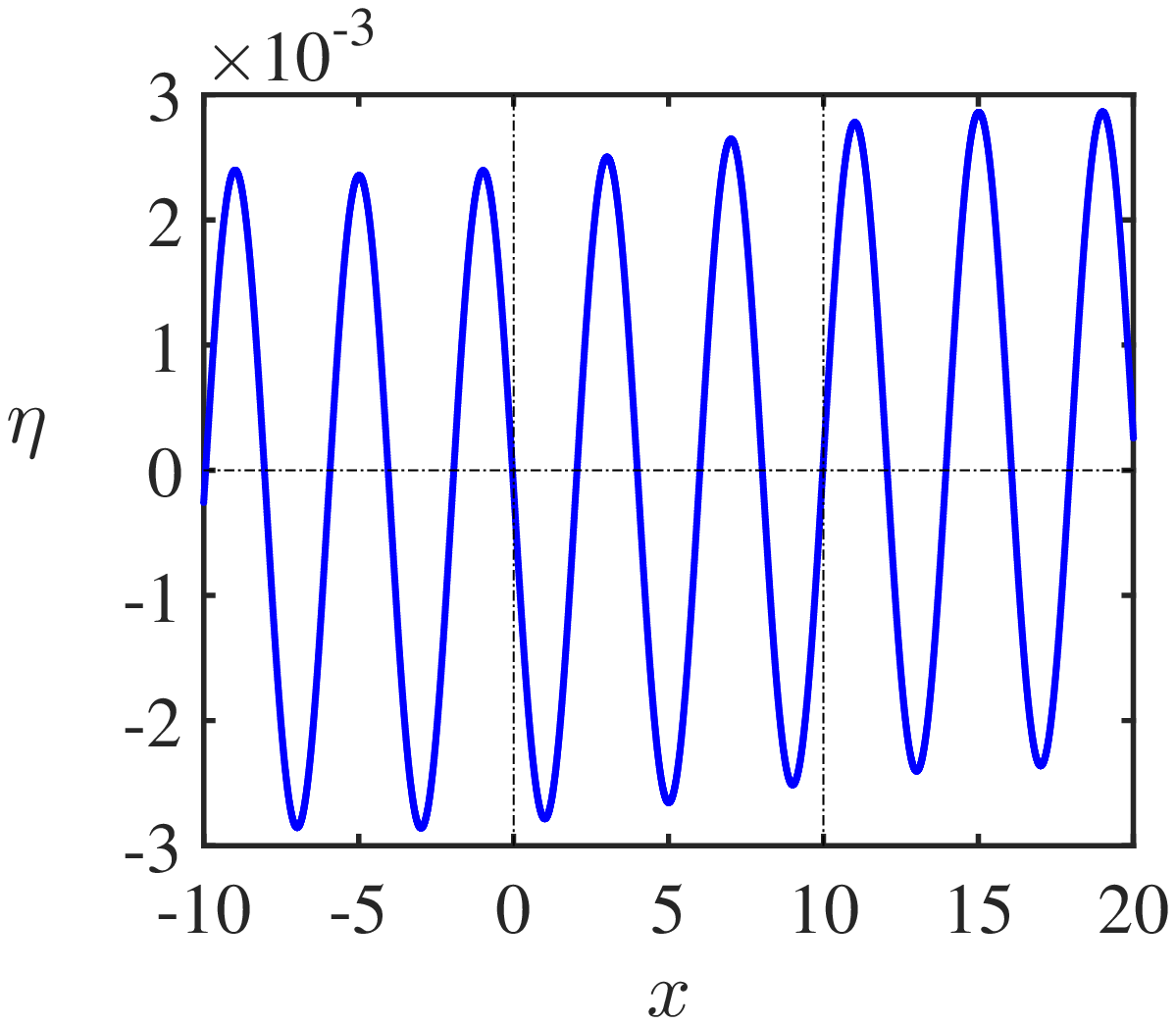} \label{fig5a}} 
\subfloat[$M=1.5=M_c$]{
\includegraphics*[width=.33\textwidth]{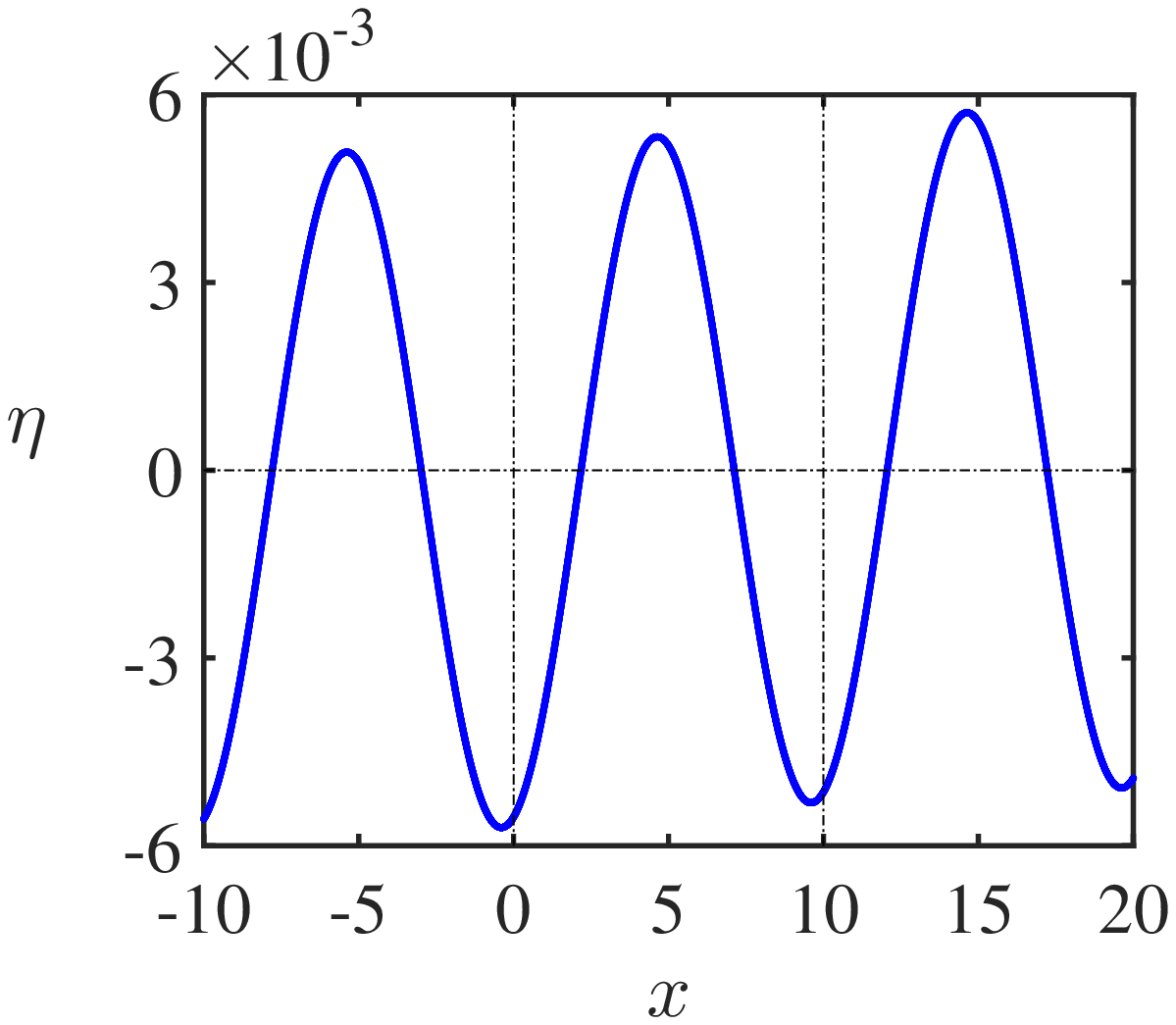} \label{fig5b}} 
\subfloat[$M=4>M_c$]{
\includegraphics*[width=.33\textwidth]{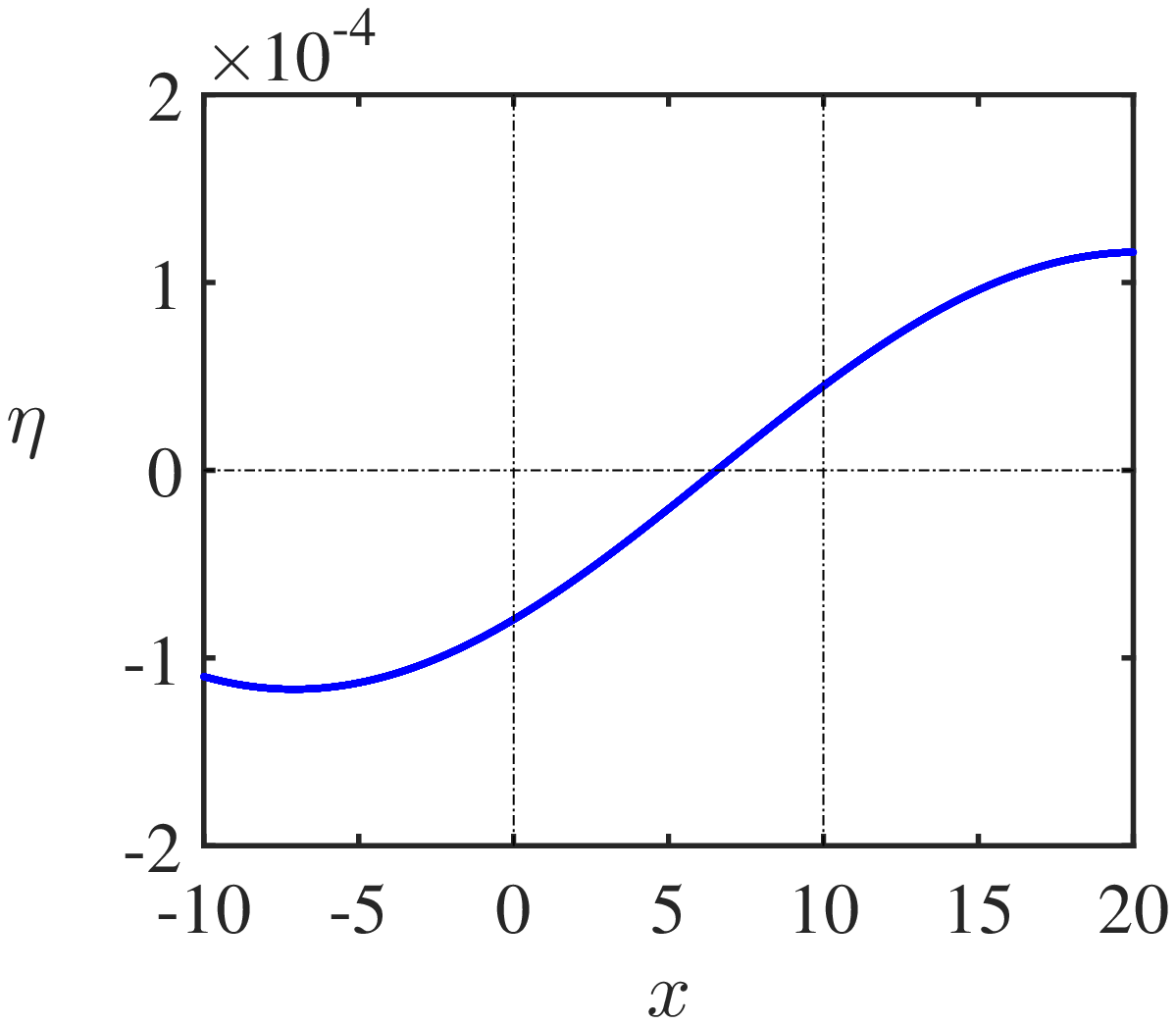} \label{fig5c}}

\caption{Surface `modes' as given by expression \rf{eta2} for a fixed $\omega$ recovered by first-order in $\alpha$ approximation \rf{omafull}, with parameters $n=1$, $M_w=1$, $\Gamma=10$, $\alpha=10^{-4}$ and $M$ according to the legend. Dashed vertical lines specify $x$-coordinates of membrane's edges. \label{surfacewaves}}
\end{figure*}

Plotting \rf{eta2} in Fig.~\ref{surfacewaves}, we observe the different regimes $1<M<M_c$, $M=M_c$ and $M>M_c$ for a certain `mode' of free surface, corresponding to the wavelengths that are shorter, equal and longer than $\Gamma$, respectively. To estimate the critical value $M=M_c$ we use the idea of phase synchronisation between modes of the membrane and the surface of the flow \cite{N1985,M1994} that in the infinite-chord membrane case gives a sufficient condition for the presence of an instability domain in the parameter space \cite{LK2020}. For the finite-chord membrane we consider the frequencies $\omega_n$ of vibration modes of the free membrane defined by equation \rf{omn} and surface waves frequencies $\omega_f$ that for simplicity we take from the shallow water dispersion relation of the decoupled system
\be{drsw}
\mathcal{D}_{SW} (\omega,\kappa) = (\omega_f - \kappa M)^2 - \kappa^2 = 0, \quad \omega_f^{\pm} = \kappa (M \pm 1).
\ee

The modes $\omega_f^- = \kappa (M-1)$ correspond to slow surface gravity waves carrying negative energy and thus exciting flutter of the infinite-chord membrane (anomalous Doppler effect) \cite{LK2020,N1985}. Choosing $\kappa=2\pi / \Gamma$ in the equality $\omega_n = \omega_f^-$   allows us to estimate the critical velocity of the flow as
\be{crit}
M_c = 1 + \frac{n M_w}{2},
\ee
which agrees with the threshold observed in Fig.~\ref{surfacewaves} for the surface modes computed from expression \rf{eta2}.

\section{Deep water limit of the finite-chord Nemtsov membrane}
\label{dwlfcnm}
Let us find the leading term of the factor at $e^{i\kappa(x-x')}$ in the integrand of the improper integral in \rf{phi} in the limit $\kappa\to +\infty$, corresponding to the deep water approximation
\be{potsdw}
 \frac{1}{\kappa^2}\frac{1 - \kappa(\sigma -  M)^2 \tanh{\kappa}}{[(\sigma -  M)^2 - \frac{\tanh{\kappa}}{\kappa}]}= -\frac{1}{\kappa}+o(\kappa^{-1}).
\ee
This yields a simplified form of the potential \rf{phi}
\be{pot_swl}
\phi_{DW}(x,\omega) = -\frac{1}{2\pi} \int_{0}^{x} \left( -i\omega + M\partial_{x'}\right)\xi(x') \int_{-\infty}^{+\infty} \frac{e^{i\kappa(x-x')} \dd\kappa}{\kappa} \, \dd x'.
\ee

For $x>x'$, the Dirichlet integral in \rf{pot_swl} can be calculated around the pole $\kappa=0$ in the sense of the Cauchy Principal Value (CPV) as follows
\be{cpv}
\text{PV}\int_{-\infty}^{+\infty} \frac{e^{i\kappa(x-x')}}{\kappa} \, \dd\kappa = i\pi,
\ee
where $\text{PV}\int := \lim\limits_{\epsilon\to 0} \, \int_{\mathbb{R}\backslash [-\epsilon , \epsilon]}$ and $\epsilon$ is the radius of the sphere enclosing the singularity \cite{AF2003}.
Inserting \rf{cpv} into \rf{pot_swl} we recover the velocity potential in the deep water approximation
\be{pot_swl2}
\phi_{DW}(x,\omega) = -\frac{1}{2} \int_{0}^{x} \left( \omega + iM\partial_{x'}\right)\xi(x') \, \dd x'.
\ee

With the potential \rf{pot_swl2} the equation \rf{ide} takes the form
\be{swl_eq}
\omega^2\xi + M_w^2\partial^2\xi+ \frac{i\alpha}{2} ( \omega + iM\partial_x )\int_{0}^{x} \left( \omega + iM\frac{\dd}{\dd x'} \right) \xi(x') \dd x' = 0.
\ee
Using the same methodology as before and taking into account the boundary condition $\xi(0)=0$, we find the Laplace transform of \rf{swl_eq}
\be{lt_swl_eq}
\left(\omega^2 + s^2M_w^2 + \frac{i\alpha}{2s} \left( \omega + isM \right)^2\right) \bar{\xi}(s) = M_w^2\xi'(0).
\ee

Following once again the procedure described in the previous section, we take $s=ip$ in the equation \rf{lt_swl_eq} and then inverse the whole expression to finally obtain the displacement as the Bromwich integral
\be{swl_xi}
\xi(x) = \frac{M_w^2}{2\pi i} \lim\limits_{T\to\infty} \int_{-T-i\nu}^{T-i\nu} \frac{ 2 \xi'(0) p  e^{ipx}}{2p ( \omega^2 - p^2M_w^2 ) + \alpha (\omega - pM)^2 } \dd p.
\ee

Evaluating \rf{swl_xi} at $x=\Gamma$ and taking into account the boundary condition $\xi(\Gamma)=0$, we recover the eigenfrequency equation
\be{eig_eq_swl}
\xi(\Gamma) = \frac{M_w^2}{2\pi i}  \oint_{\mathcal{C}_B} \frac{ 2\xi'(0)p e^{ip\Gamma}}{2p ( \omega^2 - p^2M_w^2 ) + \alpha (\omega - pM)^2 } \dd p = 0,
\ee
which can be written as follows
\be{eig_eq_swldr}
\xi(\Gamma) = M_w^2 \xi'(0) D(\omega,\alpha) = 0,
\ee
where
\be{eig_eq_swldr2}
D(\omega,\alpha)=\frac{1}{{2\pi i}}\oint_{\mathcal{C}_B} \frac{ e^{ip\Gamma}}{\mathcal{D}(\omega,\alpha,p)} \dd p
\ee
and
\be{eig_eq_swldr3}
\mathcal{D}(\omega,\alpha,p)=\omega^2 - p^2M_w^2 + \alpha\frac{(\omega - pM)^2}{2p}.
\ee

In the case of $\alpha=0$ the eigenvalue relation \rf{eig_eq_swldr} reduces to
\begin{align}
\label{D0dw}
D(\omega,0) &=\frac{1}{{2\pi i}}\oint_{\mathcal{C}_B} \frac{ e^{ip\Gamma}}{\mathcal{D}(\omega,0,p)} \dd p \nn \\
&=\frac{1}{M_w^2}\frac{1}{{2\pi i}}\oint_{\mathcal{C}_B} \frac{ -e^{ip\Gamma}}{ p^2-p_0^2} \dd p = 0,
\end{align}
where $p_0=\omega/M_w$. Applying the residue theorem to the last integral in \rf{D0dw}, we reproduce the eigenfrequencies $\omega_n$ given by equation \rf{omn}.

Simple roots $\omega(\alpha)$ of the equation $D(\omega,\alpha)=0$ can be represented as a series in $\alpha$, $0<\alpha \ll 1$, as follows \cite{LK2020,K2013dg}
\begin{align}
\label{oma2}
\omega= &\omega_n-\alpha\frac{\partial_{\alpha} D}{\partial_{\omega}D} \nn \\
&-\frac{\alpha^2}{2}\left[\frac{\partial_{\omega}^2 D}{\partial_{\omega}D} \left(\frac{\partial_{\alpha} D}{\partial_{\omega}D} \right)^2 - 2\frac{\partial_{\omega \alpha}^2 D}{\partial_{\omega}D}  \frac{\partial_{\alpha} D}{\partial_{\omega}D}+\frac{\partial_{\alpha}^2 D}{\partial_{\omega}D}\right]+o(\alpha^2),
\end{align}
where $\omega(0)=\omega_n$.

Computing the partial derivatives and evaluating them at $\alpha=0$ yields
\ba{der2}
\partial_{\alpha} D
&=& \frac{-1}{{2\pi i}}\oint_{\mathcal{C}_B} \frac{e^{i p \Gamma}(M p-\omega_n)^2}{2M_w^4p( p^2-p_{0,n}^2)^2} \dd p,\nn\\
\partial_{\omega} D
&=& \frac{-1}{{2\pi i}}\oint_{\mathcal{C}_B} \frac{2e^{i p \Gamma}\omega_n}{M_w^4( p^2-p_{0,n}^2)^2} \dd p,
\ea
where
$$
p_{0,n}=\frac{\pi n}{\Gamma}.
$$

Applying the residue theorem to the integrals in \rf{der2}, we find
\ba{der23}
\partial_{\alpha} D
&=&\frac{i M (-1)^n \Gamma^2}{2M_w^3\pi n}+\frac{M_w \Gamma^2[(-1)^n-1]}{2M_w^3\pi^2n^2},\nn\\
\partial_{\omega} D
&=&\frac{-i(-1)^n\Gamma^2}{M_w^3\pi n }.
\ea
With the derivatives \rf{der23} the series expansion \rf{oma2} takes the form
\be{newom}
\omega(\alpha)=\omega_n+\alpha \frac{M}{2}+i \alpha \frac{ M_w}{2 \pi n}  \left[(-1)^n-1\right]+o(\alpha).
\ee
Taking into account terms of the second order in $\alpha$ yields the following expression for the growth rate
\ba{imw2a}
{\rm Im}(\omega)&=&\alpha\frac{M_w}{2\pi n}[(-1)^n-1]
+\frac{\alpha^2\Gamma}{32}\left(\frac{2M_w^2(4-\pi^2 n^2)}{\pi^4 n^4}[(-1)^n-1]\right. \nn \\
&&\left.-\frac{ M^2(\pi^2 n^2+10(-1)^{n}-14)}{\pi^2 n^2}-\frac{M^4}{ M_w^2}\right)+o(\alpha^2).
\ea

For all even $n> 1$ the growth rate \rf{imw2a} is negative up to the terms of higher order than $\alpha^2$:
$$
{\rm Im}(\omega)=-\alpha^2M^2\frac{\Gamma }{32}\left(1-\frac{4}{\pi^2 n^2}+\frac{M^2}{ M_w^2}\right)+o(\alpha^2),\quad {\rm even}~~n>1.
$$

In contrast, for odd $n$ the growth rate can take positive values for some combinations of parameters. Equating the growth rate \rf{imw2a} to zero, we find the critical length of the membrane at the onset of flutter for odd $n$ in the explicit form
\begin{widetext}
\be{GMnsc}
\Gamma= \frac{32 M_w^3 \pi^3 n^3\alpha^{-1}}{4M_w^4( \pi^2 n^2-4)-M^2 M_w^2\pi^2 n^2 (\pi^2 n^2-24) -M^4 \pi^4 n^4},\quad {\rm odd}~~n\ge 1.
\ee
\end{widetext}

\begin{figure*}
\centering

\subfloat[]{
\includegraphics*[width=.5\textwidth]{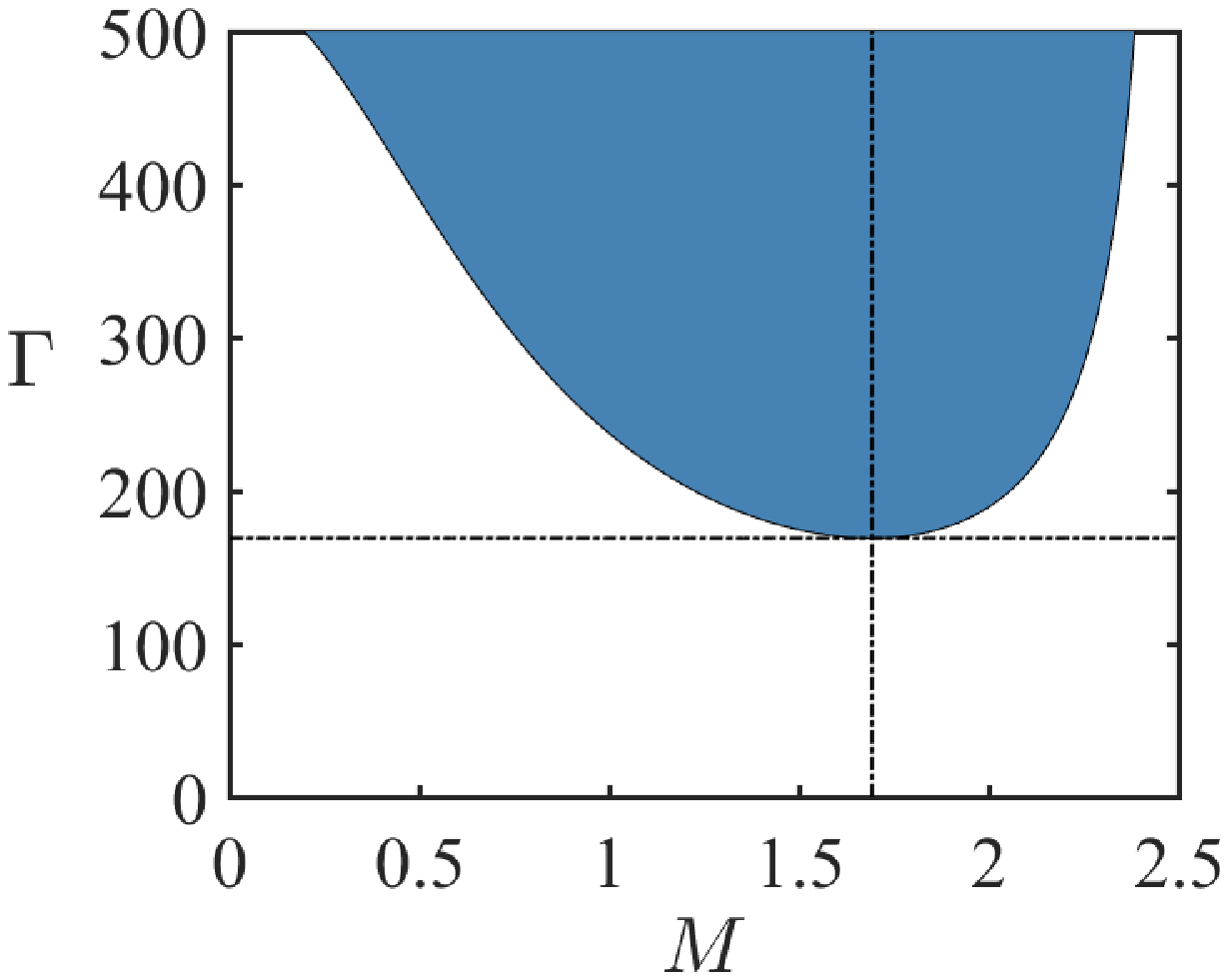} \label{fig6a}} \hspace*{-2em}
\subfloat[]{
\includegraphics*[width=.5\textwidth]{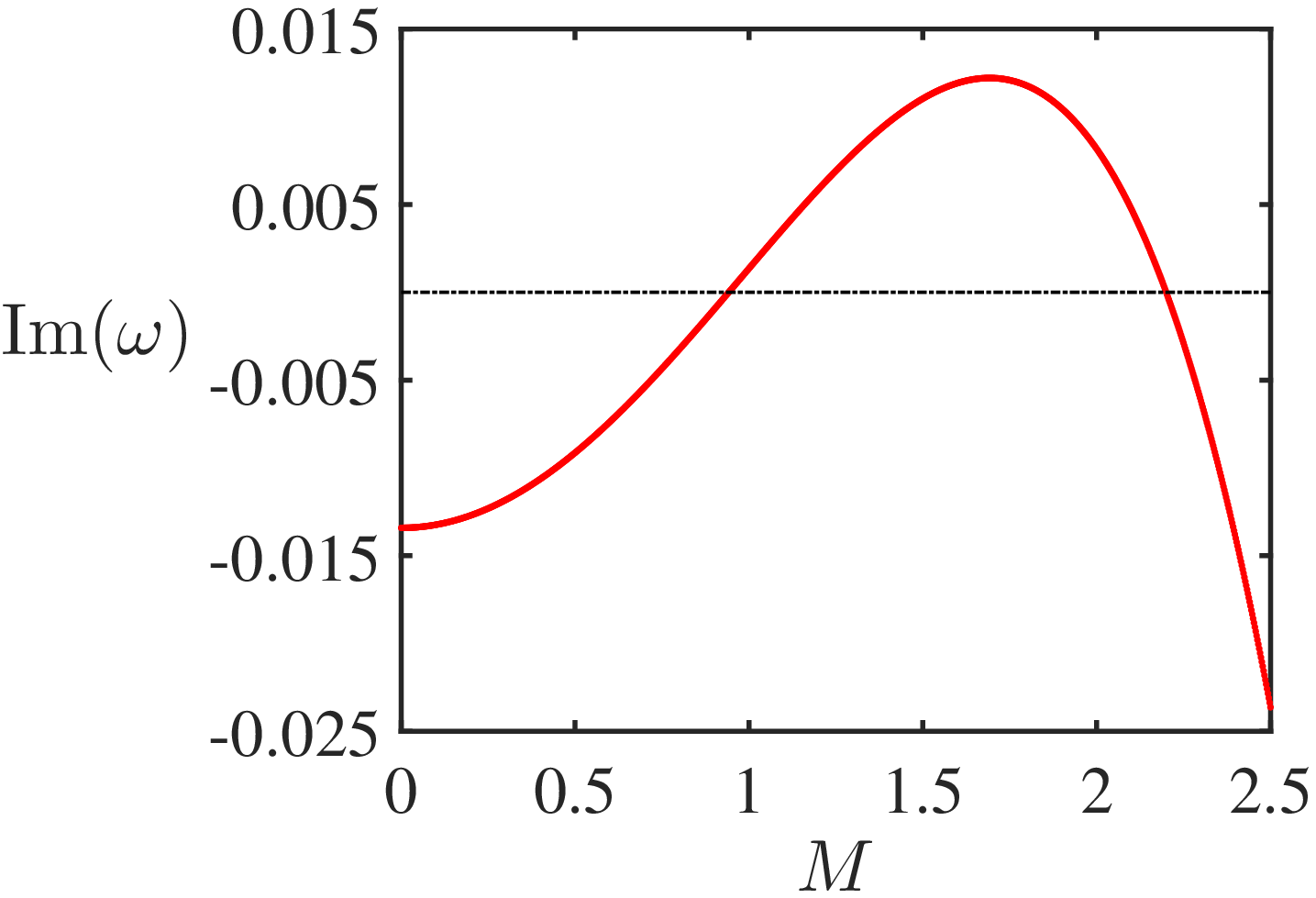} \label{fig6b}}

\caption{(a): (Blue) Instability domain computed from expression \rf{GMnsc} for $n=1$, $M_w=2$, and $\alpha=0.04$, with (dotted lines) the minimum \rf{gmin} of $\Gamma$ at the minimizer given by \rf{mmin}. (b): Growth rate for the left figure at a fixed value of $\Gamma=250$. \label{fig6}}
\end{figure*}

For instance, the minimal length of the membrane for $n=1$ below which there is no flutter, is
\be{gmin}
\Gamma_{\rm min}=\frac{128\pi^3}{\alpha M_w(\pi^4-32\pi^2+512)}
\ee
at
\be{mmin}
M_{\rm min}=M_w\frac{\sqrt{2}}{2\pi}\sqrt{24-\pi^2},
\ee
as is shown in Fig.~\ref{fig6}. Notice that $\Gamma_{\rm min}$ quickly increases as $\alpha$ is tending to zero, thus reducing chances for a finite-chord membrane to be unstable in the deep water limit, if the coupling between the membrane and the flow is vanishingly weak.

\section{The finite-chord Nemtsov membrane in the finite-depth layer}

The case of a fluid layer with finite depth requires further attention in the derivation of eigenfrequency equation than the previously considered limits of shallow and deep water. The main difficulty is the non-polynomial character of the dispersion relation for the finite depth flow \cite{LK2020}, which does not allow analytical solution via Cauchy residue theorem of the improper integral in \rf{ide} that therefore has to be now exclusively treated numerically. This major difference forces us first to count and localize the poles of the integrand using a combination of complex analysis and iterative solvers to make possible numerical implementation of the residue theorem. This allows us to discretize the integro-differential equation into an algebraic nonlinear eigenvalue problem for $\omega$ by means of Galerkin decomposition and solve the resulting equation with an appropriate numerical method. In this section, we present both the derivation of this nonlinear eigenvalue problem and the methods for its solution. We argue that our approach can easily be extended to a wide class of fluid-structure systems composed of a fixed elastic structure that interacts with a moving flow.

\subsection{Counting and localizing the poles of a non-polynomial integrand in the integro-differential equation}


For the sake of clarity in this section, we reintroduce the wave equation of the finite depth problem \rf{ide} in terms of convolution, along with its boundary conditions
\begin{align}
\label{wave_equation}
\mathcal{W}(\xi;x,\omega) = &\left( \omega^2  + M_w^2 \frac{\partial^2}{\partial x^2} \right) \xi(x) \nn \\
&+ \alpha \left( -i\omega + M \frac{\partial}{\partial x} \right) \left( u*v \right)(x) = 0  , \nn \\
&\xi(0)=\xi(\Gamma)=0,
\end{align}
where we have
\begin{align}
u(x,\omega) &= \left( -i\omega + M\frac{\partial}{\partial x} \right) \xi(x) , \label{u} \\
v(x,\omega) &= \frac{1}{2\pi} \int_{-\infty}^{+\infty} \frac{[\kappa - (\omega - \kappa M)^2 \tanh{\kappa}] e^{i\kappa x} }{\kappa[(\omega - \kappa M)^2 - \kappa \tanh{\kappa}]} \dd\kappa \nn \\
&= \frac{1}{2\pi} \int_{-\infty}^{+\infty} F(\kappa) \dd\kappa. \label{v}
\end{align}

As stated before, the main issue in solving \rf{wave_equation} is the presence of the improper integral \rf{v}. Indeed, in contrast with the shallow- and deep water cases, the difficulty in direct application of the Cauchy residue theorem to this integral is that it requires knowledge of the poles of the meromorphic function $F(\kappa)$, or equivalently, the zeros of the analytic function
\be{fk}
f(\kappa) = \left( \omega - \kappa M \right)^2 - \kappa \tanh{\kappa},
\ee
since the trivial pole $\kappa = 0$ of $F(\kappa)$ is already known. As one can notice, expression \rf{fk} is nothing else but the dispersion relation of the surface gravity waves travelling along a fluid layer with finite depth and infinite extension, moving uniformly along a rigid bottom \cite{LK2020}.

Since the characteristic equation \rf{fk} is not polynomial in $\kappa$ we cannot say a priori how many zeroes it has in the complex $\kappa$-plane. We determine this number numerically using the standard expression \cite{AF2003,DL1967}
\be{nu}
\nu = \frac{1}{2\pi i} \oint_{\mathcal{C}} \frac{f'(\kappa)}{f(\kappa)} \dd\kappa = \frac{R}{2\pi} \int_{0}^{2\pi} \frac{f'\left(Re^{i\theta}\right)}{f\left(Re^{i\theta}\right)} e^{i\theta} \dd\theta
\ee
that relates the integer number $\nu$ of zeros of $f(\kappa)$ in the complex $\kappa$-plane inside an arbitrary closed contour $\mathcal{C}$ which, for numerical purposes, we choose to be a circle of radius $R$ that is centered at the origin.

\begin{figure*}
\centering

\subfloat[$(\omega,M)=(0.5+0.1i,1.5)$]{
\includegraphics*[width=.5\textwidth]{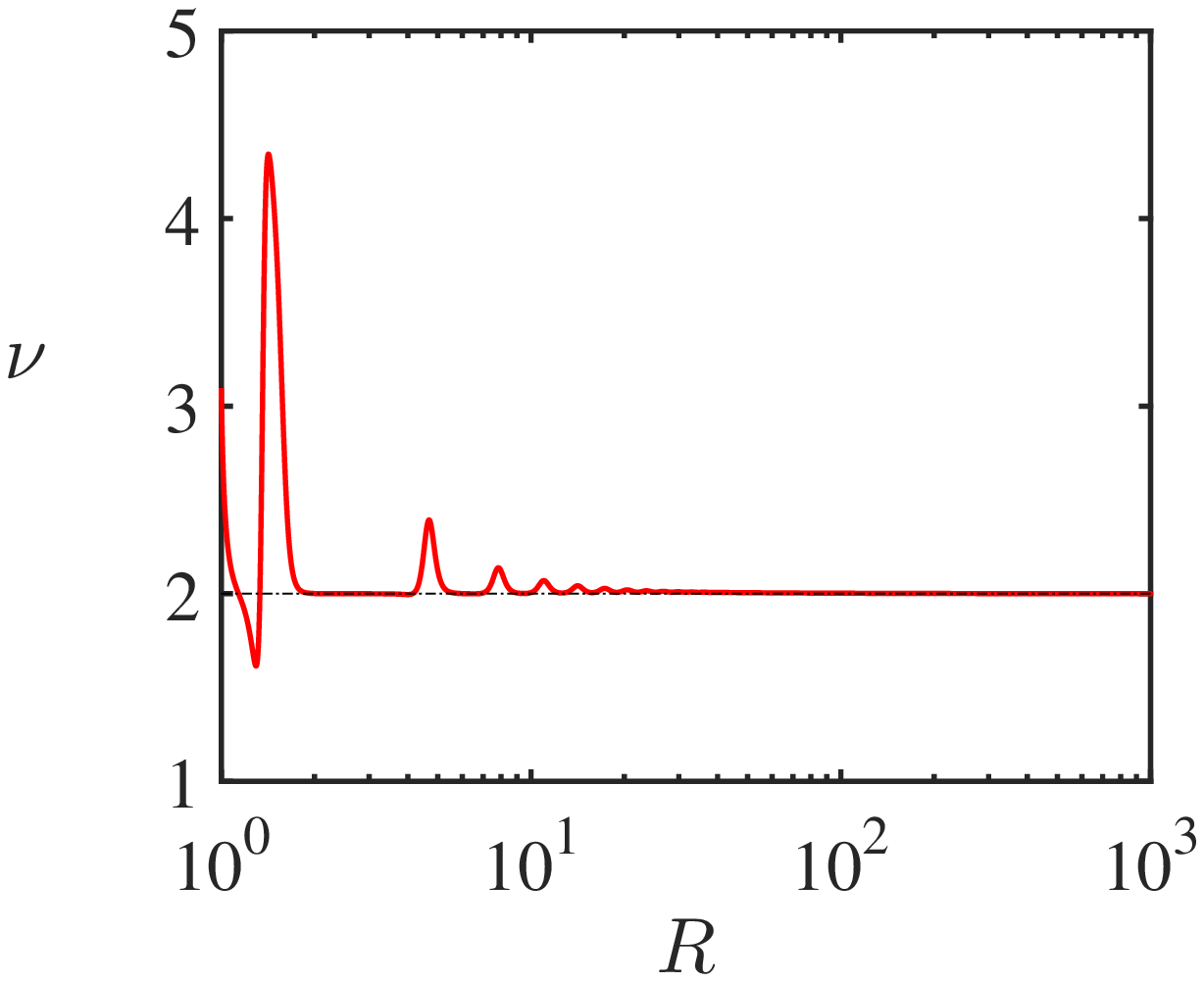} \label{nu1}} 
\subfloat[$(\omega,M)=(3\pi -4i,4.5)$]{
\includegraphics*[width=.5\textwidth]{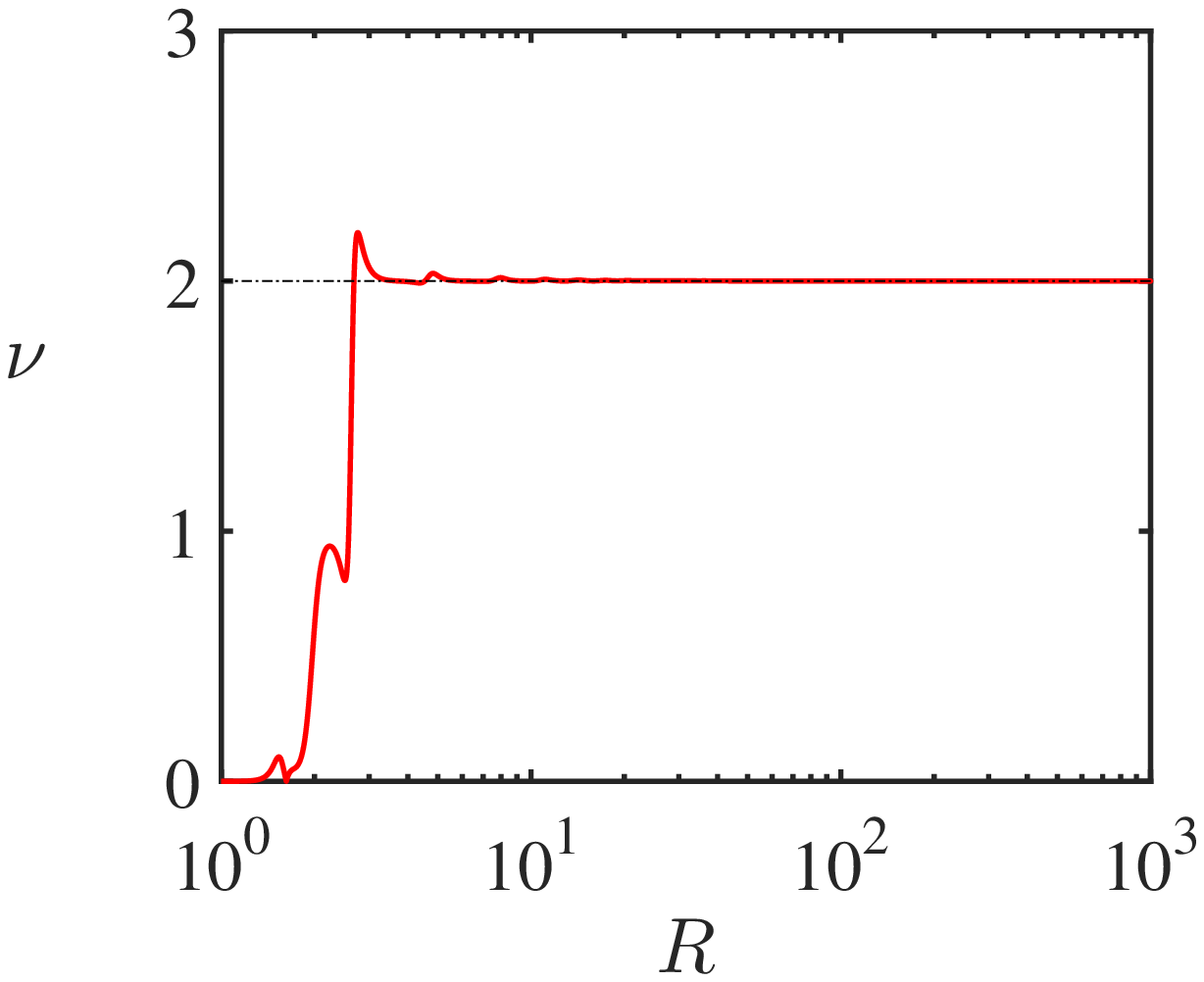} \label{nu2}}

\caption{Convergence of expression \rf{nu} over the radius $R$ for arbitrary pair of complex eigenfrequencies and Mach numbers. We used $N=20$ nodes for the trapezoidal quadrature. \label{cvnu}}
\end{figure*}

Solving \rf{nu} numerically using a standard trapezoidal quadrature with $\omega\in \mathbb{C}$ and $M\in\mathbb{R}$ for increasing values of $R$, demonstrates convergence to $\nu=2$ with a good accuracy that is evident in Fig.~\ref{cvnu}. This result is supported by the fact that both in the shallow water limit (when at $\kappa\to 0$ we have $\tanh(\kappa)\approx \kappa)$ and in the deep water limit (when $\tanh(\kappa)=1$ as $\kappa\to +\infty$), dispersion relation \rf{fk} reduces to a quadratic polynomial in $\kappa$ which therefore has only $2$ distinct roots.

\begin{figure*}
\centering

\subfloat[$(M_w,M)=(1,1.5)$]{
\includegraphics*[width=.5\textwidth]{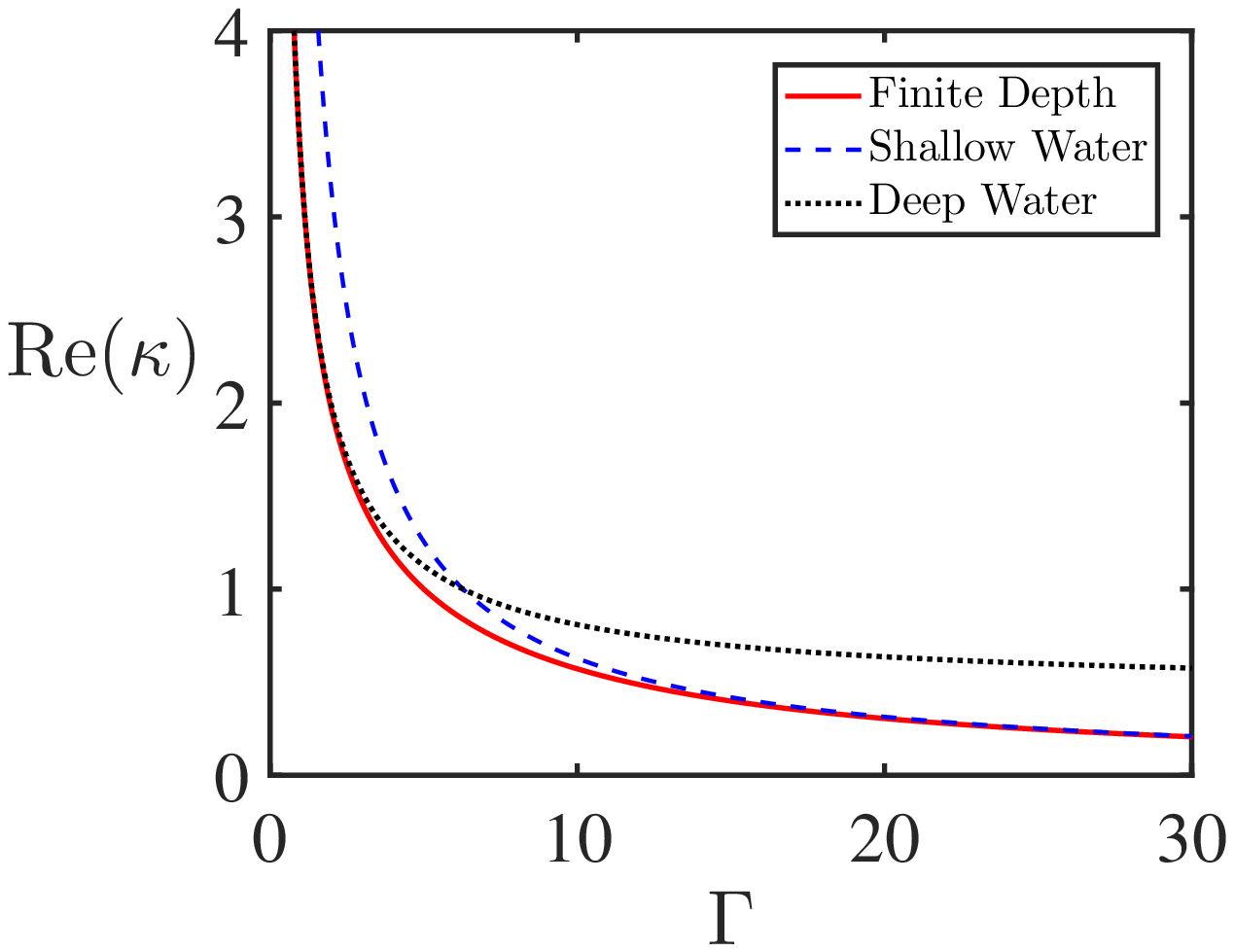} \label{r1}} 
\subfloat[$(M_w,M)=(0.6,2.4)$]{
\includegraphics*[width=.5\textwidth]{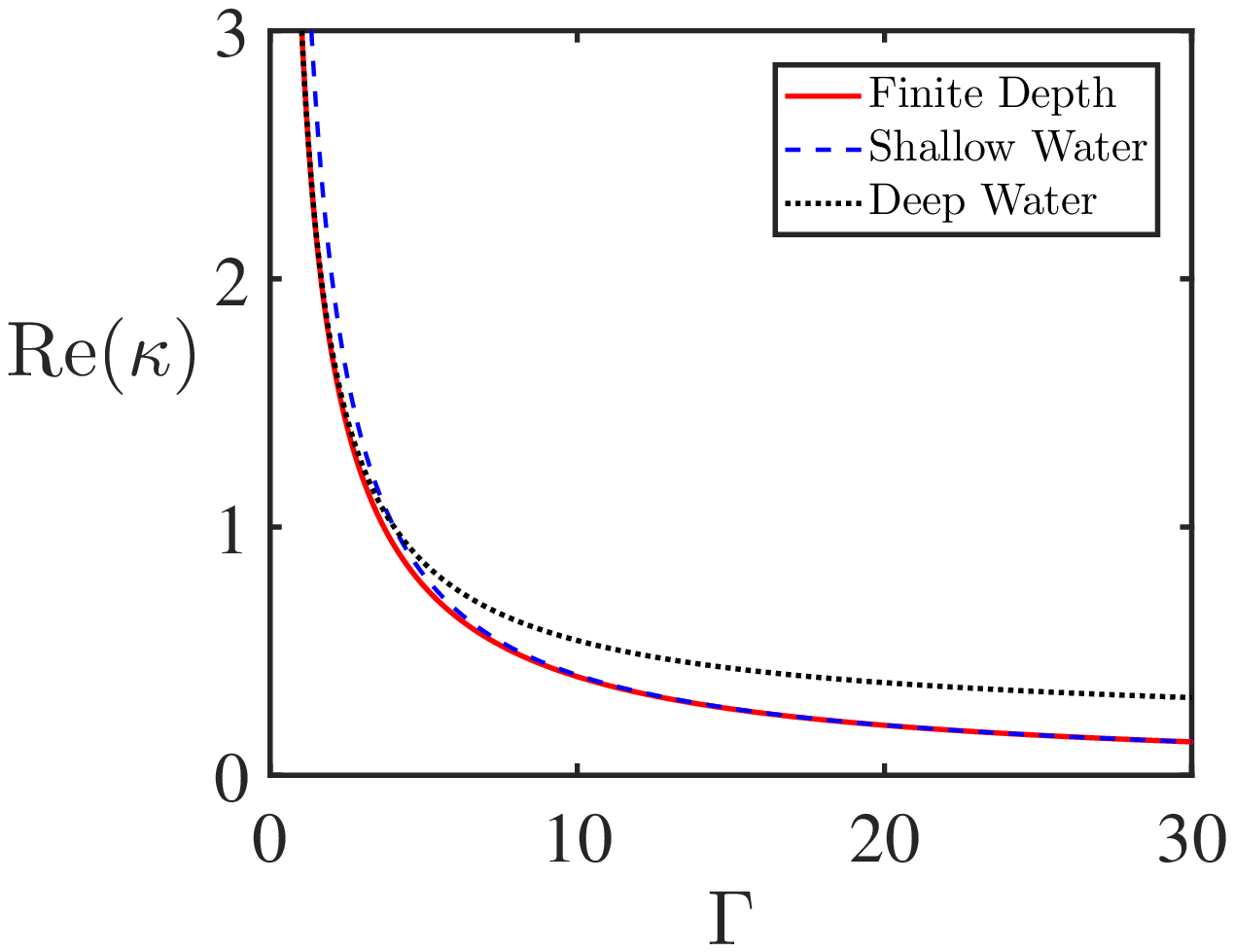} \label{r2}}

\caption{Convergence of zeros of \rf{fk} to the roots \rf{rsd} over $\Gamma$, for $\omega=\omega_n+\omega_i$ with fixed values of imaginary parts (a): $\omega_i=0.001$ and (b): $\omega_i=-0.1$. \label{cvroot}}
\end{figure*}

Now, when the number of zeros in expression \rf{fk} is established, we can use an iterative algorithm such as the standard Newton-Raphson method to locate precisely where these zeros lie in the complex $\kappa$-plane. In order for the algorithm to initiate, we need to provide an initial guess that is close enough to the exact value of the desired root. A natural choice is to use the roots of the dispersion relation \rf{fk} either in the shallow- or in the deep water limit 
\be{rsd}
\kappa^{SW}_{\pm} = \frac{\omega}{M \pm 1}, \quad \kappa^{DW}_{\pm} = \frac{2\omega M + 1 \pm \sqrt{4\omega M + 1}}{2M^2},
\ee
which depends on the length $\Gamma$ of the membrane. Indeed, since $\omega = \omega_n + i\omega_i$, where $\omega_n=n\pi M_w/\Gamma$ is the free membrane frequency \rf{omn} and $\omega_i\in\mathbb{R}$ is fixed, we have a clear numerical evidence that the zeros of \rf{fk} tend either to $\kappa^{SW}_{\pm}$ as $\Gamma \to \infty$ or to $\kappa^{DW}_{\pm}$ as $\Gamma \to 0$, see Fig.~ \ref{cvroot}.

Since the location of the poles of the integrand in \rf{v} is now determined, we continue our investigation by applying the residue theorem to this improper integral in the same way as we did in the shallow water limit. This time, due to the presence of a pole at the origin, the contour we consider is decomposed as
\begin{align}
\label{contour2}
\mathcal{C} = [-R , -\epsilon ] \cup \Delta_{\epsilon} \cup &[ \epsilon , R] \cup \Delta_R, \quad \Delta_{\epsilon}= \{ \epsilon e^{it}, -\pi \leq t \leq 0 \} , \nn \\
&\Delta_R= \{ Re^{it},0 \leq t \leq \pi \} ,
\end{align}
where $0 < \epsilon \ll 1$ and $R \gg 1$.

Using the oriented curve \rf{contour2} in the contour integral \rf{v} yields
\begin{align}
\label{sum_int}
\oint_{\mathcal{C}} F(\kappa) \dd\kappa = &\int_{-R}^{-\epsilon} F(\kappa) \dd\kappa + \int_{\Delta_{\epsilon}} F(\kappa) \dd\kappa \nn \\
&+ \int_{\epsilon}^{R} F(\kappa) \dd\kappa + \int_{\Delta_R} F(\kappa) \dd\kappa .
\end{align}

Adopting the same argument as in the shallow water case and taking into account that $F(z)$ is a continuous function for $z \in \Delta_R$ and that $\lim_{R\to +\infty} \vert g(Re^{i\theta}) \vert =0$, where $g(\kappa) = F(\kappa) e^{-i\kappa x}$ and $\theta \in [0,\pi]$, we find that the contribution of the arc integral over $\Delta_R$ vanishes as we enlarge the radius, according to Jordan's lemma \cite{AF2003}.

The contribution of the arc integral over $\Delta_{\epsilon}$ in \rf{sum_int} also vanishes as $\epsilon\to 0$ because
\be{arc_int_eps}
\lim\limits_{\epsilon\to 0} \int_{\Delta_{\epsilon}} F(\kappa) \dd\kappa = \lim\limits_{\epsilon\to 0} i\epsilon \int_{-\pi}^{0} e^{i\theta} F(\epsilon e^{i\theta}) \dd \theta = 0.
\ee

Finally, taking the two limits of $R\to +\infty$ and $\epsilon\to 0$ simultaneously and then applying the residue theorem to \rf{sum_int}, we obtain
\begin{widetext}
\be{res_th}
\lim\limits_{\substack{\mathllap{R} \to \mathrlap{+\infty} \\ \mathllap{\epsilon} \to \mathrlap{0}}} \quad \left[ \int_{-R}^{-\epsilon} F(\kappa) \dd\kappa + \int_{\epsilon}^{R} F(\kappa) \dd\kappa \right] = \text{PV} \int_{-\infty}^{+\infty} F(\kappa) \dd\kappa = 2\pi i\sum_{j=1}^{2} \text{res} \left( F(\kappa) , \kappa_j \right) ,
\ee
\end{widetext}
where the improper integral has to be taken in the sense of Cauchy Principal Value and where $\kappa_j$ are the zeros of \rf{fk} lying in the upper-half plane. 

A similar argument works as well for an oriented contour $\mathcal{C}$ in the lower half of the complex $\kappa$-plane, to enclose the poles with the negative imaginary parts. 

Now we are prepared to recover the function $v(x,\omega)$ from the general expression \rf{v} as
\be{vres}
v(x,\omega) = i\sum_{j=1}^{2} \text{res} \left( F(\kappa) , \kappa_j \right), \quad \kappa_j \in \mathbb{C},
\ee
which constitutes the main result of this section.

Note that owing to the fact that the poles are computed numerically and that $F$ in \rf{vres} cannot be expanded in the Laurent series, we need to calculate the residues in \rf{vres} with an alternative, however equivalent, expression to that used for obtaining \rf{res_lwl}. Namely, if we consider a meromorphic function $M(z)=a(z)/b(z)$ with a simple pole $z_1$ that is a root of $b(z)$, then the residue for $M(z)$ at $z_1$ reads as \cite{AF2003}
\be{res1}
\text{res} \left( M(z) , z_1 \right) = \lim\limits_{z \to z_1} \frac{a(z)}{b'(z)} .
\ee
If the function $M(z)$ has a double pole $z_2$, we find similarly that \cite{AF2003}
\be{res2}
\text{res}\left( M(z) , z_2 \right) = \lim\limits_{z \to z_2} \frac{6a'(z)b''(z) - 2a(z)b'''(z)}{3\left[b''(z)\right]^2}.
\ee
Expressions \rf{res1} and \rf{res2} will be utilized in the numerical treatment of equation \rf{wave_equation}, for instance, in the computation of the coresponding Jacobian.

\subsection{Galerkin discretization and reduction to an algebraic nonlinear eigenvalue problem}

The last step in solving the integro-differential equation \rf{wave_equation} numerically is to introduce a modal form for the displacement $\xi(x)$ that respects the boundary conditions $\xi(0)=\xi(\Gamma)=0$. For this purpose, we introduce the following Galerkin decomposition, based on a superposition of modes of a free membrane vibrating in vacuum
\be{galerkin}
\xi(x) = \sum^{N}_{j=1} \gamma_j \xi^{(0)}_j, \quad \xi^{(0)}_j = \sin{\left( \frac{j\pi x}{\Gamma} \right)}.
\ee

Substituting \rf{galerkin} into \rf{wave_equation} and using the orthogonality of the modes \rf{galerkin} while integrating over the membrane chord, we find
\be{Fij}
\pazocal{F}_{ij} (\omega) \gamma_j = \sum^{N}_{j=1} \gamma_j  \int_{0}^{\Gamma} \mathcal{W}(\xi_j^{(0)};x,\omega) \xi_i^{(0)} \dd x = 0.
\ee

The expression \rf{Fij} represents the $j$-th scalar equation of the algebraic nonlinear eigenvalue problem in $\omega$, which matrix pencil can be written as
\be{nep}
\pazocal{F}(\omega) = -\omega^2 \pazocal{I} + \alpha \pazocal{P} (\omega) + \pazocal{K} ,
\ee
where the matrices in \rf{nep} can be explicitly recovered through the following expressions \cite{V2012}
\ba{}
\pazocal{I}_{ij} &=& \delta_{ij}, \nn \\
\pazocal{K}_{ij} &=& \left( \frac{j\pi M_w}{\Gamma} \right)^2 \delta_{ij}, \nn \\
\pazocal{P}_{ij} &=& \frac{2}{\Gamma} \int_{0}^{\Gamma} P (\xi^{(0)}_j;x,\omega) \xi^{(0)}_i \dd x \label{pij} ,
\ea
with $\delta_{ij}$ standing for the Kronecker delta. The function $P$ follows from \rf{wave_equation} after application of the Leibniz rule \rf{leib}:
\be{P}
P(\xi(x);x,\omega) = i\omega \left( u*v \right) - M \left[ \left( u * \frac{\partial v}{\partial x} \right) + u(x,\omega) v(0,\omega) \right],
\ee
where $v(x)$ is given by \rf{vres} and 
\be{dvdx}
\frac{\partial v}{\partial x} = \frac{i}{2\pi} \int_{-\infty}^{+\infty} \frac{[\kappa - (\omega - \kappa M)^2 \tanh{\kappa}] e^{i\kappa x} }{(\omega - \kappa M)^2 - \kappa\tanh{\kappa}} \dd\kappa
\ee
should be computed separately with the similar approach. Indeed, integral \rf{dvdx} is nothing else but a slightly modified version of expression \rf{v}, without the pole located at the origin. Therefore, a similar analysis  applied to \rf{dvdx} reduces it to 
\be{dvdxa}
\frac{\partial v}{\partial x}=iv(x)=-\sum_{j=1}^{2} \text{res} \left( F(\kappa) , \kappa_j \right), \quad \kappa_j \in \mathbb{C}.
\ee
With the expressions \rf{vres} and \rf{dvdxa} we can recover an explicit form of all the matrices constituting the matrix pencil \rf{nep} of the nonlinear eigenvalue problem by direct numerical computation.

\subsection{Jacobian of the nonlinear matrix pencil $\pazocal{F}(\omega) $}

Nonlinear eigenvalue problems constitute nowadays a challenging and ongoing research topic for a whole community of mathematicians. State-of-the-art reviews \cite{MV2004,BH2015,GT2017} identify and classify different classes of methods to solve them, depending on the nonlinearity. Most of the known methods are either based on the Newton-Raphson iterative process or use contour integration \cite{BH2015}. In the present paper we prefer the former.

The Newton-Raphson iterative process needs the derivative of the pencil \rf{nep} with respect to $\omega$, or Jacobian of the system. In order to reach an acceptable convergence rate for the method, we compute the analytical form of the Jacobian instead of approximating it numerically. 

From \rf{nep} we find
\be{jacobian}
\frac{\dd \pazocal{F}}{\dd\omega} = -2\omega\pazocal{I} + \alpha \frac{\dd \pazocal{P}}{\dd\omega},
\ee
where the derivative $\frac{\dd \pazocal{P}}{\dd\omega}$ involves the derivative of the function $P(x,\omega)$ as defined in \rf{P}. Applying the Leibniz rule \rf{leib} once again, we obtain
\begin{widetext}
\be{dP}
\frac{\partial P}{\partial\omega} = i \left( u*v \right) + i\omega\left[  \left( \frac{\partial u}{\partial \omega} * v \right) +  \left( u * \frac{\partial v}{\partial \omega} \right) \right]
- M \left[ \left( \frac{\partial u}{\partial\omega} * \frac{\partial v}{\partial x} \right) +  \left( u * \frac{\partial^2 v}{\partial x\partial\omega} \right) + \frac{\partial u}{\partial\omega}(x,\omega)v(0,\omega) + u(x,\omega) \frac{\partial v}{\partial\omega}(0,\omega) \right],
\ee
\end{widetext}
where $u$ is given by \rf{u}, $v$ by \rf{vres}, and $\frac{\partial v}{\partial x}$ by \rf{dvdxa}.

\begin{figure*}
\centering

\subfloat[$n=1$]{
\includegraphics*[width=.5\textwidth]{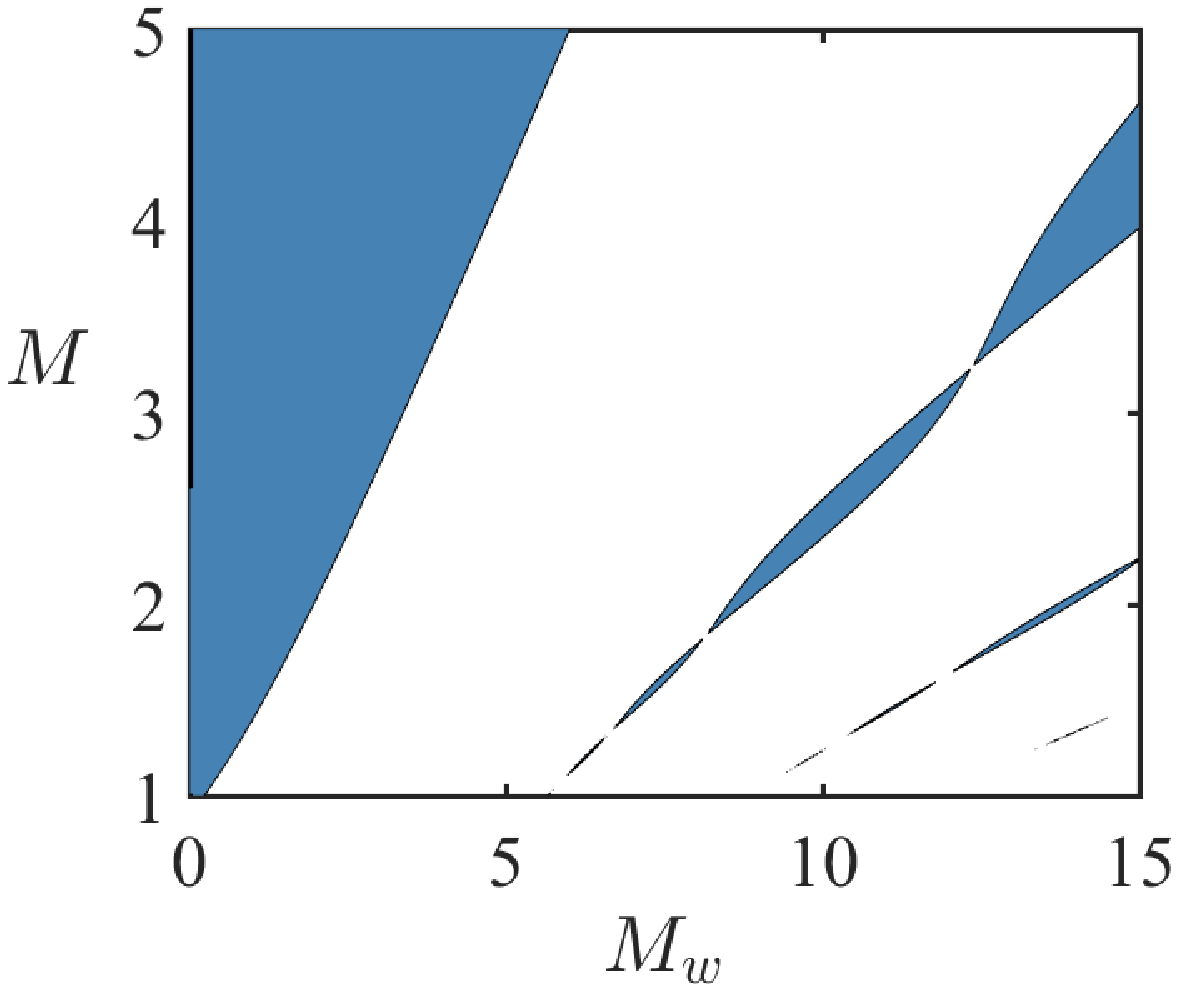}} \hspace*{-3em}
\subfloat[$n=2$]{
\includegraphics*[width=.5\textwidth]{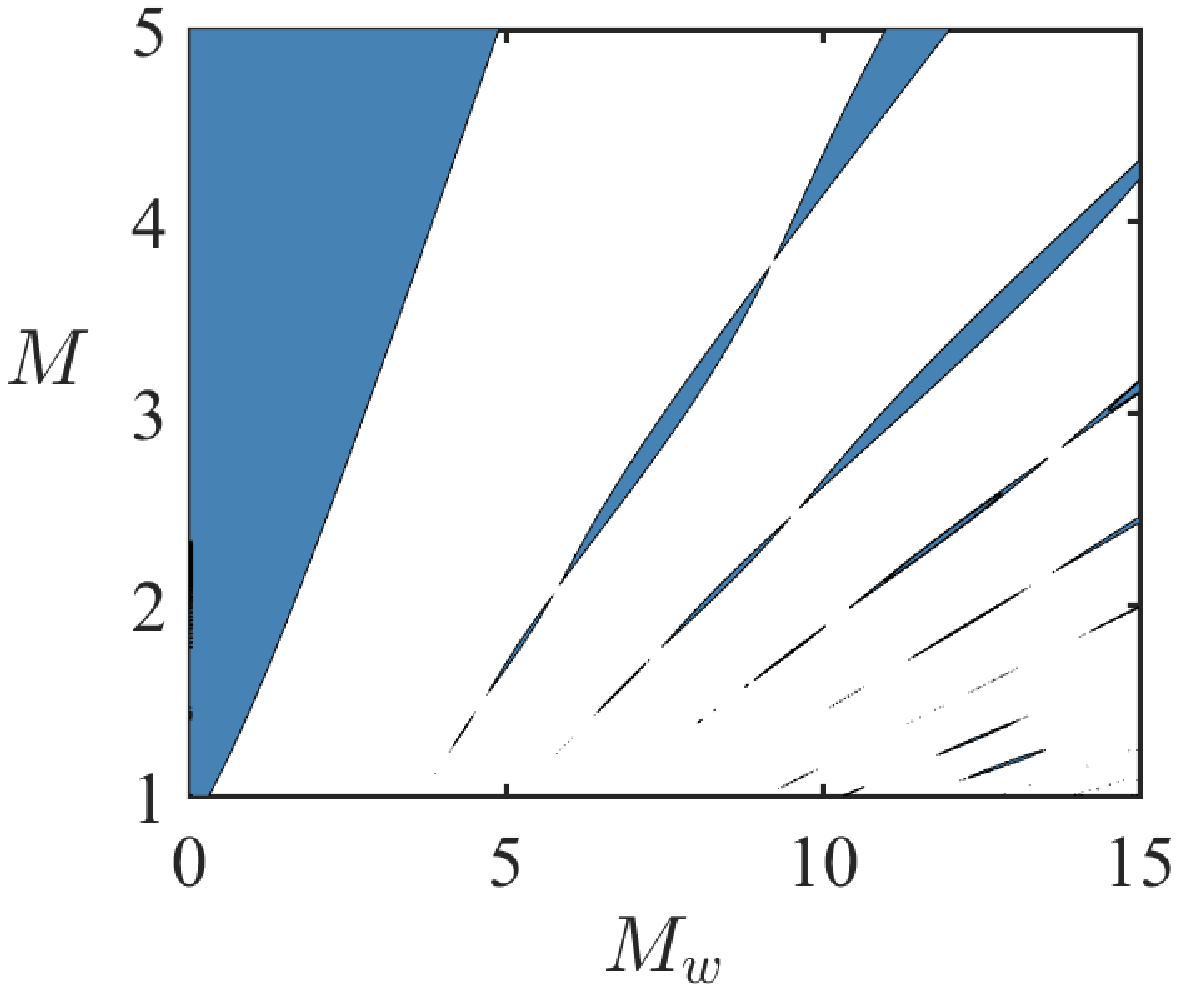}} \\ \vspace*{-1.5em}
\subfloat[$n=3$]{
\includegraphics*[width=.5\textwidth]{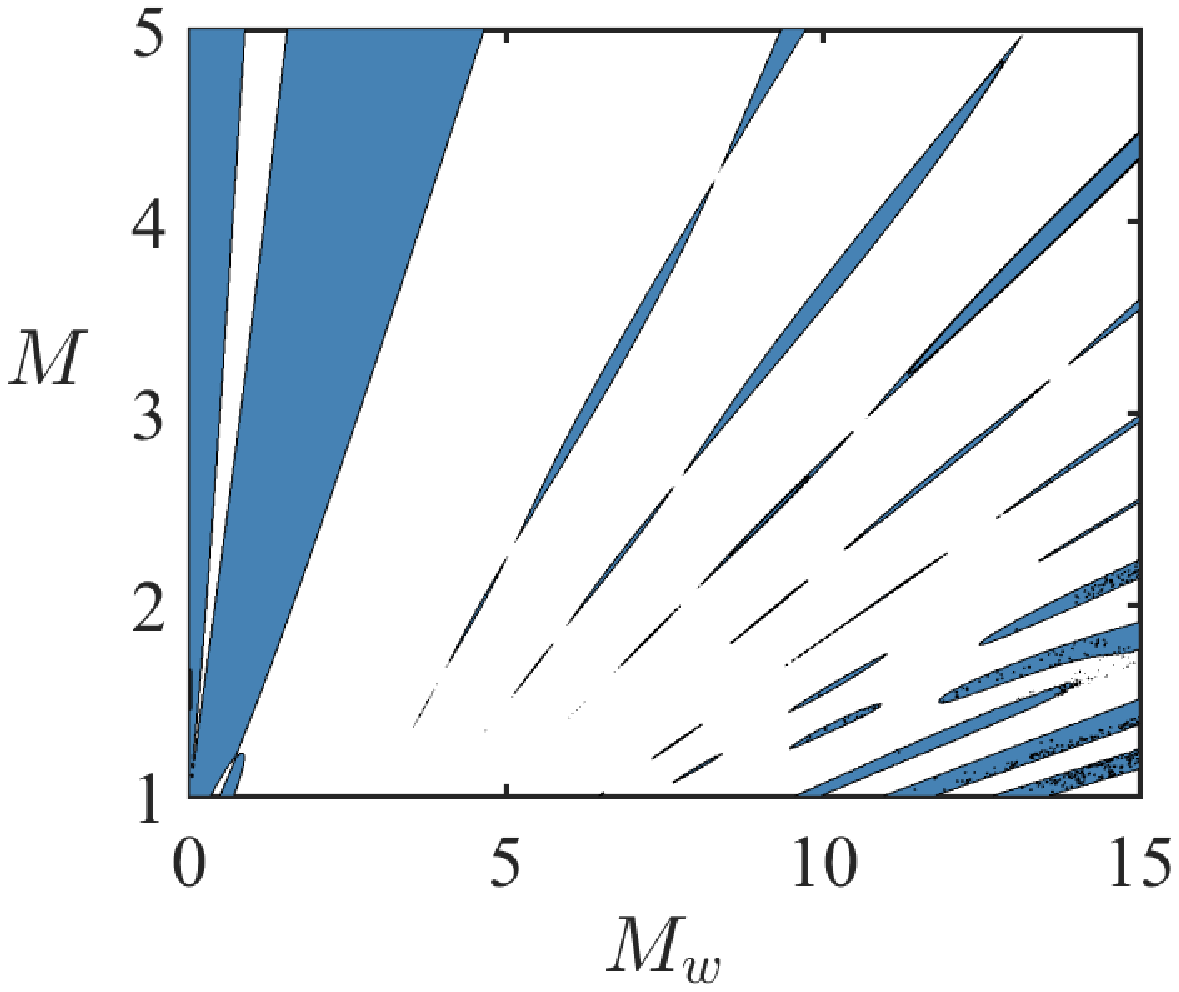}} \hspace*{-3em}
\subfloat[$n=4$]{
\includegraphics*[width=.5\textwidth]{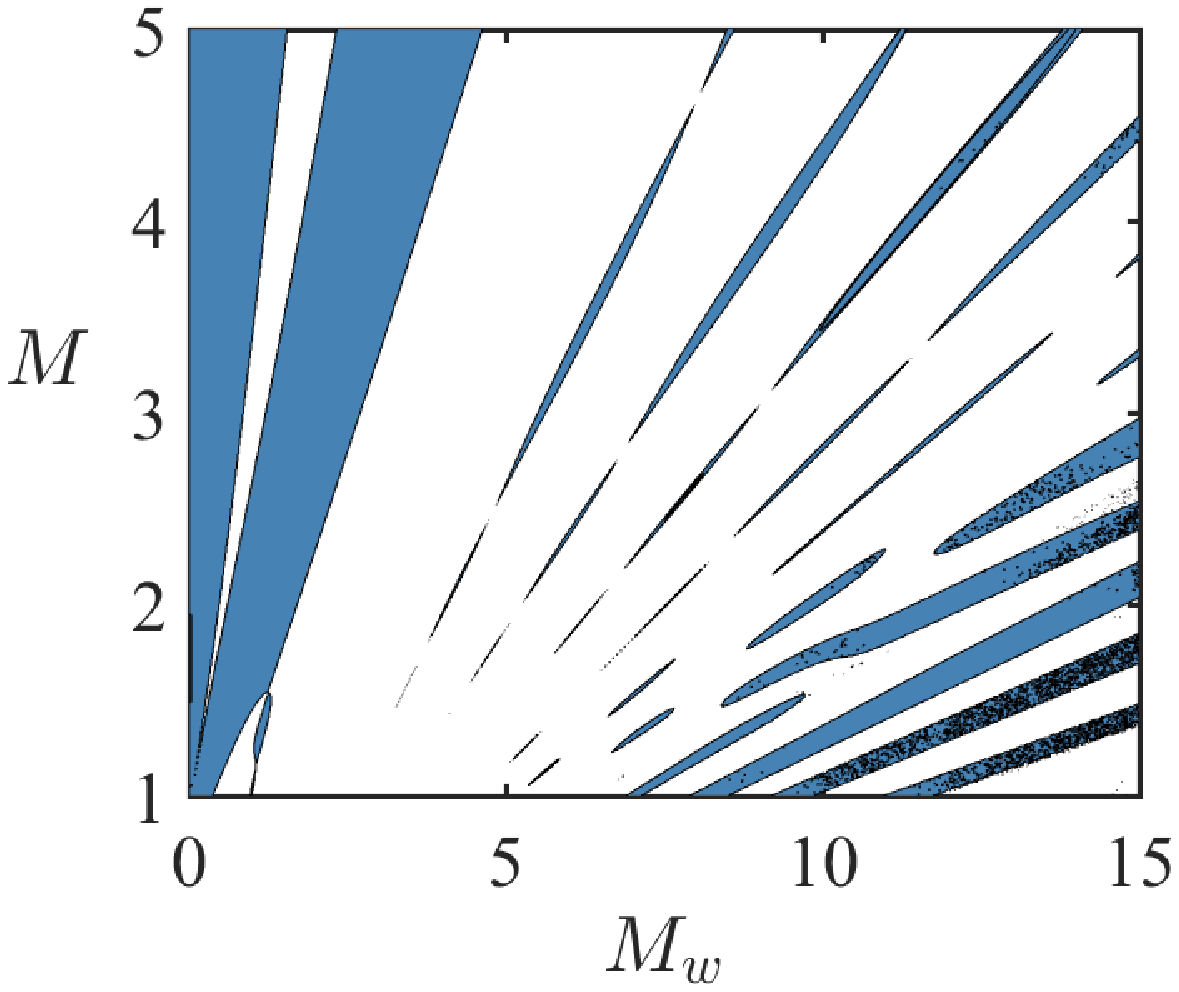}}

  \caption{Stability maps in the $(M_w, M)$-plane from the numerical solution of the matrix pencil \rf{nep} for $\Gamma=10$, $\alpha=10^{-4}$ and $n$ according to the legend. \label{FD_MMw}}
\end{figure*}

As in the previous case, the higher-order derivatives of \rf{v} involve slightly different improper integrals. Their explicit expressions are recovered as
\ba{dvdw}
\frac{\partial v}{\partial \omega} &=& \frac{1}{\pi} \int_{-\infty}^{+\infty} \frac{(\omega-\kappa M)(\tanh^2{\kappa} - 1) e^{i\kappa x} }{[(\omega - \kappa M)^2 - \kappa\tanh{\kappa}]^2} \dd\kappa , \nn \\
\frac{\partial^2 v}{\partial x \partial \omega} &=& \frac{i}{2\pi} \int_{-\infty}^{+\infty} \frac{\kappa(\omega-\kappa M)(\tanh^2{\kappa} - 1) e^{i\kappa x} }{[(\omega - \kappa M)^2 - \kappa\tanh{\kappa}]^2} \dd\kappa .
\ea

Despite the numerator in the integrand of \rf{dvdw} is different from \rf{v}, the poles remain identical to \rf{dvdx}, with the only difference that they are of second order. Therefore, we can apply the same procedure involving the residue theorem as we did in the derivation of \rf{vres} (with a particular attention to the pre-factors of the integrals). As the poles are no longer simple, we need to use the expression \rf{res2} for the residues when computing derivatives \rf{dvdw}.

\subsection{Numerical evaluation of the integral in \rf{pij}}

In our approach, the integrals in the expression \rf{pij} and its derivative, as well as in all the convolutions, are integrated using numerical quadrature. Since the domain of integration is one-dimensional and we want to keep high accuracy in our numerical scheme, we prefer to use the Legendre-Gauss-Lobatto quadrature rule to approximate every integral. The nodes and weights of this spectral collocation method are recovered using the Golub-Welsch algorithm \cite{GW1969}, which is based on the inversion of a linear system obtained from the three-term recurrence relation for Legendre polynomials. The computed eigenvalues correspond to the quadrature points, while the eigenvectors are used to recover the weights. In this method, the nodes are defined on an interval $[-1,1]$ before being mapped, using a linear transformation, to the interval of integration $[0,\Gamma]$. This spectral quadrature is able to reach computer accuracy with less than $20$ nodes of discretization of the interval $[0,\Gamma]$ and is thus used all over our code.

\begin{figure*}
\centering

\subfloat[$n=1$]{
\includegraphics*[width=.5\textwidth]{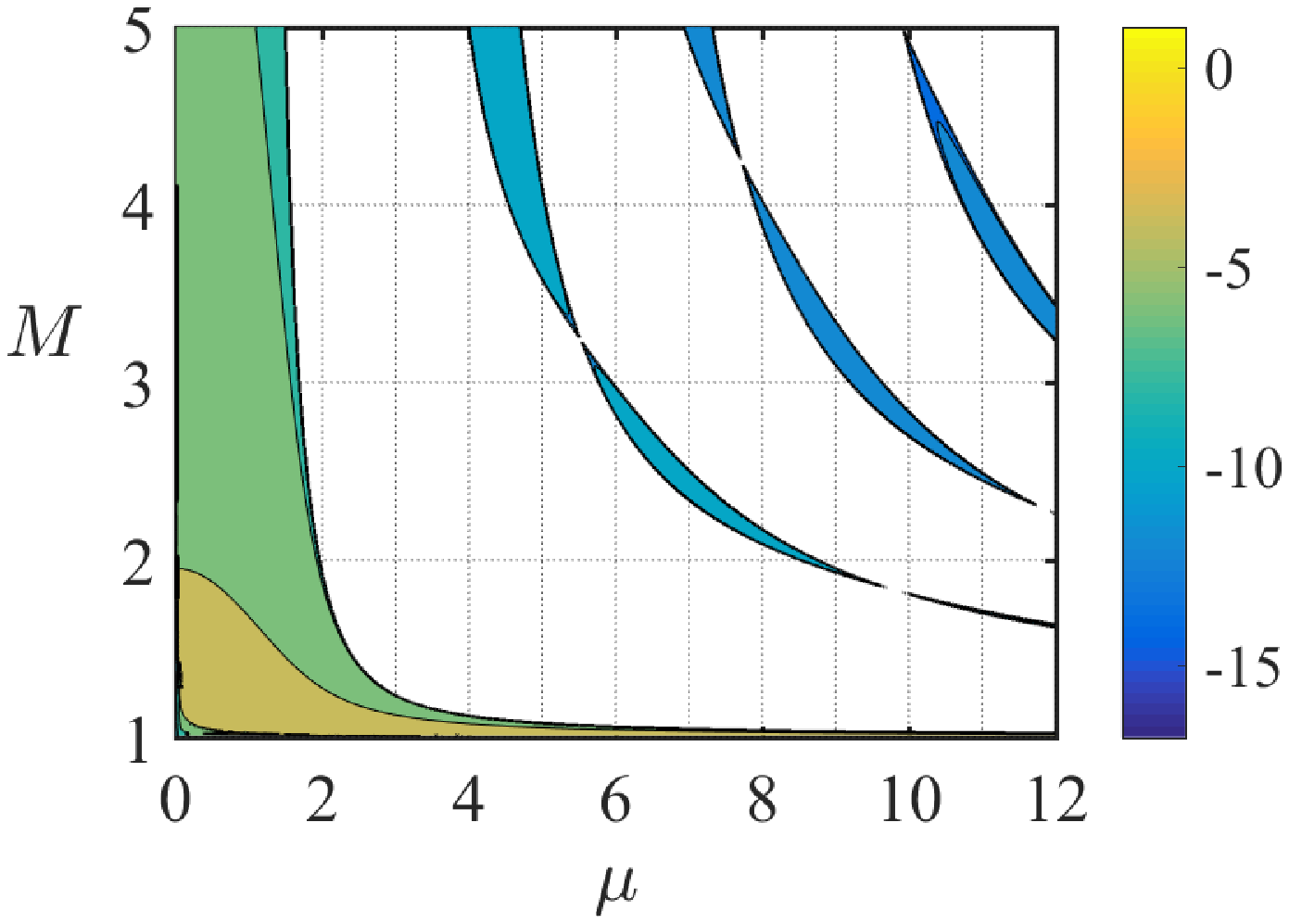}} \hspace*{-1em}
\subfloat[$n=2$]{
\includegraphics*[width=.5\textwidth]{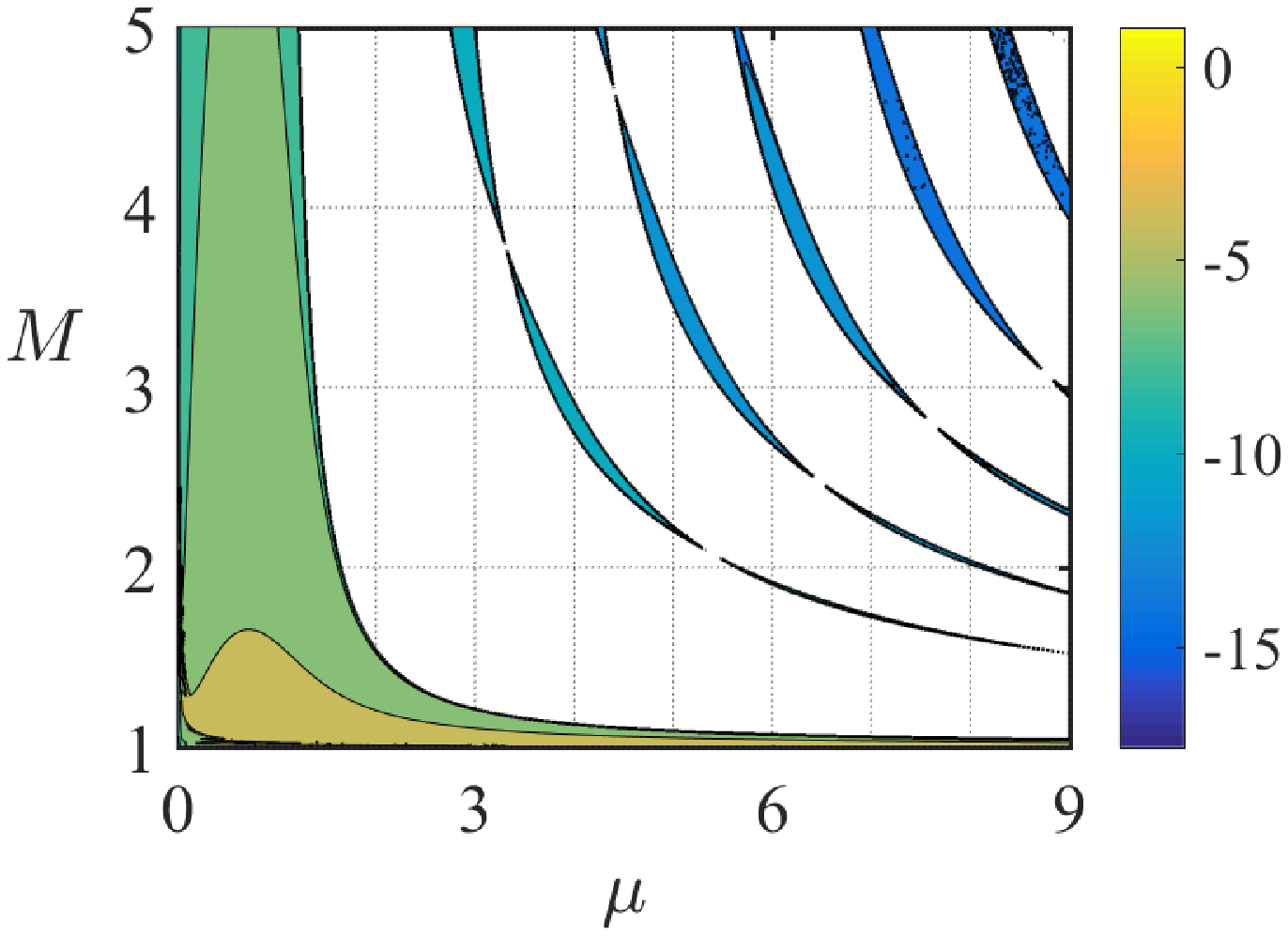}} \\ \vspace*{-1.5em}
\subfloat[$n=3$]{
\includegraphics*[width=.5\textwidth]{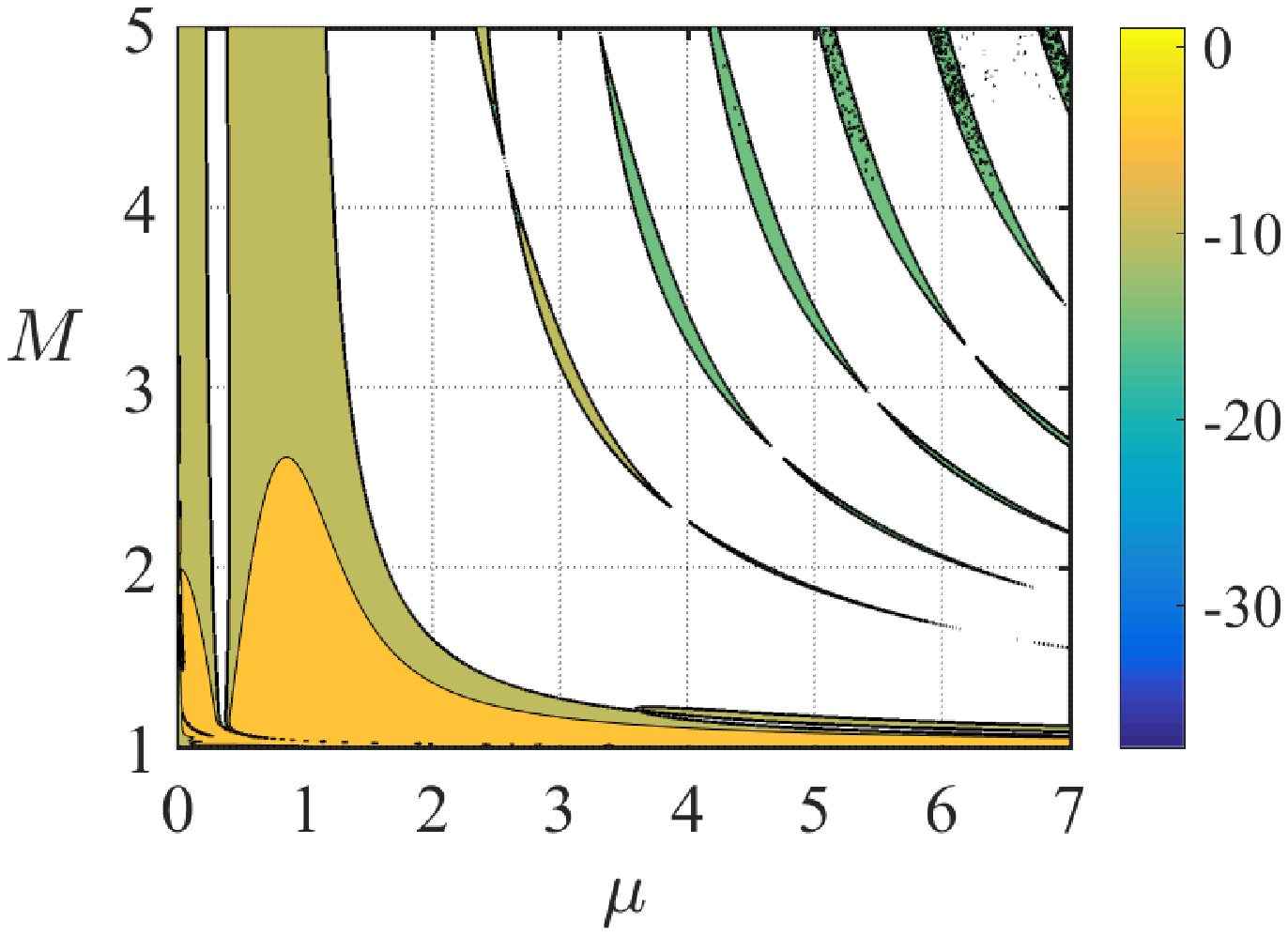}} \hspace*{-1em}
\subfloat[$n=4$]{
\includegraphics*[width=.5\textwidth]{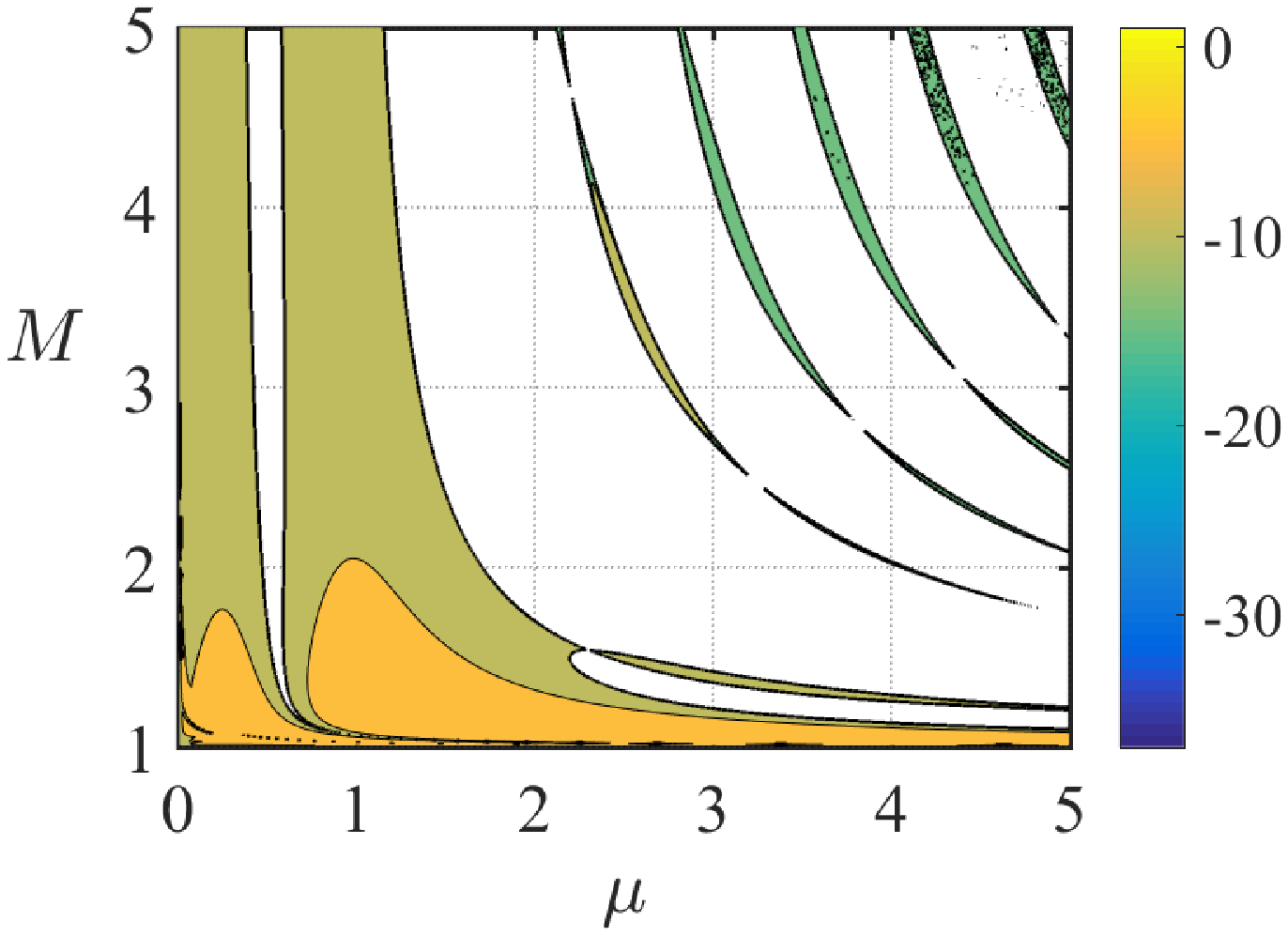}}

  \caption{Stability maps in the $(\mu, M)$-plane from the numerical solution of the matrix pencil \rf{nep} for $\Gamma=10$, $\alpha=10^{-4}$ and $n$ according to the legend. We use a logarithmic scale for the growth rate.
   \label{FD_Mmu}}
\end{figure*}

\subsection{Newton-like numerical method for solving the nonlinear eigenvalue problem}

Now when every element of the matrix pencil \rf{nep} and its derivative \rf{jacobian} can be recovered by direct numerical computation, we shall introduce the method that we will be using throughout the stability analysis of the nonlinear eigenvalue problem. This numerical method is introduced in the recent review \cite{MV2004} and is designed to solve the characteristic equation $\det{\pazocal{F}(\omega)}=0$ from the inversion of successive linear problems. As always in Newton-like methods, it has to start with an initial guess that is close enough to the exact solution $\omega$. In our case, due to strong nonlinearity in $\omega$ in the term \rf{P}, we shall restrict ourselves to  reasonably low values of $\alpha$. Indeed, this parameter acts as a linear factor at the nonlinear in $\omega$ operator and keeping it sensibly small prevents us from departing too far from the free membrane solution (corresponding to $\alpha=0$). This allows us to initiate our algorithm by choosing the eigenfrequency of the free membrane \rf{omn} as an initial starting point. With this first guess $\omega^{(0)}=\omega_n$, the method of successive linear problems is an iterative routine, where the $p$-th iteration requires to solve a linear eigenvalue problem 
\be{slp}
\pazocal{F}(\omega^{(p)}) \bm{u} = \theta \pazocal{F}'(\omega^{(p)}) \bm{u} .
\ee
After inversion of expression \rf{slp}, we re-initiate the method as follows
\be{sli}
\omega^{(p+1)} = \omega^{(p)} - \theta ,
\ee
where $\theta$ is chosen to be the smallest eigenvalue of \rf{slp} in the absolute value. As expected from the method, we can easily reach quadratic convergence in $\omega$.

The computational method presented through this section has been fully implemented and parallelized in MATLAB using the Parallel Computing Toolbox available from the software. All stability maps in the parameter spaces that will be presented in the next section are recovered from the direct computation of growth rates ${\rm Im}(\omega)$ of the nonlinear matrix pencil \rf{nep}.

\subsection{Stability maps for the finite-chord Nemtsov membrane in the finite-depth flow}

\begin{figure*}
\centering

\subfloat[$n=1$]{
\includegraphics*[width=.33\textwidth]{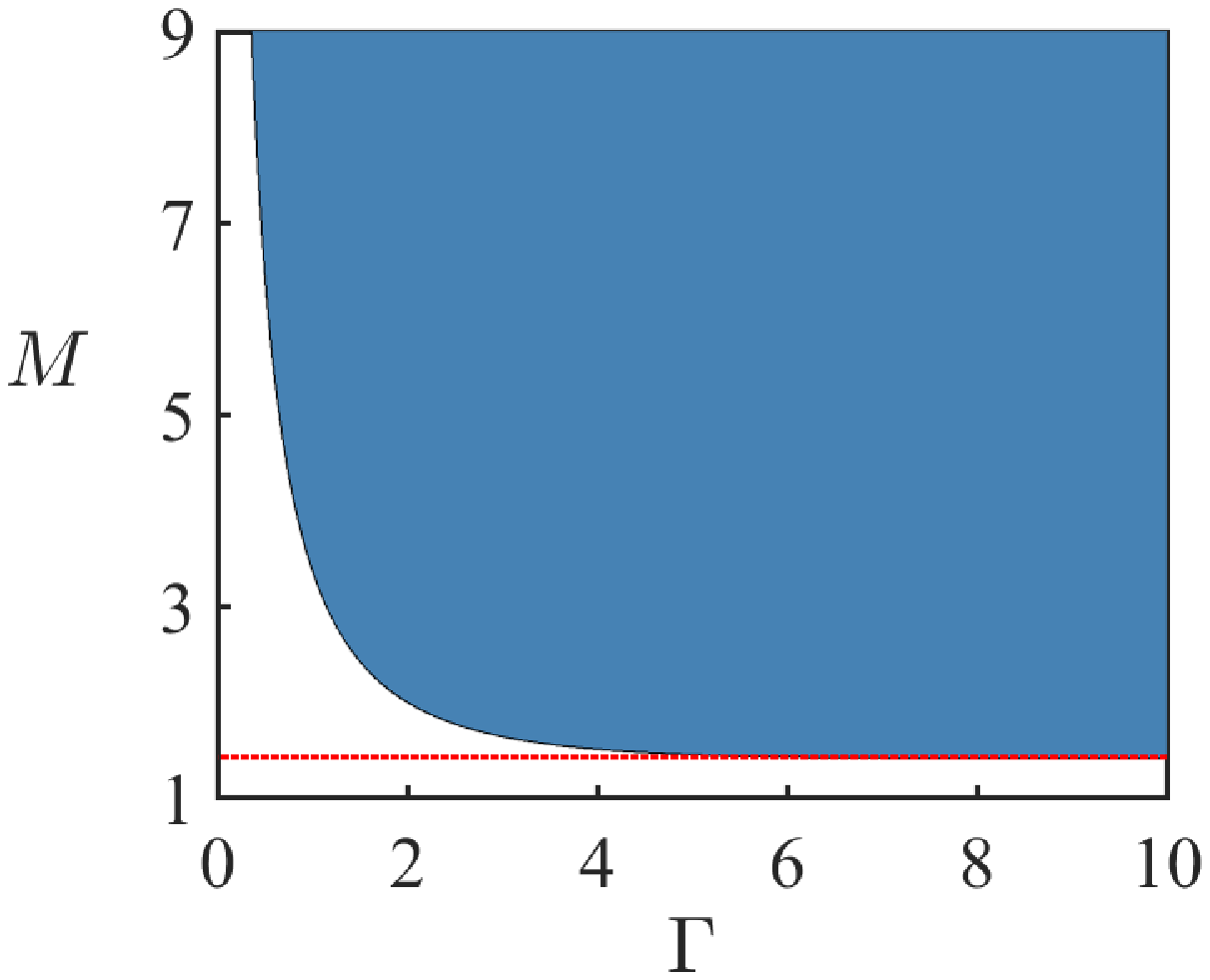} \label{FD_n12a}} \hspace*{-3em}
\subfloat[$n=2$]{
\includegraphics*[width=.33\textwidth]{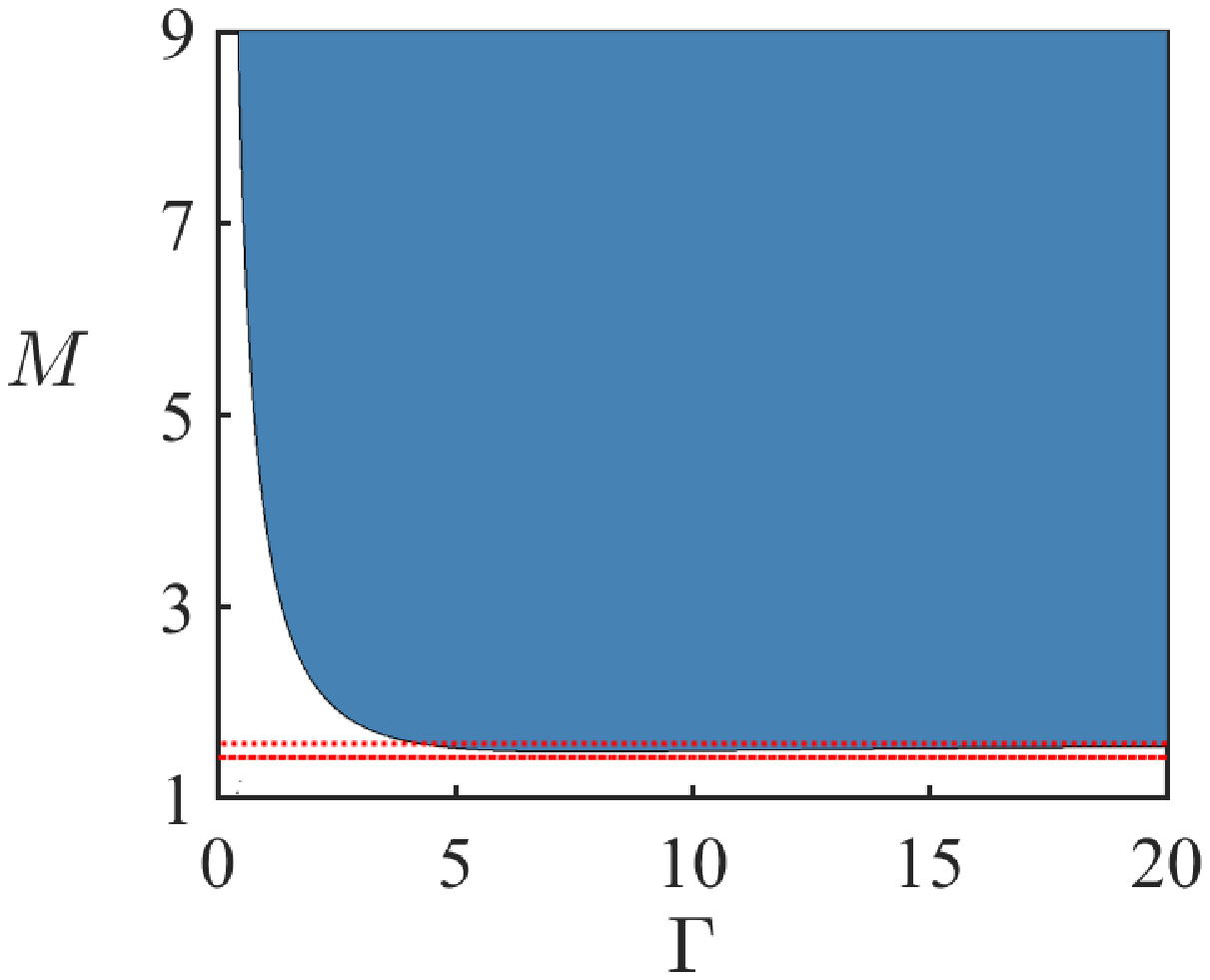} \label{FD_n12b}} \hspace*{-3em}
\subfloat[$n=3$]{
\includegraphics*[width=.33\textwidth]{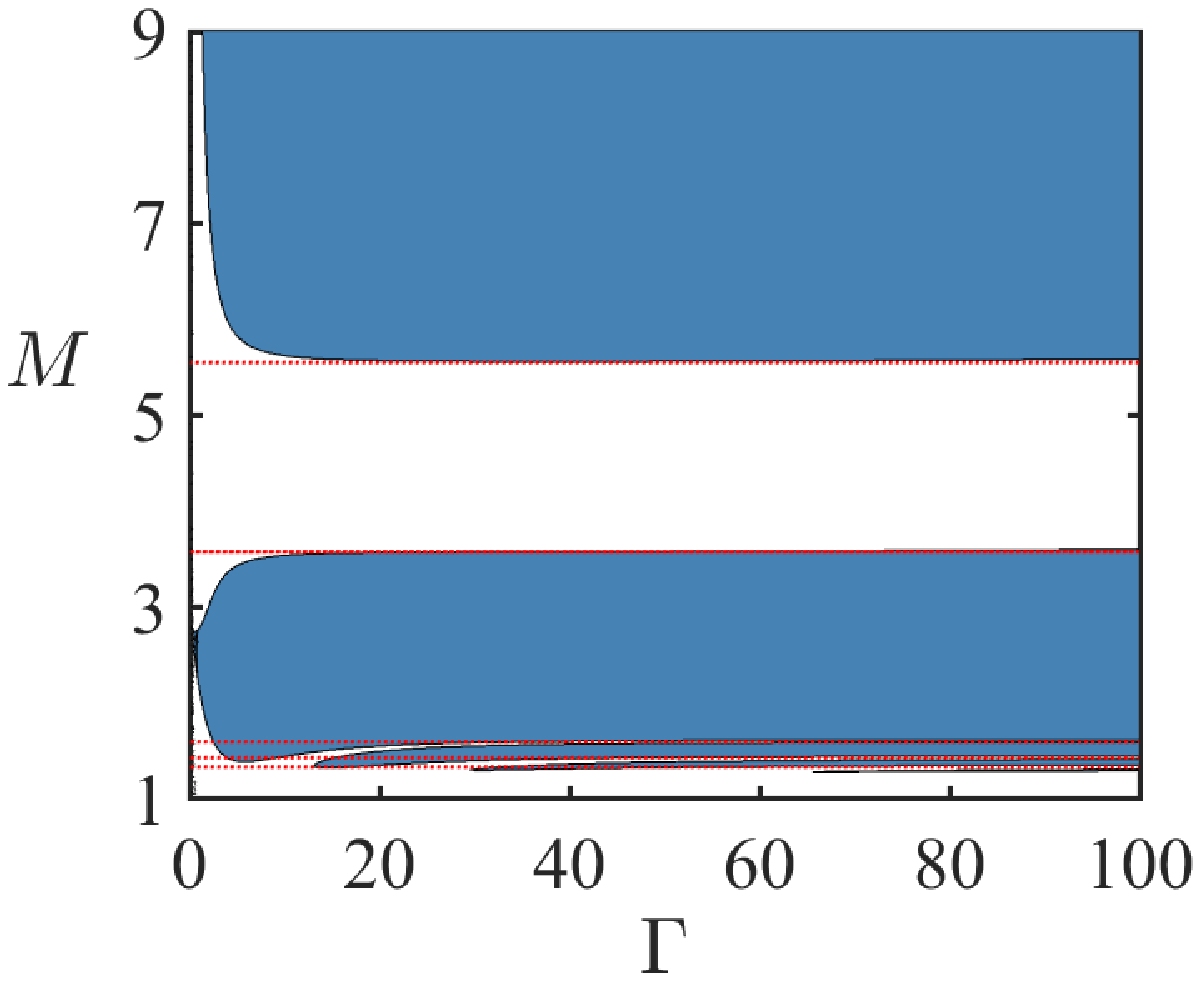} \label{FD_n3a}} \\ \vspace*{-1.5em}
\subfloat[$n=3$]{
\includegraphics*[width=.33\textwidth]{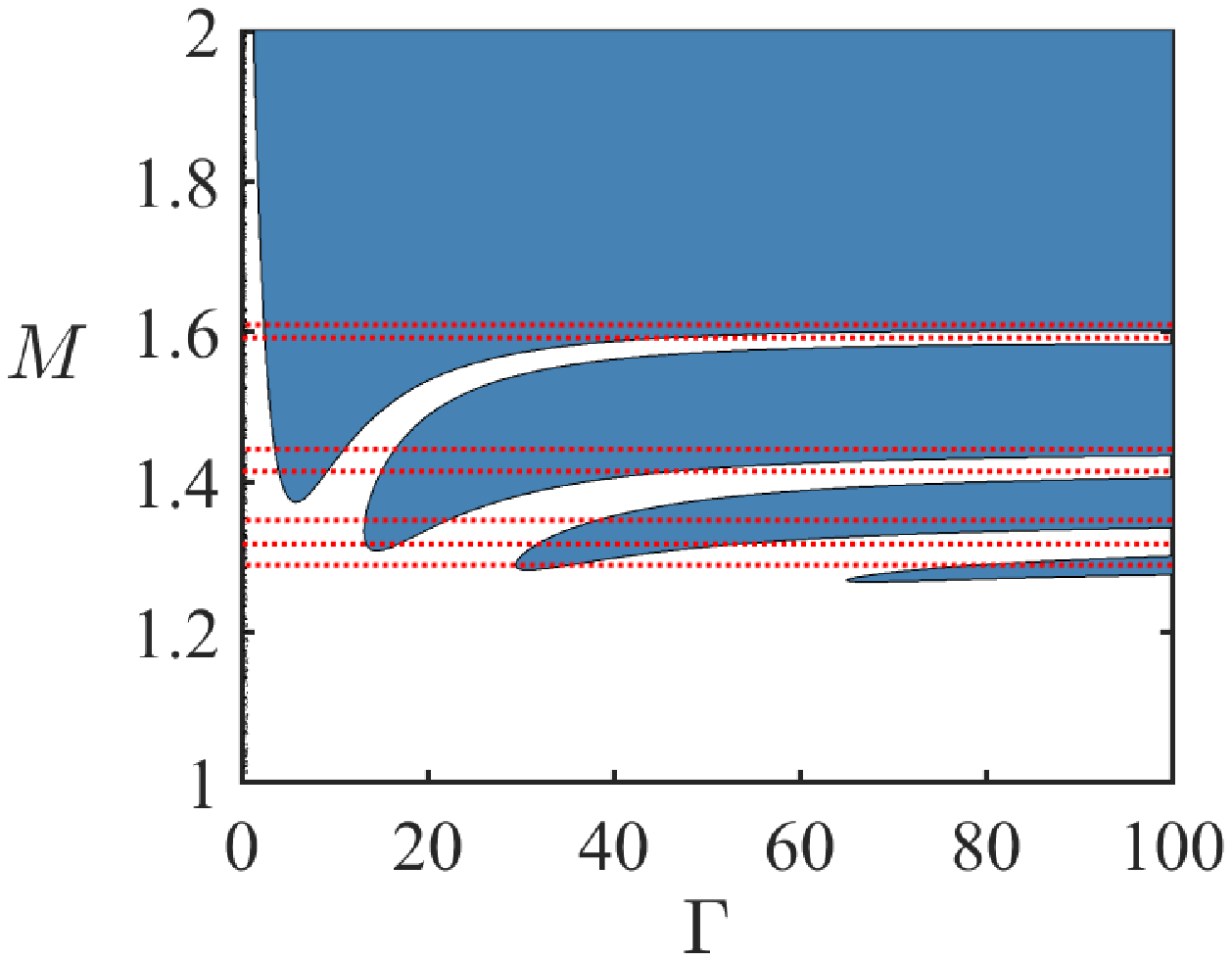} \label{FD_n3b}} \hspace*{-2em}
\subfloat[$n=3$]{
\includegraphics*[width=.33\textwidth]{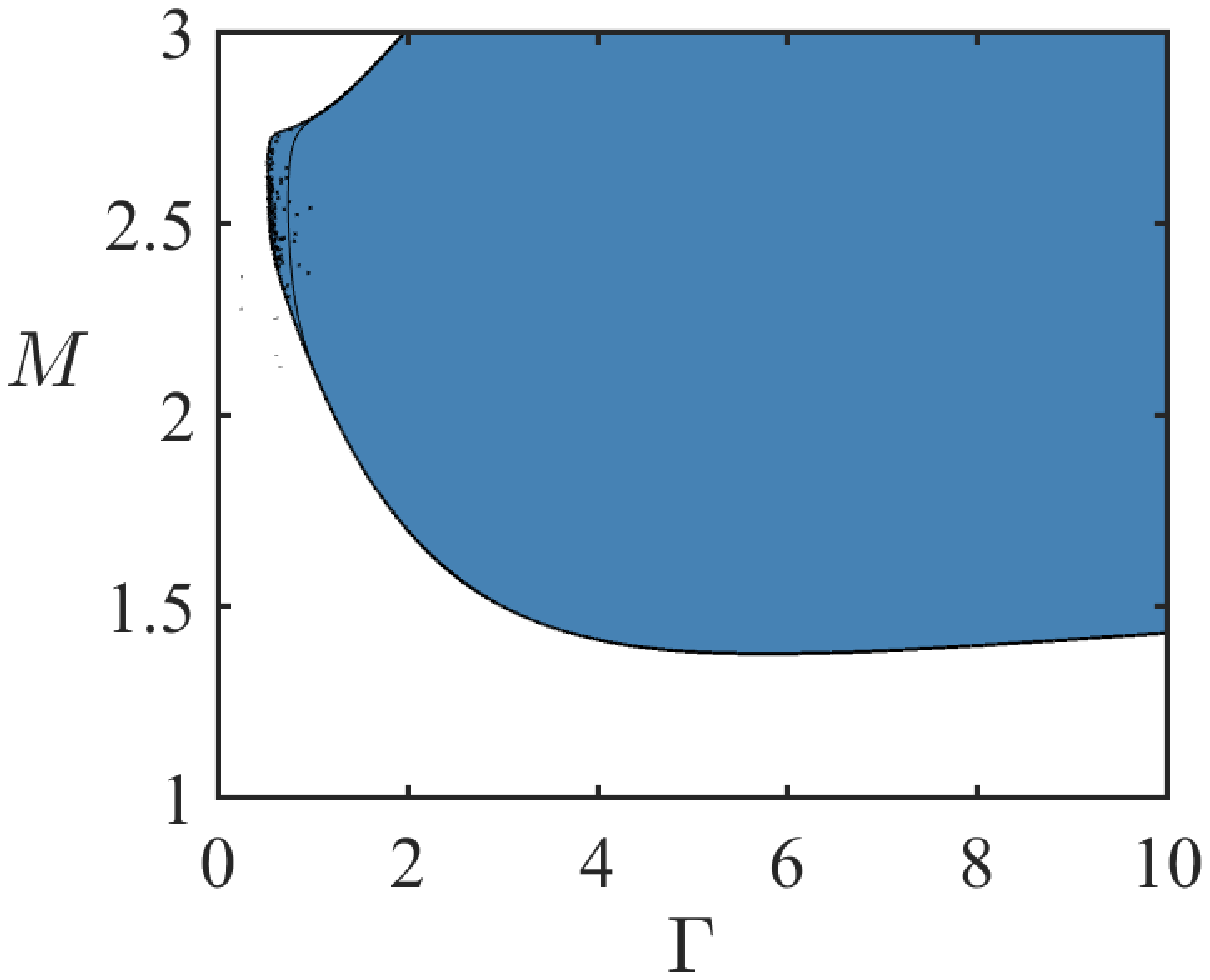}}

\caption{Instability domains (blue) of the finite depth system and finite-chord membrane for the eigenfrequencies $\omega$ recovered from the matrix pencil \rf{nep}. Parameters used are $M_w=1$, $\alpha = 10^{-4}$ and $n$ according to the legend.  Neutral stability line obtained from the first-order expansion of the shallow water growth rate \rf{gr_lwl} are shown as dotted red.  \label{FD_n12}}
\end{figure*}
\begin{figure*}
\centering
\includegraphics*[width=.7\textwidth]{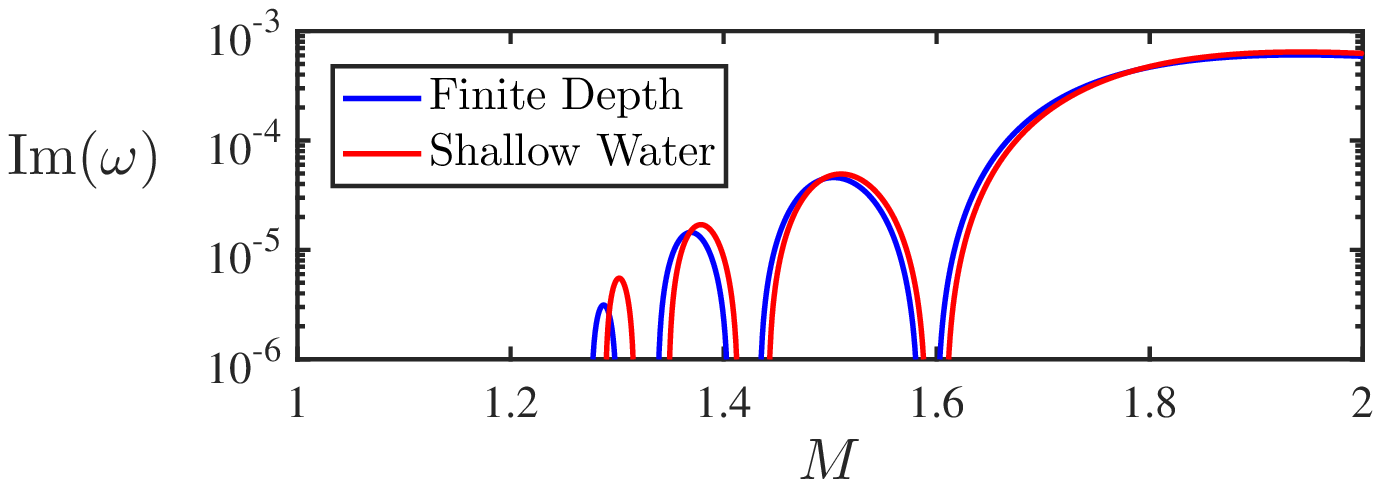} 
\label{FD_n3c}

\caption{ Finite depth growth rates converge to the shallow water ones as $\Gamma\to\infty$ for $M_w=1$, $\alpha = 10^{-4}$, $n=3$, and $\Gamma=100$. \label{FD_n3}}
\end{figure*}

Applying the computational method described above to the nonlinear eigenvalue problem defined by the matrix pencil \rf{nep}, we find frequencies and growth rates of the finite-chord Nemtsov membrane coupled to the free surface flow of finite depth. First, we are benchmarking our method against the analytical solution \rf{gr_lwl} in the shallow-water limit, corresponding to $\Gamma \to \infty$, and shown in Fig.~\ref{fig2}. In Fig.~\ref{FD_MMw} we show an analogue of Fig.~\ref{fig2} for the membrane with the chord length $\Gamma=10$ and $\alpha=10^{-4}$ that demonstrates all the structural characteristics that are present in the shallow water stability map including the tongues of intertwining flutter regions and the wide stability gap. Nevertheless, one can observe that in the case of the finite-chord membrane in the finite-depth flow some of the tongues are either separated to individual instability islands or merged into continuous instability belts. These new effects are caused by the finite values of the chord length of the membrane and the finite depth of the fluid layer.

In a similar way, we produce stability maps in the $(\mu, M)$-plane computed from the numerical solution of the algebraic nonlinear matrix pencil \rf{nep} for $\Gamma=10$ and $\alpha=10^{-4}$. Comparing the results shown in Fig.~\ref{FD_Mmu} with the analytical solutions in the shallow water limit that are visualized in Fig.~\ref{fig4}, we notice good qualitative and quantitative agreement of the two approaches.

A drawback of our first-order in $\alpha$ analytical expression \rf{gr_lwl} for the growth rates in the shallow water approximation was that the size of the membrane $\Gamma$ in it was only playing the role of a scaling factor and thus did not change the shape of the instability domains as it is the case for the finite depth solution. In contrast, the numerical solution of the nonlinear eigenvalue problem with the pencil \rf{nep} allows us to investigate zones of the radiation-induced flutter in a broad range of variation of the chord length $\Gamma$.

\begin{figure*}
\centering

\subfloat[]{
\includegraphics*[width=.33\textwidth]{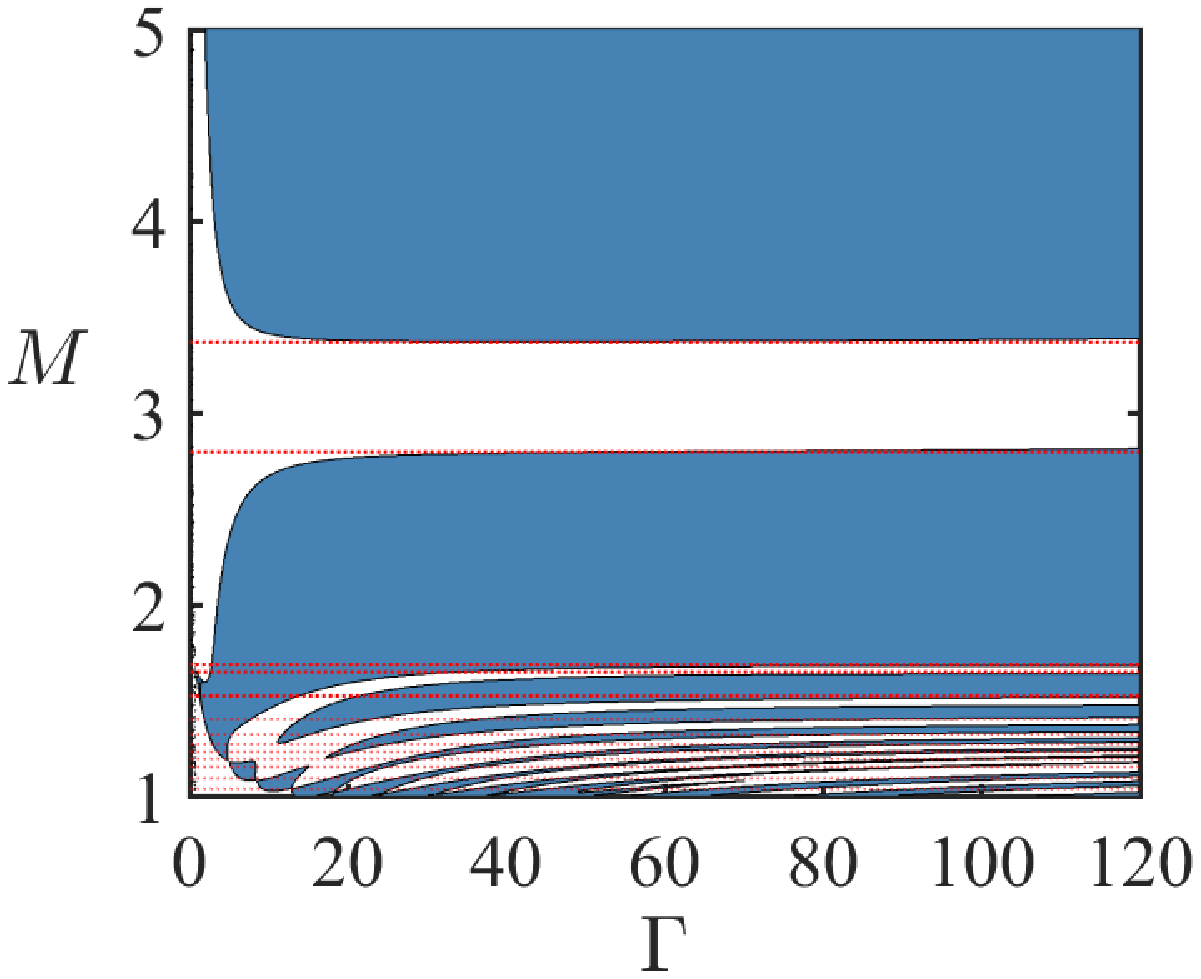} \label{FD_n4a}} \hspace*{-2em}
\subfloat[]{
\includegraphics*[width=.33\textwidth]{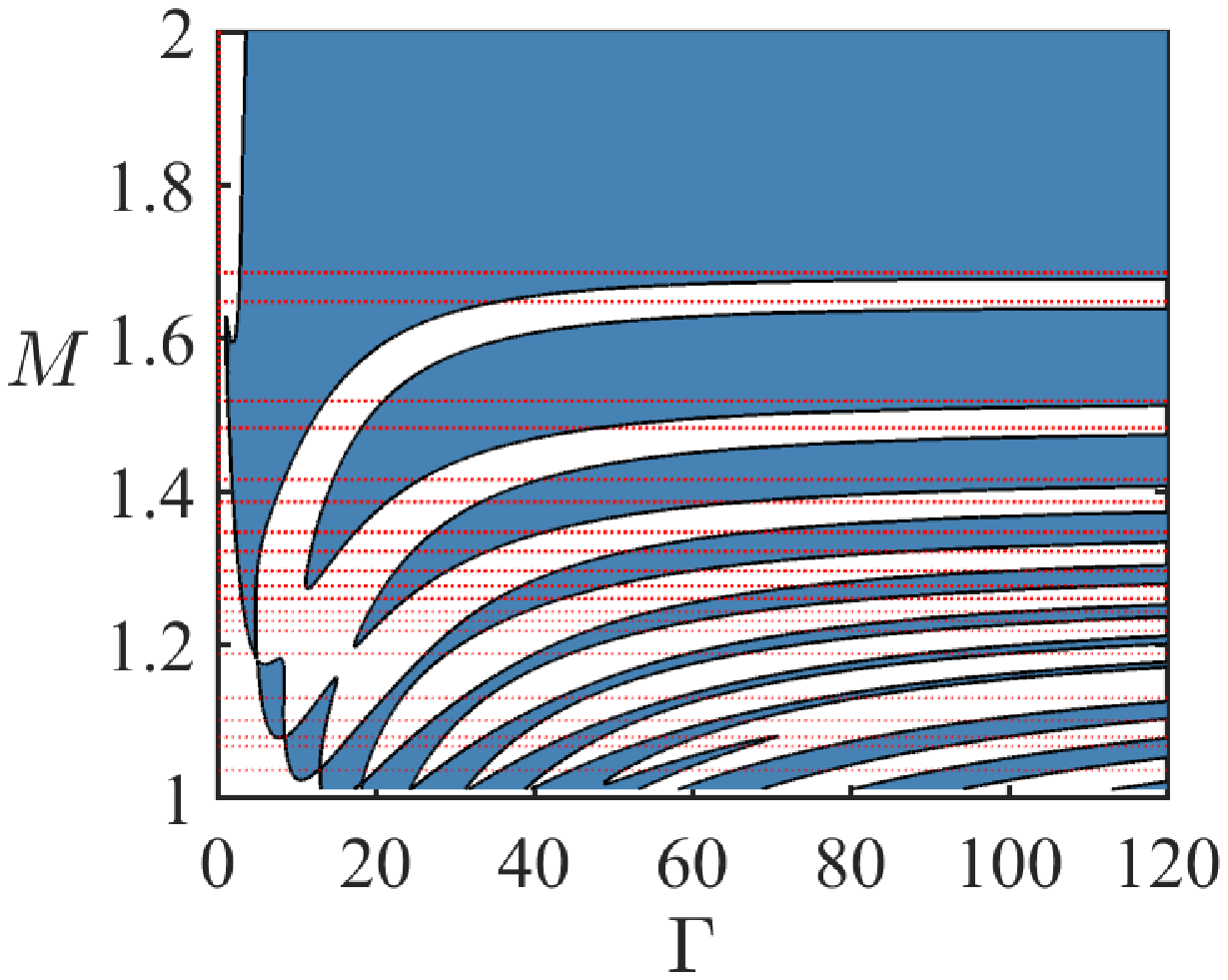} \label{FD_n4b}} \hspace*{-2em}
\subfloat[]{
\includegraphics*[width=.33\textwidth]{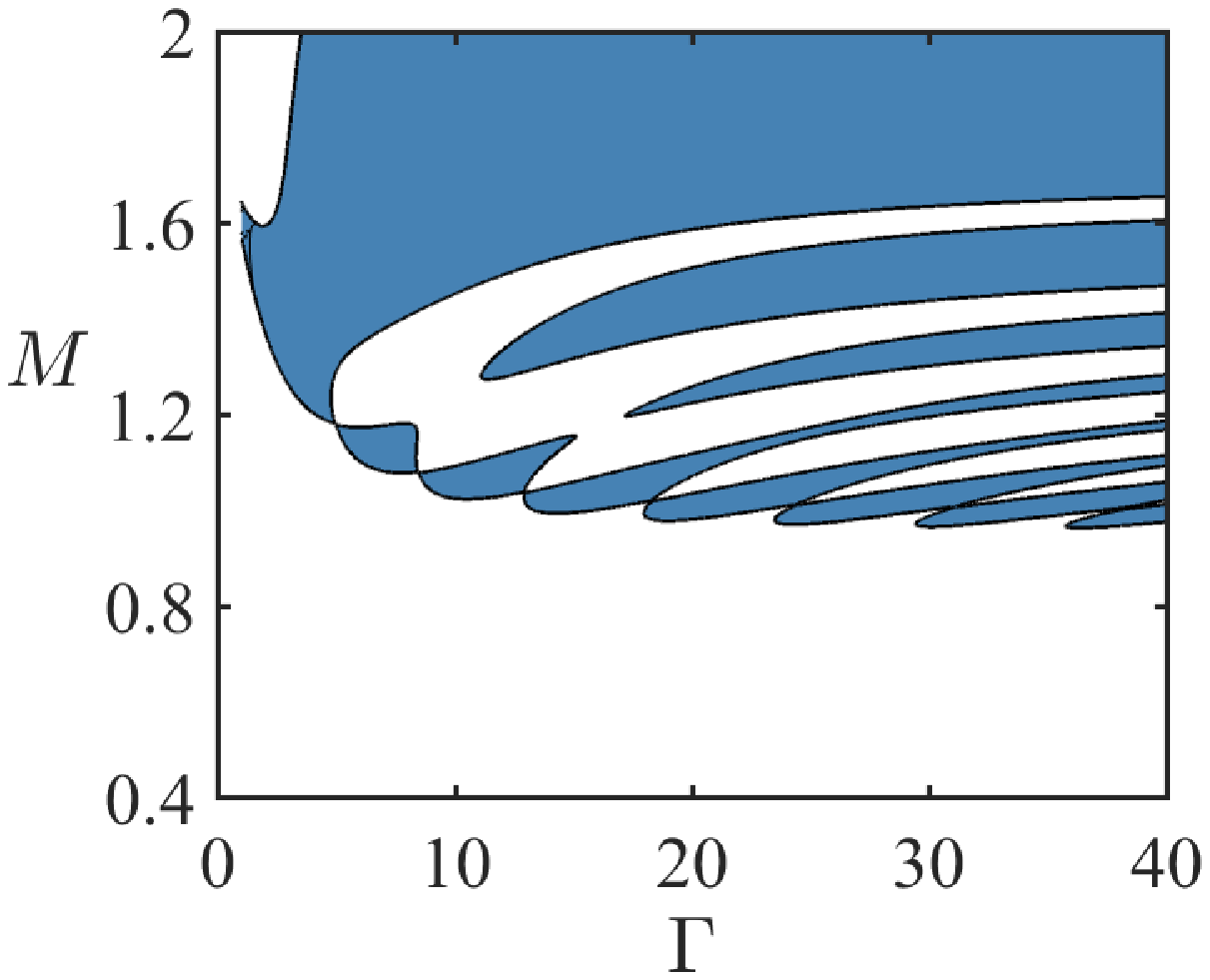}} \\ \vspace*{-1em}
\subfloat[]{
\includegraphics*[width=.6\textwidth]{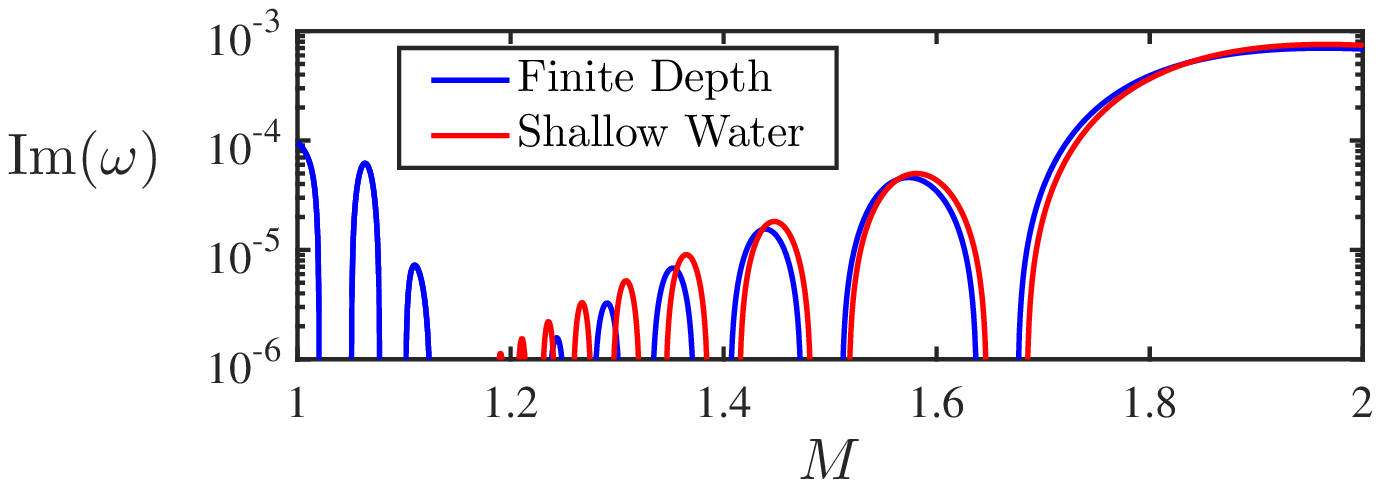} \label{FD_n4c}}

\caption{(a) Instability domain and (b,c) its close views (blue) for $M_w=1$, $\alpha = 10^{-4}$, and $n=4$. (c) The leftmost tongue never touches the axis $\Gamma = 0$. (d) The lower panel represents growth rates computed at a fixed value of $\Gamma=120$. \label{FD_n4}}
\end{figure*}

In Fig.\ref{FD_n12}(a,b) we present stability maps in the $(\Gamma, M)$-plane for the membrane modes with $n=1,2$ that show stability close to the critical values $M=1$ and $\Gamma=0$ with instability (blue domain) everywhere else. Notice that as $\Gamma\to\infty$, the lower boundary of the flutter domain for the finite depth layer tends to the horizontal neutral stability curve (shown as a red dotted line) that follows from the first-order in $\alpha$ expansion of the growth rate
\rf{gr_lwl} in the shallow water approximation. 
On the other hand, in the opposite limit of $\Gamma\to0$, corresponding to the deep water approximation, the finite-chord-length Nemtsov membrane is stable, in accordance with the perturbation analysis in section \ref{dwlfcnm}. These observations confirm that our numerical results for the finite depth problem are in a very good agreement with the analytical treatment presented in the previous sections and that our numerical procedure converges to correct eigenvalues. 

\begin{figure*}
\centering

\subfloat[$n=1$]{
\includegraphics*[width=.33\textwidth]{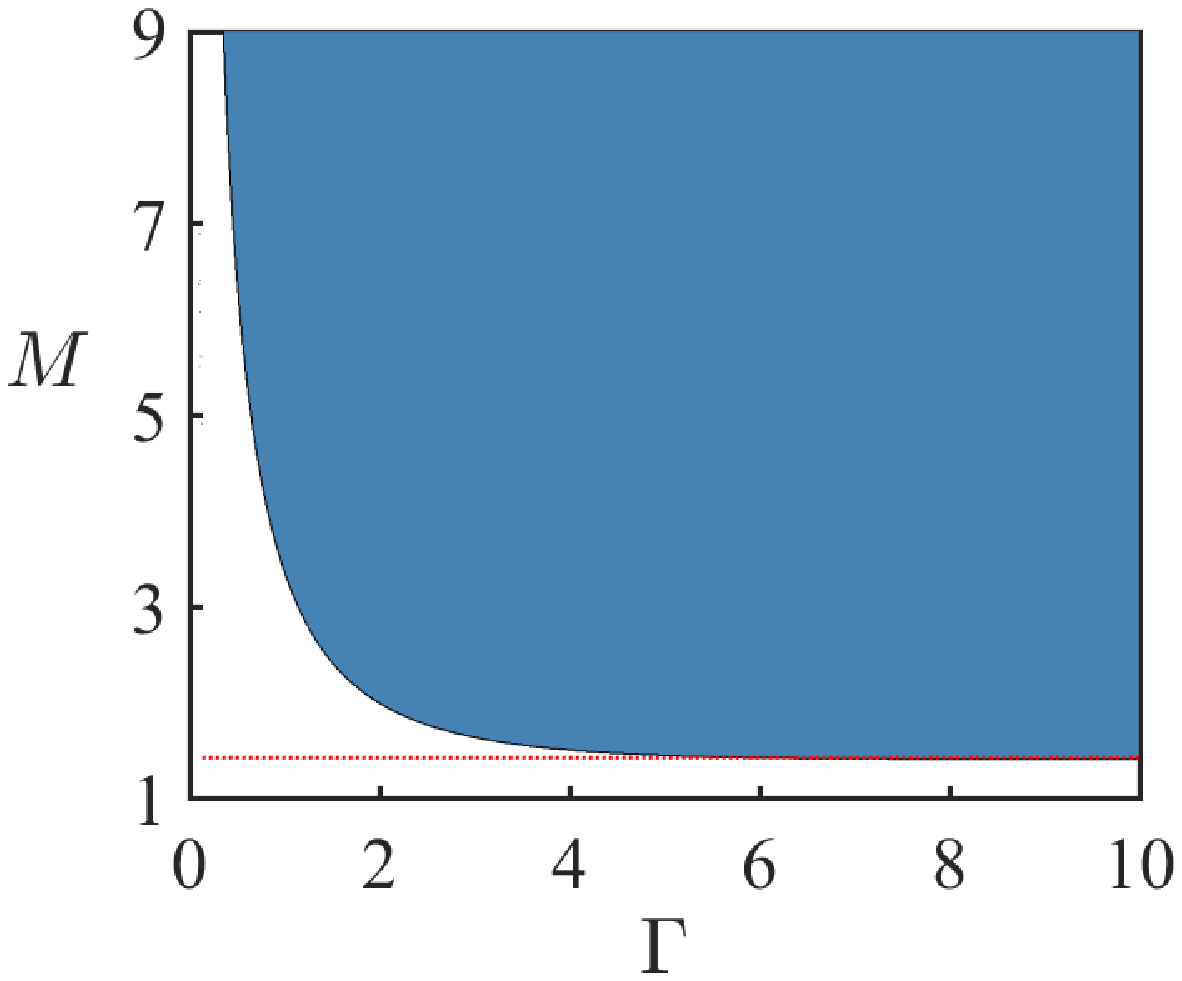} \label{FD_n1234a}} \hspace*{-3em}
\subfloat[$n=3$]{
\includegraphics*[width=.33\textwidth]{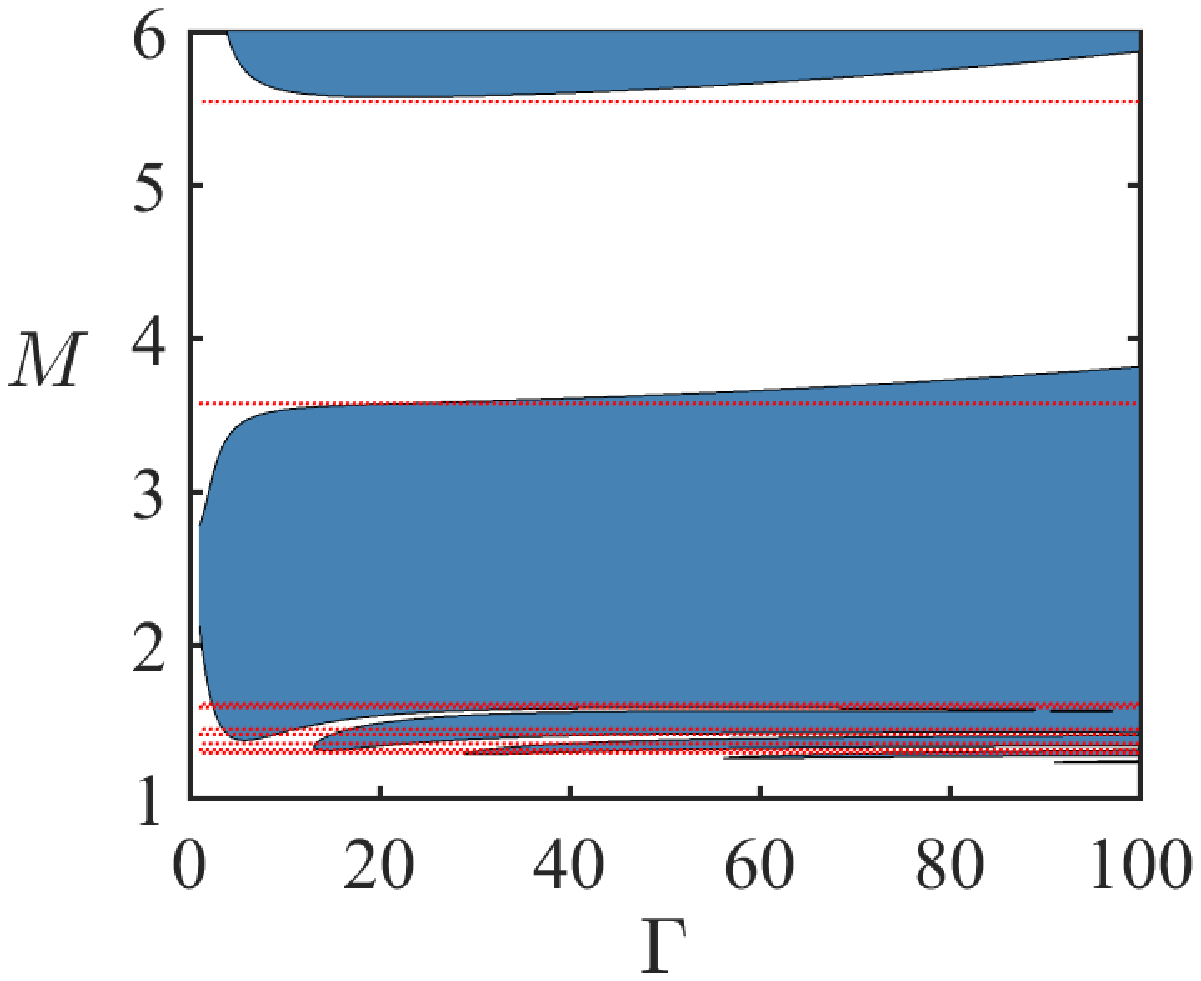} \label{FD_n1234c}} \hspace*{-3em}
\subfloat[$n=4$]{
\includegraphics*[width=.33\textwidth]{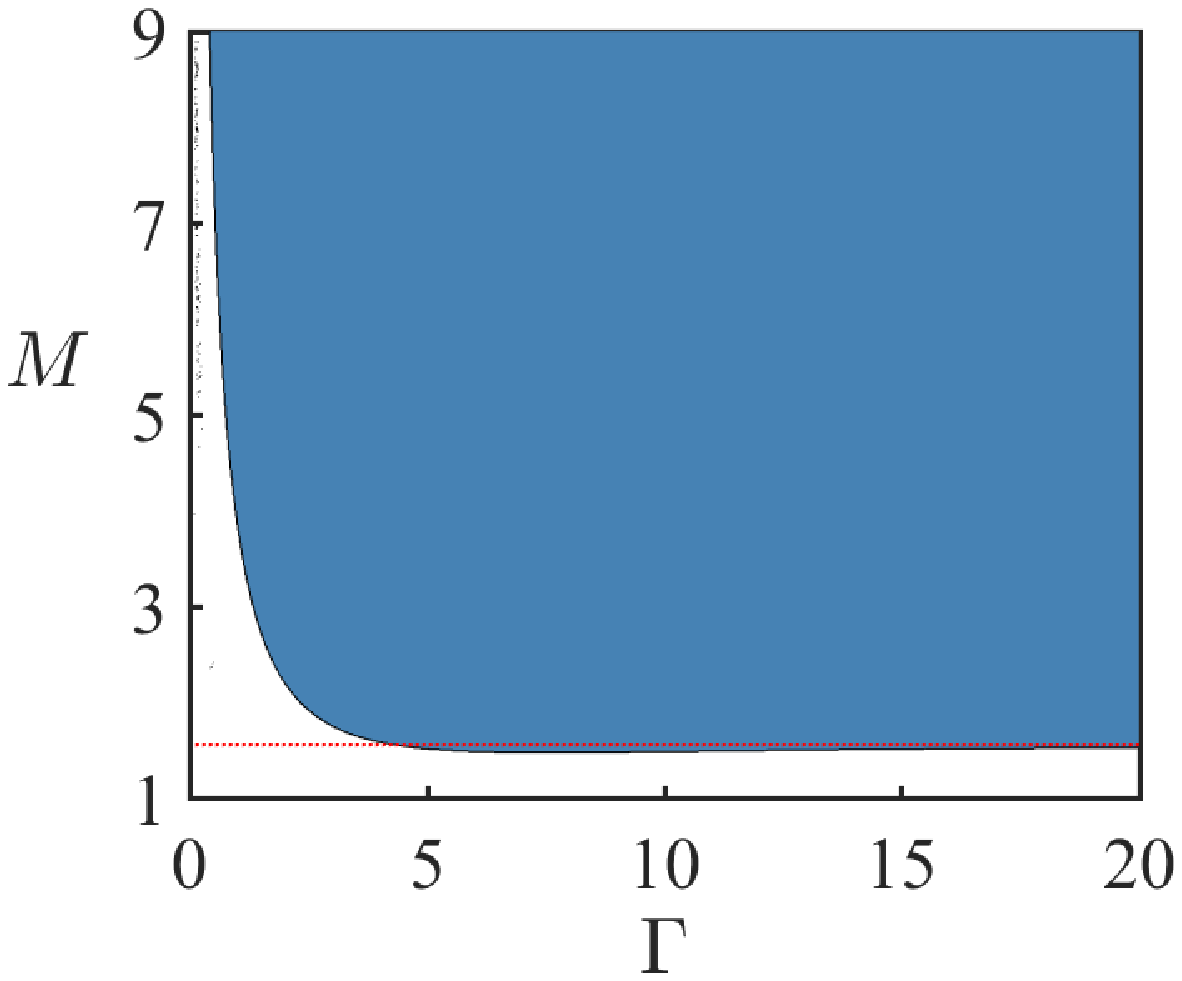} \label{FD_n1234d}} \\ \vspace*{-1em}
\subfloat[$n=2$]{
\includegraphics*[width=.33\textwidth]{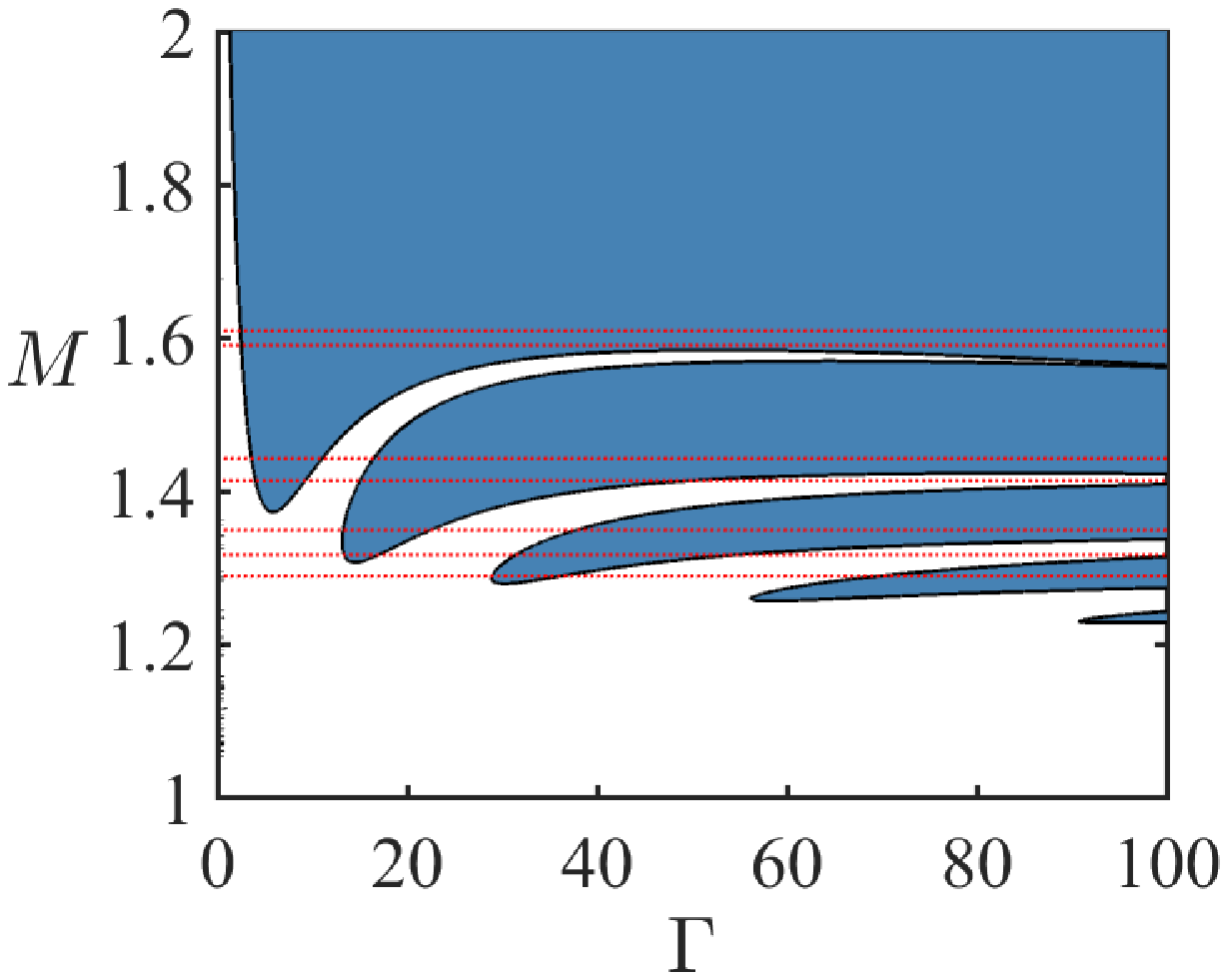} \label{FD_n1234b}} \hspace*{-2.5em}
\subfloat[$n=3$]{
\includegraphics*[width=.33\textwidth]{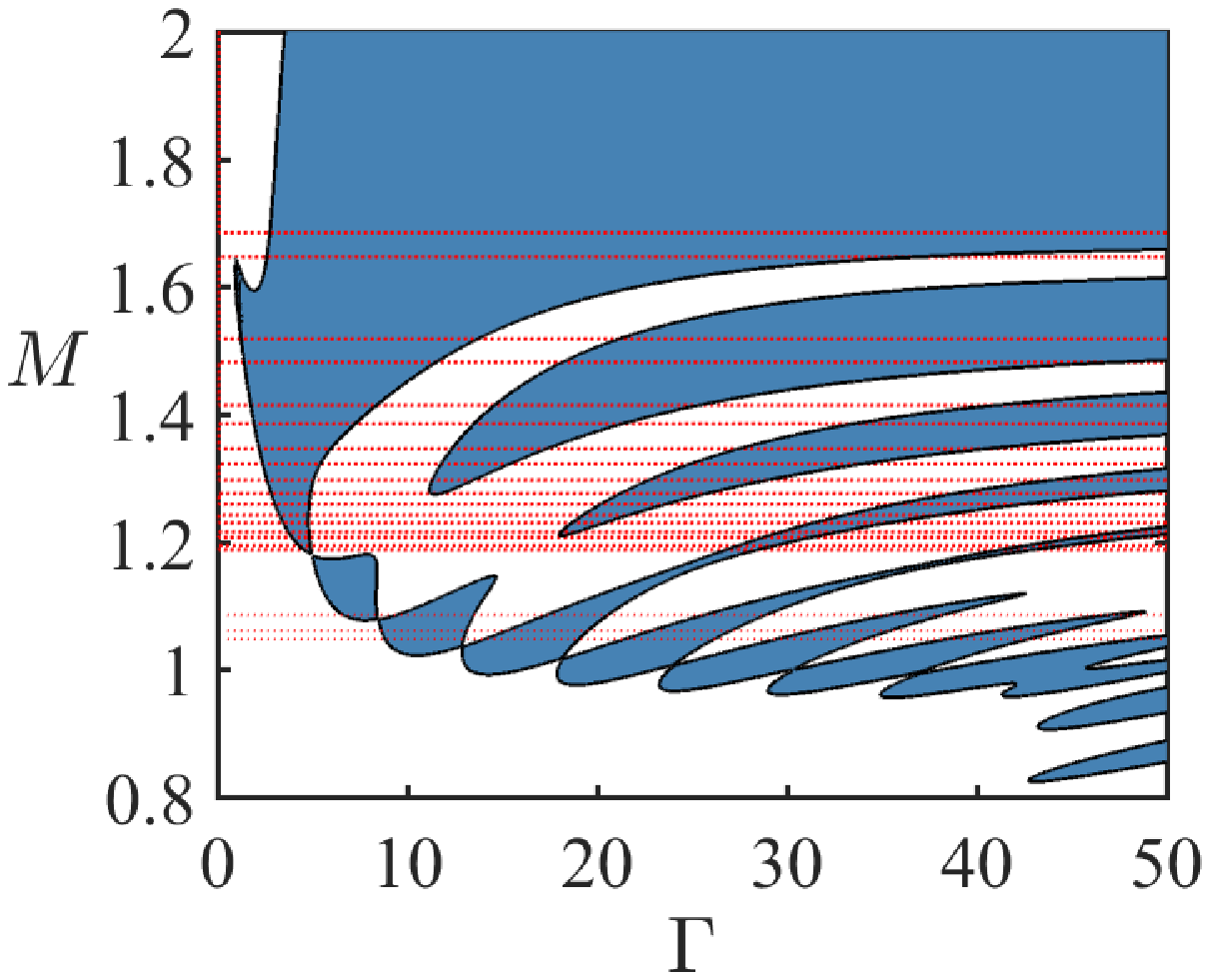} \label{FD_n1234dd}}
 
\caption{Instability domains (blue) of the finite depth system and finite-chord membrane for the eigenfrequencies $\omega$ recovered from the matrix pencil \rf{nep}. Parameters used are $M_w=1$, $\alpha = 10^{-3}$ and $n$ according to the legend. The dotted red lines represent the neutral stability curves obtained from the first-order expansion of the shallow water growth rate \rf{gr_lwl}. \label{FD_n1234} }
\end{figure*}

Exploring the stability map in the $(\Gamma,M)$-plane further for the higher-order membrane modes with $n=3$ unveils even more intriguing pattern shown in Fig.\ref{FD_n12}(c,d). First, Fig.\ref{FD_n12}(c) highlights a curious structure of two prominent subdomains in the parameter plane with the stability gap corresponding to a sector of stability also visible in the left part of Fig.~\ref{FD_n12}(c). Second, the lower subdomain in Fig.\ref{FD_n12}(c) decomposes to a bunch of instability tongues spreading along the $\Gamma$-axis, see Fig.~\ref{FD_n12}(d). Third, for every tongue, even for that commencing very close to the $M$-axis (Fig.~\ref{FD_n12}(e)), there is a critical value of the chord length $\Gamma$ such that the shorter membranes are stable, quite in accordance with the analytical results for the deep water reported in section~\ref{dwlfcnm}, cf. Fig.~\ref{fig6}. Fourth, and, probably the most rewarding, is the evidence that in the limit of $\Gamma\to\infty$ the boundaries of the instability tongues of Fig.~\ref{FD_n12}(d) converge to the shallow water solution shown by red dotted horizontal lines in Fig.~\ref{FD_n12}(d), which is also confirmed by the convergence of the corresponding growth rates shown in Fig.~\ref{FD_n3}. Therefore, the pattern of instability tongues that we discovered first in the shallow water approximation manifests itself also in the general case of the finite-chord membrane in the finite-depth fluid flow. With the increase in $n$, the instability tongues start to break and intertwine, making the pattern even more intriguing, see Fig.~\ref{FD_n4} for the stability maps corresponding to $n=4$.

As is evident in Fig.\ref{FD_n1234} modification of the added mass ratio parameter $\alpha$ by an order of magnitude from $\alpha=10^{-4}$ to $\alpha=10^{-3}$ deforms the pattern of instability regions. Although all the qualitative features remain in place, the behaviour of the stability boundaries at large values of $\Gamma$ does not demonstrate a perfect convergence to the shallow water solution as one can see in the closer views of the stability domains in Fig.\ref{FD_n1234}(e,f). Notice, however, that the red dotted lines are obtained from the first order in $\alpha$ perturbation expansion \rf{gr_lwl} of the full dispersion relation \rf{eig_eq_lwldr}.  This fact explains the discrepancy and suggests that sensitivity of stability of the Nemtsov membrane to $\alpha$ has been proven to be rather important, quite in accordance with the remark of Barbone and Crighton \cite{BC1994} that the intermediate values of the fluid-loading parameter appear to be the most complicated as it is impossible to label the mode of the fluid-solid system as corresponding solely to a `solid mode' or a `fluid mode'. This complication manifests itself in the fact that $\alpha$ is a factor of the nonlinear and non-polynomial in $\omega$ operator in \rf{nep}. By this reason, as soon as $\alpha$ departs from the origin, the choice of the free membrane mode \rf{omn} is less suitable as a first guess to initiate the Newton-like iterative process \rf{sli}, which therefore works less stably at larger values of $\alpha$. Parameter continuation proposed, e.g. in \cite{A2008} in a different setting, only slightly improved performance of our method, reflecting the fact that nonlinear eigenvalue problems of fluid-structure interaction are notoriously hard. Nevertheless, our method allowed to obtain new results in the classical problem in the unprecedentedly broad range of all other important parameters, including, first of all the chord length, dimensionless velocity of the flow, and speed of propagation of elastic waves in the membrane.

 \section{Concluding Remarks}

In this paper we studied conditions for the onset of a radiation-induced instability of the Nemtsov membrane in a uniform flow with free surface. In contrast to previous works \cite{LK2020,N1985} that were limited either by the shallow water approximation or by the assumption that the membrane has infinite chord length, we consider the problem in its entirety and take into account both the finite chord of the membrane and the finite depth of the fluid layer.

First, we derive a new integro-differential equation for the deflection of the finite-chord membrane that is coupled to the finite-depth flow.

Then, we develop an analytical procedure allowing to find the eigenvalues of the membrane that is weakly coupled to the flow in the shallow- and deep water approximations. Our original contribution is a systematic procedure that combines Laplace transform, residue calculus, and perturbation of eigenvalues. The analytical solution allowed us to plot detailed stability maps and find a new pattern of intertwining instability tongues that to the best of our knowledge has not been previously reported in the literature. Furthermore, we were able to find analytically the geometrical structure that governs position, orientation and self-intersections of the instability tongues.

Next, we developed an original numerical method to treat the finite-chord membrane in the finite-depth uniform flow with the free surface. With a combination of complex analysis and Galerkin discretization we reduced the boundary eigenvalue problem for an integro-differential equation to an algebraic non-polynomial nonlinear eigenvalue problem and solved it with a Newton-like method. This approach allowed us to explore the onset of instability with respect to the chord length of the membrane, velocity of the flow, and speed of elastic waves propagating along the membrane at small but finite values of the added mass ratio parameter that plays a role of an effective damping due to radiation of surface gravity waves.

We believe we have made a convincing case that the Nemtsov membrane is able to play the same paradigmatic role for understanding radiation-induced instabilities as the famous Lamb oscillator coupled to a string has played for understanding radiation damping. We believe that our procedure is applicable to a broad class of fluid-structure interaction problems that require solving nonlinear eigenvalue problems. An extension of it allowing for a significant continuation with respect to the coupling parameter has proven to be a harder topic that we leave beyond the scope of this paper.

\section*{Aknowledgments}
We thank Nabil Achour and J\'er\^ome Mougel from IMFT Toulouse for communicating to us their recent numerical results that, in particular, independently confirm our analysis. We are grateful to Prof. M. L. Overton from the Courant Institute for helpful discussions. We thank the London Mathematical Society for
supporting Prof. Overton's visit to Northumbria through the Scheme 4 Research in Pairs grant No 41820. J.~L. was supported by a Ph.D. Scholarship from Northumbria University that also provided him with an opportunity to run the code on the High Performance Cluster. The research
of O.~N.~K. was supported in part by the Royal Society Grant No. IES{\textbackslash}R1{\textbackslash}211145.

\end{document}